\documentclass[]{aa}  

\usepackage{graphicx}
\usepackage{color}
\usepackage{txfonts}
\begin{document}

   \title{Signatures of the non-Maxwellian $\kappa$-distributions in optically thin line spectra}

   \subtitle{I. Theory and synthetic Fe IX--XIII spectra}

   \author{J. Dud\'ik\inst{1,2}\thanks{RS Newton International Fellow}
          \and G. Del Zanna\inst{1}
	  \and H. E. Mason\inst{1}
	  \and E. Dzif\v{c}\'akov\'a\inst{3}
          }

   \institute{Department of Applied Mathematics and Theoretical Physics, CMS, University of Cambridge, Wilberforce Road,
	      Cambridge CB3 0WA, United Kingdom
              \email{J.Dudik@cam.ac.uk}
         \and Department of Astronomy, Physics of the Earth and Meteorology, Faculty of Mathematics, Physics and Informatics,
	      Comenius University, 842 48 Bratislava, Slovak Republic
         \and Astronomical Institute, Academy of Sciences of the Czech Republic, 25165 Ond\v{r}ejov, Czech Republic,\\
               \email{elena@asu.cas.cz}              
             }

   \date{Received ; accepted }

  \abstract
   {}
   {We investigate the possibility of diagnosing the degree of departure from the Maxwellian distribution using single-ion spectra originating in astrophysical plasmas in collisional ionization equilibrium.}
   {New atomic data for excitation of \ion{Fe}{IX}\,--\,\ion{Fe}{XIII} are integrated under the assumption of a $\kappa$-distribution of electron energies. Diagnostic methods using lines of a single ion formed at any wavelength are explored. Such methods minimize uncertainties from the ionization and recombination rates, as well as the possible presence of non-equilibrium ionization. Approximations to the collision strengths are also investigated.}
   {The calculated intensities of most of the \ion{Fe}{IX}\,--\,\ion{Fe}{XIII} EUV lines show consistent behaviour with $\kappa$ at constant temperature. Intensities of these lines decrease with $\kappa$, with the vast majority of ratios of strong lines showing little or no sensitivity to $\kappa$. Several of the line ratios, especially involving temperature-sensitive lines, show a sensitivity to $\kappa$ that is of the order of several tens of per cent, or, in the case of \ion{Fe}{IX}, up to a factor of two. Forbidden lines in the near-ultraviolet, visible, or infrared parts of the spectrum are an exception, with smaller intensity changes or even a reverse behaviour with $\kappa$. The most conspicuous example is the \ion{Fe}{X} 6378.26\AA~red line, whose intensity incerases with $\kappa$. This line is a potentially strong indicator of departures from the Maxwellian distribution. We find that it is possible to perform density diagnostics independently of $\kappa$, with many \ion{Fe}{XI}, \ion{Fe}{XII}, and \ion{Fe}{XIII} line ratios showing strong density-sensitivity and negligible sensitivity to $\kappa$ and temperature. We also tested different averaging of the collision strengths. It is found that averaging over 0.01 interval in log$(E [\mathrm{Ryd}])$ is sufficient to produce accurate distribution-averaged collision strengths $\Upsilon(T,\kappa)$ at temperatures of the ion formation in ionization equilibrium.}
   {}

   \maketitle
%
\section{Introduction}
\label{Sect:1}

The central assumption when analyzing spectra originating from optically thin astrophysical plasmas in collisional ionization equilibrium is that the electron distribution function is Maxwellian and therefore that the equilibrium is always ensured locally. This may, however, not be true if there are correlations between the particles in the system. These correlations may be induced by any long-range interactions in the emitting plasma \citep{Collier04,Livadiotis09,Livadiotis10,Livadiotis11a,Livadiotis13}, such as wave-particle interactions, shocks, or particle acceleration and associated streaming of fast, weakly collisional particles from the reconnection site. Under such conditions, the distribution function may depart from the Maxwellian one.

Although still debatable \citep{Storey14a,Storey14b}, a claim has been made that the $\kappa$-distributions, characterized by a high-energy tail, can explain the observed spectra of planetary nebulae \citep{Binette12,Nicholls12,Nicholls13,Dopita13}. The deviations from the Maxwellian distribution required to explain the observed spectra of ions such as \ion{O}{III} or \ion{S}{III} spectra were small, with $\kappa$\,$\approx$\,20. These values are much larger than the $\kappa$ measured in situ in the solar wind, which is observed to be strongly non-Maxwellian, typically with $\kappa$\,$>$\,2.5 \citep[e.g.][]{Collier96,Maksimovic97a,Maksimovic97b,Zouganelis08,Livadiotis10,LeChat11}.

Since $\kappa$-distributions are observed in the solar wind, a question arises whether these could originate in the solar corona. Although attempts to detect high-energy particles have been made \citep{Feldman07,Hannah10}, their presence is still unclear. \citet{Feldman07} used bi-Maxwellian spectral modelling of the line intensities of He-like lines observed by SUMER \citep{Wilhelm95}. The second Maxwellian was assumed to have a temperature of 10\,MK. The authors argued that the presence of this high-temperature Maxwellian was not neccessary to explain the observed spectra. However, the analysis was limited to Maxwellian distributions and did not include the effects of $\kappa$-distributions on the spectra. \citet {Hannah10} used the RHESSI instrument \citep{Lin02} to observe an off-limb quiet-Sun region and derived strong constraints on the number of particles at energies of several keV. However, the derived limits on the emission measure of the plasma at temperatures of several MK are still somewhat large, and increase with increasing $\kappa$.

On the other hand, \citet{Scudder13} argue that stellar coronae above 1.05$R_\odot$ should be strongly non-Maxwellian. \citet{Dzifcakova11} analyzed the \ion{Si}{III} line intensities reported by \citet{Pinfield99} and concluded that the observed spectra can be explained by $\kappa$-distributions with $\kappa$\,$\approx$\,7 for the active region, and $\kappa$\,$\approx$\,11--13 for the quiet Sun and coronal hole. The analysis also works under the assumption of a differential emission measure. If these diagnosed values are correct, it could be expected that the solar corona should have values of $\kappa$ between these and those observed in the solar wind. Analysis of the effect of $\kappa$-distributions on the intensities of lines observed by the Hinode/EIS instrument \citep{Culhane07} have been made for Fe lines by \citet{Dzifcakova10} and for non-Fe lines by \citet{Mackovjak13}. These authors used the atomic data corresponding to CHIANTI v5.3 \citep{Dere97,Landi06} and v7 \citep{Landi12}, with the excitation cross sections for $\kappa$-distributions being recovered using an approximate parametric method \citep{DzifcakovaMason08}. Although some indications of the presence of $\kappa$-distributions were obtained from the \ion{O}{IV}--\ion{O}{V} and \ion{S}{X}--\ion{S}{XI} lines, the authors could not exclude the presence of multithermal effects. Futhermore, the diagnostics had to include lines formed at neighbouring ionization stages, because the wavelength range of the EIS instrument is limited (170\AA\,--\,211\AA~and 246\AA\,--\,292\AA). The observed lines of a single ion have by neccessity similar excitation thresholds, which makes their behaviour with $\kappa$ similar. Including the lines formed in neighbouring ionization stages increases the sensitivity of the diagnostics to $\kappa$, but complicates the analysis in the multithermal case. Moreover, the assumption of ionization equilibrium is a possible additional source of uncertainties. The diagnostics of $\kappa$ performed by these authors was also limited by density effects. Nevertheless, \citet{Dzifcakova10} showed that some density-sensitive line ratios can be used for diagnostics of electron density independently of the $\kappa$ value.

A successful diagnostic of $\kappa$-distributions has been performed in flares. \citet{Kasparova09} showed that some coronal sources observed by RHESSI \citep{Lin02} in partially occulted flares can be fitted with a $\kappa$-distribution. \citet{Oka13} showed that $\kappa$-distributions provide a good fit to the observed high-energy tail, and dispense with the need for a low-energy cut-off. However, the authors showed that a near-Maxwellian component is also present at lower energies in the flare studied.

In recent years, significant progress has been made in the calculation of atomic data for collisional excitation of astrophysically important ions \citep[e.g.][]{Liang09,Liang10,Liang12,ODwyer12,DelZanna10a,DelZanna10b,DelZanna11,DelZanna11b,DelZanna12a,DelZanna12b,DelZanna13,DelZanna14}. These calculations represent significant improvements to previous atomic data, since they include a large number of levels and the associated cascading and resonances. These cross sections for electron excitation are in several cases significantly different from the previous ones. These atomic data will be implemented in the next version 8 of the CHIANTI database. Significant progress has also been made on line identifications of complex coronal Fe ions \citep[e.g.][]{DelZanna04,DelZanna14,DelZanna05,Young09b,DelZanna10c,DelZanna11,DelZanna12d,DelZanna12e}.

For these reasons, in this paper we use the original cross sections, which allows us to dispense with the approximative parametric method of \citet{DzifcakovaMason08}. This paper is a first in a series of papers exploring the effect of non-Maxwellian $\kappa$-distributions on the spectra arising from optically thin plasma in collisional ionization equilibrium. Here, we focus on the ions formed in solar and stellar coronae at temperatures of $\approx$1--2\,MK. The atomic data calculations are exploited by calculating the synthetic spectra for \ion{Fe}{IX} -- \ion{Fe}{XIII} for $\kappa$-distributions throughout the entire wavelength range of spectral line formation. This is done to include the strong forbidden transitions in the visible or infrared part of the electromagnetic spectrum \citep[see e.g.][]{Habbal13}. The \ion{Fe}{IX}--\ion{Fe}{XIII} ions are chosen since they produce some of the strongest lines that are observed even during minima of the solar cycle, i.e. in absence of active regions \citep{Landi10}.

The paper is structured as follows. The $\kappa$-distributions are described in Sect. \ref{Sect:2}. The method used to calculate the relative level populations and synthetic spectra using generalized distribution-averaged collision strengths is described in Sect. \ref{Sect:3}. Averaging of the collision strengths is investigated in Sect. \ref{Appendix:Omega}. Synthetic spectra are presented in Sect. \ref{Sect:4}, with the influence of $\kappa$-distributions on density-sensitive line ratios studied in Sect. \ref{Sect:5}. Section \ref{Sect:6} explores the possibilities of diagnosing $\kappa$ simultaneously with $T$ using lines of only one ion. Conclusions are given in Sect. \ref{Sect:7}.

%
\begin{figure}
	\centering
	\includegraphics[width=8.8cm]{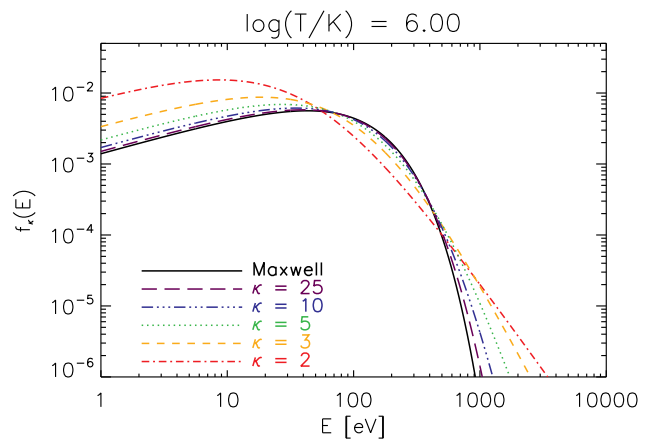}
\caption{The $\kappa$-distributions with $\kappa$\,=\,2, 3, 5, 10, 25, and the Maxwellian distribution plotted for log($T$/K)\,=\,6.0. Different colors and line styles correspond to different values of $\kappa$.
\label{Fig:kappa}}
\end{figure}
%
%
\begin{figure}
	\centering
	\includegraphics[width=8.8cm]{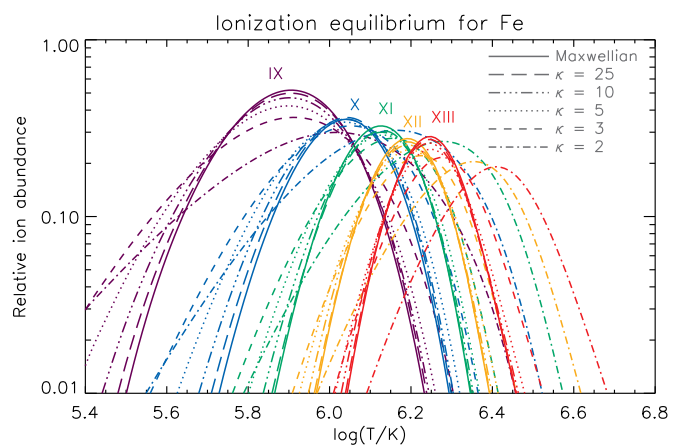}

\caption{Ionization equilibrium for the $\kappa$-distributions. Relative ion abundances for \ion{Fe}{IX}--\ion{Fe}{XIII} are plotted as a function of $T$ for $\kappa$\,=\,2, 3, 5, 10, 25, and the Maxwellian distribution.
\label{Fig:Ioneq}}
\end{figure}
%

\section{The non-Maxwellian $\kappa$-distributions}
\label{Sect:2}

The $\kappa$-distribution (Fig. \ref{Fig:kappa}, \textit{top}) is a distribution of electron energies characterized by a power-law high-energy tail \citep{Owocki83,Livadiotis09,Livadiotis13}
        \begin{equation}
                f(E,\kappa) \mathrm{d}E = A_{\kappa} \frac{2}{\sqrt{\pi} (k_\mathrm{B}T)^{3/2}} \frac{E^{1/2}\mathrm{d}E}{\left 			  (1+ \frac{E}{(\kappa - 3/2) k_\mathrm{B}T} \right)^{\kappa+1}}\,,
                \label{Eq:Kappa}
        \end{equation}
where ${\cal A}_{\kappa}$\,=\,$\Gamma(\kappa+1)$/$\left(\Gamma(\kappa-1/2) (\kappa-3/2)^{3/2}\right)$ is the normalization constant and $k_\mathrm{B}$\,=\,1.38 $\times 10^{-16}$ erg\,K$^{-1}$ is the Boltzmann constant; $T$ and $\kappa$\,$\in$\,$\left(3/2,+\infty\right)$ are the parameters of the distribution. The parameter $\kappa$ changes the shape of distribution function from $\kappa$\,$\rightarrow$\,3/2 corresponding to the highest deviation from Maxwellian distribution, to $\kappa$\,$\rightarrow$\,$\infty$ corresponding to the Maxwellian distribution.

The mean energy $\left< E \right> = {3k_\mathrm{B}T}/{2}$ of a $\kappa$-distribution is independent of $\kappa$, so that $T$ can be defined as the temperature also for $\kappa$-distributions. The reader is referred to the paper of \citet{Livadiotis09} for a more detailed discussion of why $T$ can be properly defined as the thermodynamic temperature in the framework of the generalized Tsallis statistical mechanics \citep{Tsallis88,Tsallis09}. 

%
%
\section{Method}
\label{Sect:3}
%
\subsection{Spectral line emissivities}
\label{Sect:3.1}

The emissivity $\varepsilon_{ji}$ of an optically thin spectral line arising from a transition $j \to i$, $j > i$, in a $k$-times ionized ion of the element $X$, is given by \citep[e.g.][]{Mason94,Phillips08}
	\begin{eqnarray}
		\varepsilon_{ji} &=& \frac{hc}{\lambda_{ji}} A_{ji} n(X_j^{+k}) = \frac{hc}{\lambda_{ji}} \frac{A_{ji}}{n_\mathrm{e}} \frac{n(X_j^{+k})}{n(X^{+k})} \frac{n(X^{+k})}{n(X)} A_X n_\mathrm{e} n_\mathrm{H}\\
		\nonumber &=& A_X G_{X,ji}(T,n_\mathrm{e},\kappa) n_\mathrm{e} n_\mathrm{H}\,,
		\label{Eq:line_emissivity}
	\end{eqnarray}
where $h$\,$\approx$\,6.62\,$\times$\,10$^{-27}$ erg\,s is the Planck constant, $c$\,$\approx$\,3\,$\times$10$^{10}$\,cm\,s$^{-1}$ represents the speed of light, $\lambda_{ji}$ is the line wavelength, $A_{ji}$ the corresponding Einstein coefficient for the spontaneous emission, $n(X_j^{+k})$ is the density of the ion $+k$ with electron on the excited upper level $j$, $n(X^{+k})$ the total density of ion $+k$, $n(X)$\,$\equiv$\,$n_X$ the total density of element $X$ whose abundance is $A_X$, and $n_\mathrm{H}$ the hydrogen density. The function $G_{X,ji}(T,n_\mathrm{e},\kappa)$ is the contribution function for the line $\lambda_{ji}$. The observed intensity $I_{ji}$ of the spectral line is then given by the integral of emissivity along the line of sight $l$, i.e.
	\begin{equation}
		I_{ji} = \int A_X G_{X,ji}(T,n_\mathrm{e},\kappa) n_\mathrm{e} n_\mathrm{H} \mathrm{d}l\,,
		\label{Eq:line_intensity}
	\end{equation}
where the quantity $EM$\,=\,$\int n_\mathrm{e} n_\mathrm{H} \mathrm{d}l$ is called the emission measure of the plasma.

%
\subsection{Ionization rates and ionization equilibrium}
\label{Sect:3.2}

The ratio $n(X^{+k})/n_X$ is the relative abundance of ion $n(X^{+k})$, and is therefore given by the ionization equilibrium. Since the ionization and recombination rates are a function of $T$ and $\kappa$, the ratio $n(X^{+k})/n_X$ is also a function of $T$ and $\kappa$ \citep[e.g.][]{Dzifcakova92,Dzifcakova02,Wannawichian03,Dzifcakova13}. Here, we use the latest available ionization equilibria for $\kappa$-distributions calculated by \citet{Dzifcakova13}. These use the same atomic data for ionization and recombination as the ionization equilibrium for the Maxwellian distribution available in the CHIANTI database, v7.1 \citep{Landi13,Dere07,Dere97}.

Figure \ref{Fig:Ioneq} shows the behaviour of the relative ion abundances of \ion{Fe}{IX} -- \ion{Fe}{XIII} with $\kappa$. The ionization peaks are in general wider for lower $\kappa$, and are shifted to higher $T$ for low $\kappa$\,=\,2--3 \citep{Dzifcakova13}. Compared to the Maxwellian distribution, the ionization peak can be shifted up to $\Delta$log($T$/K)\,$\approx$\,0.15 for $\kappa$\,=\,2. Therefore, the \ion{Fe}{IX} -- \ion{Fe}{XIII} line intensities are expected to peak at higher log ($T$/K) if $\kappa$\,=\,2--3.

%
\subsection{Excitation rates and excitation equilibrium}
\label{Sect:3.3}

The relative level populations, described by the ratios $n(X_j^{+k})/n(X^{+k})$ are obtained here under the assumption of an equilibrium situation. This assumption means that the total number of transitions from the level $j$ to any other level $m$ must be balanced by the total number of transitions from the other levels $m$ to the level $j$. In coronal conditions, the transitions are due to collisional excitation and de-excitation by an impacting electron, as well as spontaneous radiative decay. If we denote the rates of the collisional excitation and de-excitation as $C_{jm}^\mathrm{e}$ and $C_{jm}^\mathrm{d}$, the excitation equilibrium can be obtained by solving the following equations for each level $j$,
	\begin{eqnarray}
		\nonumber \sum\limits_{j>m} n(X^{+k}_j) n_\mathrm{e} C_{jm}^\mathrm{d} + \sum\limits_{j<m} n(X^{+k}_j) n_\mathrm{e} C_{jm}^\mathrm{e} + \sum\limits_{j>m} n(X^{+k}_j) A_{jm} = \\
		= n(X^{+k}_j) \left( \sum\limits_{j<m} n_\mathrm{e} C_{mj}^\mathrm{d} + \sum\limits_{j>m} n_\mathrm{e} C_{mj}^\mathrm{e} + \sum\limits_{j<m} A_{mj} \right)\,, 
		\label{Eq:Excit_eq}
	\end{eqnarray}
where the left-hand side of Eq. (\ref{Eq:Excit_eq}) contains transitions $j$\,$\to$\,$m$, and the right-hand side contains transitions $m$\,$\to$\,$j$. Equation (\ref{Eq:Excit_eq}) must be supplemented with the condition that
	\begin{equation}
		\sum\limits_{j} n(X^{+k}_j) = n(X^{+k})\,.\\	 
		\label{Eq:Excit_sum}
	\end{equation}

The electron collisional excitation rate for the upward transition $i$\,$\to$\,$j$ is given by the expression
	\begin{equation}
		C_{ij}^\mathrm{e}(T,\kappa) = \sqrt{\frac{2}{m_\mathrm{e}}}\frac{\pi a_0^2}{\omega_i} I_H \int\limits_0^{+\infty} \Omega_{ij}(E_i) E_i^{-1/2} f_\kappa(E_i) \mathrm{d}E_j\,, \\
		\label{Eq:Excit_rate}
	\end{equation}
where $a_0$ is the Bohr radius, $m_\mathrm{e}$ is the electron mass, $I_H$\,$\approx$\,13.6\,eV\,$\equiv$\,1\,Ryd is the hydrogen ionization energy, $\omega_{i}$ is the statistical weight of the level $i$, $E_i$ is the incident electron energy energy, and $E_j$\,=\,$E_i$\,$-$\,$\Delta E_{ji}$ is the final electron energy. The quantity $\Omega_{ji}(E_j)$\,=\,$\Omega_{ij}(E_i)$ is the collision strength (non-dimensionalized cross section), given by
	\begin{equation}
		\Omega_{ji}(E_j) = \omega_j \frac{E_j}{I_H} \frac{\sigma_{ji}^\mathrm{d}(E_j)}{\pi a_0^2} = \omega_i \frac{E_i}{I_H} \frac{\sigma_{ij}^\mathrm{e}(E_i)}{\pi a_0^2}\,, \\
		\label{Eq:Omega}
	\end{equation}
where the $\sigma_{ji}^\mathrm{e}$ and $\sigma_{ij}^\mathrm{d}$ are the cross sections for the electron impact excitation and de-excitation, respectively. The collisional de-excitation is essentially a reverse process, so that
	\begin{equation}
		C_{ji}^\mathrm{d}(T,\kappa) = \sqrt{\frac{2}{m_\mathrm{e}}}\frac{\pi a_0^2}{\omega_j} I_H \int\limits_0^{+\infty} \Omega_{ji}(E_j) E_j^{-1/2} f_\kappa(E_j) \mathrm{d}E_i\,. \\
		\label{Eq:Deexcit_rate}
	\end{equation}
%

%
\subsection{Collision strengths datasets}
\label{Sect:3.4}

In this paper, we use the collision strength data obtained by R-matrix calculations \citep{Badnell97,Badnell11}. The individual atomic data were calculated by the APAP team\footnote{www.apap-network.org} for \ion{Fe}{IX}, \ion{Fe}{X}, \ion{Fe}{XII}, and \ion{Fe}{XIII} by \citet{DelZanna14}, \citet{DelZanna10b}, \citet{DelZanna12a}, and \citet{DelZanna12b}, respectively. For \ion{Fe}{XI}, the atomic data of \citet{DelZanna13} are used, except transitions involving levels 37, 39, and 41 ($J$\,=\,1 levels in the 3s$^2$\,3p$^3$\,3d configuration), for which the target failed to provide accurate energies and collision strengths. We use the earlier atomic data of \citet{DelZanna10a} for the transitions involving these three levels. This has to be done, because using incorrect collision strengths for transitions that have strong observed intensities \citep[see][Table 2 therein]{DelZanna10a} will affect the calculated intensities of other lines through Eq. (\ref{Eq:Excit_eq}).

An example of the collision strength for the \ion{Fe}{XI} 257.57\,\AA~line is provided in Fig. \ref{Fig:Averaged}, \textit{top right} in Appendix \ref{Appendix:Omega}. Typically, the collision strengths are calculated in non-uniform grid of incident energy $E_i$ that covers all of the excitation tresholds and resonance regions. It spans from $E_i$\,=\,0 up to several tens of Ryd as needed, and contains up to ten thousand $E_i^{(q)}$ points. The high-energy limit ($E_i$\,$\to$\,$+\infty$) is also provided. 

The $A_{ji}$ values required to calculate the relative level populations (Eq. \ref{Eq:Excit_eq}) for each ion are also provided \citep{DelZanna14,DelZanna10b,DelZanna13,DelZanna10a,DelZanna12a,DelZanna12b}.

%
\subsection{Calculation of the distribution-averaged collision strengths for $\kappa$-distributions}
\label{Sect:3.5}

It is advantageous to define the generalized distribution-averaged collision strengths Upsilon and Downsilon, $\Upsilon_{ij}(T,\kappa)$ and \rotatebox[origin=c]{180}{$\Upsilon$}$_{ji}(T,\kappa)$ \citep{Bryans06}
	\begin{eqnarray}
		\Upsilon_{ij}(T,\kappa) = \frac{\sqrt{\pi}}{2} \mathrm{exp}\left(\frac{\Delta E_{ji}}{k_\mathrm{B}T}\right)  \int\limits_0^{+\infty} \Omega_{ij}(E_i) \left(\frac{E_i}{k_\mathrm{B}T}\right)^{-\frac{1}{2}} f_\kappa(E_i) \mathrm{d}E_j\,,
		\label{Eq:Upsilon} \\
		\rotatebox[origin=c]{180}{$\Upsilon$}_{ji}(T,\kappa) = \frac{\sqrt{\pi}}{2} \int\limits_0^{+\infty} \Omega_{ji}(E_j) \left(\frac{E_j}{k_\mathrm{B}T}\right)^{-\frac{1}{2}} f_\kappa(E_j) \mathrm{d}E_i\,,
		\label{Eq:Downsilon}
	\end{eqnarray}
so that the collision excitation and de-excitation rates are given by
	\begin{eqnarray}
		C_{ij}^\mathrm{e}(T,\kappa) = \left(\frac{2\pi}{m_\mathrm{e}k_\mathrm{B}T}\right)^{1/2} \frac{2 a_0^2}{\omega_i} I_H \mathrm{exp}\left(-\frac{\Delta E_{ji}}{k_\mathrm{B}T}\right) \, \Upsilon_{ij}(T,\kappa)\,, \label{Eq:Excit_rate_Upsilon} \\
		C_{ji}^\mathrm{d}(T,\kappa) = \left(\frac{2\pi}{m_\mathrm{e}k_\mathrm{B}T}\right)^{1/2} \frac{2 a_0^2}{\omega_j} I_H \, \rotatebox[origin=c]{180}{$\Upsilon$}_{ji}(T,\kappa)\,. \\
		\label{Eq:Deexcit_rate_Downsilon}
	\end{eqnarray}
After substituting Eq. (\ref{Eq:Kappa}) to Eqs. (\ref{Eq:Upsilon}) and (\ref{Eq:Downsilon}), the expressions for $\Upsilon_{ij}(T,\kappa)$ and \rotatebox[origin=c]{180}{$\Upsilon$}$_{ji}(T,\kappa)$ can be written as
	\begin{eqnarray}
		\Upsilon_{ij}(T,\kappa) = A_\kappa \mathrm{exp}\left(\frac{\Delta E_{ji}}{k_\mathrm{B}T}\right)  \int\limits_0^{+\infty} \frac{\Omega_{ji}(E_j)}{ \left(1+ \frac{E_j +\Delta E_{ji}}{(\kappa-3/2)k_\mathrm{B}T}\right)^{\kappa+1}} \,\mathrm{d}\left(\frac{E_j}{k_\mathrm{B}T}\right)\,,
		\label{Eq:Upsilon_kappa} \\
		\rotatebox[origin=c]{180}{$\Upsilon$}_{ji}(T,\kappa) = A_\kappa \int\limits_0^{+\infty} \frac{\Omega_{ji}(E_j)}{\left(1+ \frac{E_j}{(\kappa-3/2)k_\mathrm{B}T}\right)^{\kappa+1}} \,\mathrm{d}\left(\frac{E_j}{k_\mathrm{B}T}\right)\,.
		\label{Eq:Downsilon_kappa}
	\end{eqnarray}

These quantities can then be used to calculate the corresponding excitation and de-excitation rates for $\kappa$-distributions (Eqs. (\ref{Eq:Excit_rate}) and (\ref{Eq:Deexcit_rate})) in a similar manner to the $\Upsilon_{ij}(T)$ commonly used for the Maxwellian distribution \citep[][]{Seaton53,Burgess92,Mason94,Bradshaw13}, for which the equality $\Upsilon_{ij}(T)$\,$\equiv$\,\rotatebox[origin=c]{180}{$\Upsilon$}$_{ji}(T)$ holds.

In the CHIANTI database \citep{Dere97,Landi13}, the $\Upsilon$s are fed directly to the \textit{pop\_solver.pro} routine used to calculate the relative level populations. The \textit{pop\_solver.pro} routine utilizes a matrix solver in conjunction with the supplied collisional and radiative rates to calculate the relative level populations $n(X_j^{+k})/n(X^{+k})$. This routine was modified for $\kappa$-distributions in conjunction with the $\Upsilon$s and \rotatebox[origin=c]{180}{$\Upsilon$}s obtained by numerical calculation of Eqs. (\ref{Eq:Upsilon_kappa}) and (\ref{Eq:Downsilon_kappa}). Excitation by proton-ion collisions was not taken into account, as it is negligible compared to the excitation by electron-ion collisions.

The numerical calculation of $\Upsilon_{ij}(T,\kappa)$ proceeds as follows. For each transition, the integral is approximated by the sum over the $N$ invididual energy points $E_j^{(q)}$ at which the collision strength $\Omega_{ji}(E_j)$ is calculated (0\,$\leqq$\,$q$\,$\leqq$\,$N$\,$-$\,1). A substitution $u$\,=\,$\left(E_j +\Delta E_{ji}\right) / \left((\kappa-3/2)k_\mathrm{B}T \right)$ is then performed. Next, following \citet{Burgess92} and \citet{Bryans06}, the $\Omega_{ji}(u)$ is approximated by a straight line between points $u(E_j^{(q)})$ and $u(E_j^{(q+1)})$, so that
	\begin{equation}
		\Omega_{ji}\left(u^{(q)} \leqq u \leqq u^{(q+1)}\right) = w_1^{(q)} u +w_0^{(q)}\,,\\
		\label{Eq:Omega_straight}
	\end{equation}
where $w_{0,1}^{(q)}$ are constants within the energy interval $E_j^{(q)}$\,$\leqq$\,$E_j$\,$\leqq$\,$E_j^{(q+1)}$. Finally, Eq. (\ref{Eq:Upsilon_kappa}) is analytically integrated to obtain
	\begin{eqnarray}
		\nonumber \Upsilon_{ij}(T,\kappa) =  A_\kappa (\kappa-3/2) \mathrm{exp}\left(\frac{\Delta E_{ji}}{k_\mathrm{B}T}\right) \times \hspace{3.0cm} \\
		\times \sum\limits_{q=0}^{N-1} \left[-\frac{1}{\kappa} \left(w_1^{(q)}u+w_0^{(q)}\right)(1+u)^{-\kappa} - \frac{w_1^{(q)}}{\kappa(\kappa-1)}  (1+u)^{1-\kappa} \right]_{E_j^{(q)}}^{E_j^{(q+1)}}\,.
		\label{Eq:Upsilon_num}
	\end{eqnarray}
The \rotatebox[origin=c]{180}{$\Upsilon$}$_{ji}(T,\kappa)$ can be obtained in a similar manner.

The high-energy limit $E_j$\,$\to$\,$+\infty$ is in the numerical calculations set to $10^5$\,Ryd. Between the last energy point $E_j$ and the value of $E_j$\,=\,10$^5$\,Ryd, $\Omega_{ji}$ is approximated by a straight line in the scaled \citet{Burgess92} domain. We note that the type of the scaling depends on the type of the transition \citep{Burgess92}. Additional energy points are added to fill the space between the last $E_j^{(q)}$ point and the $E_j$\,=\,10$^5$\,Ryd. We found that about $\approx$10$^3$ points are required. We tested that the $\Upsilon_{ij}(T,\kappa)$ and \rotatebox[origin=c]{180}{$\Upsilon$}$_{ji}(T,\kappa)$ calculated are not sensitive either to the choice of the high-energy limit or to the number of additional points. This is because the sub-integral expression in Eqs. (\ref{Eq:Upsilon_kappa}) and (\ref{Eq:Downsilon_kappa}) decreases steeply with $E_j$ for any $\kappa$\,$>$\,3/2.

\begin{figure*}[!ht]
   \centering
   \includegraphics[width=8.8cm,bb= 0 48 498 283,clip]{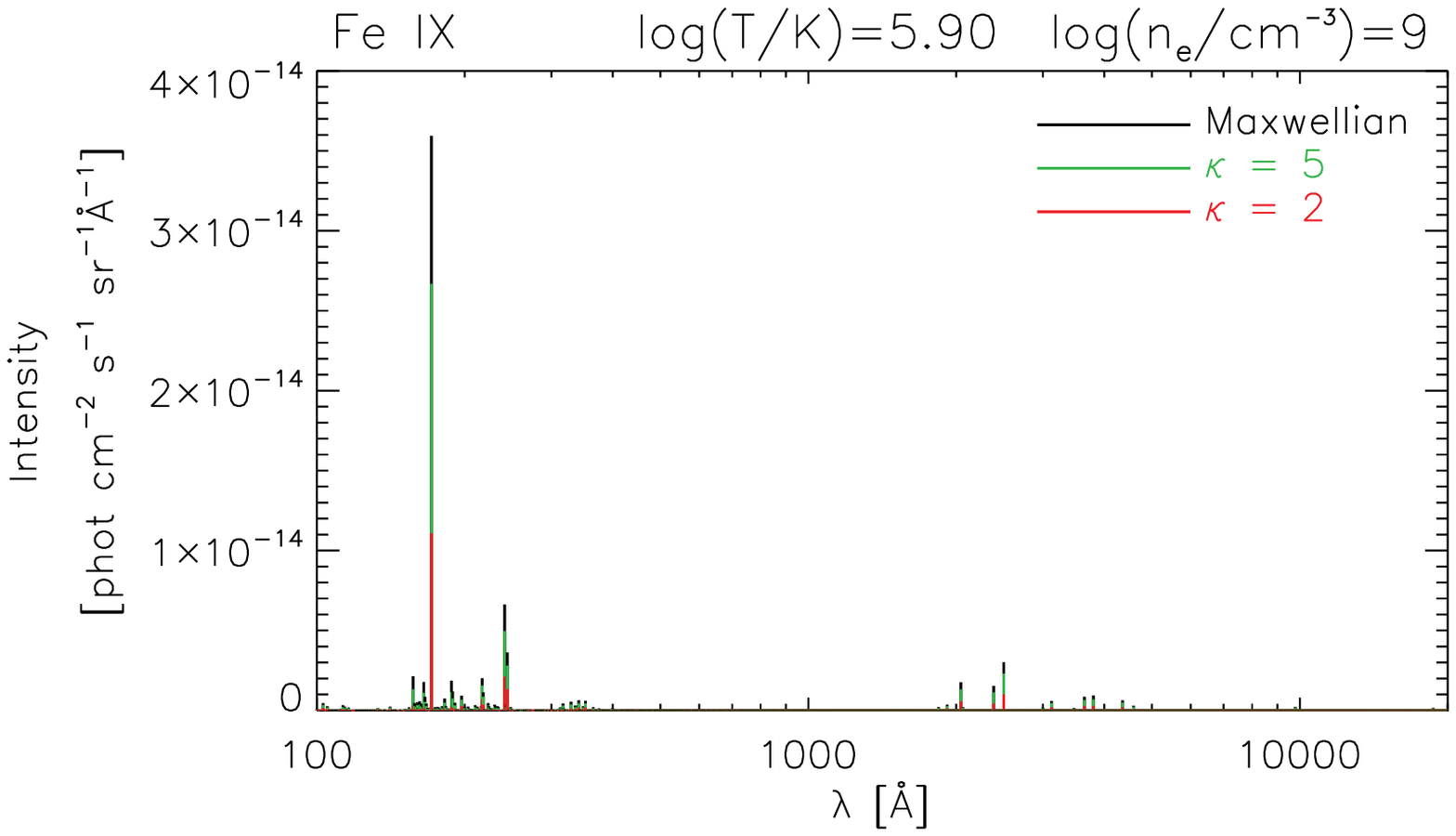}
   \includegraphics[width=8.8cm,bb= 0 48 498 283,clip]{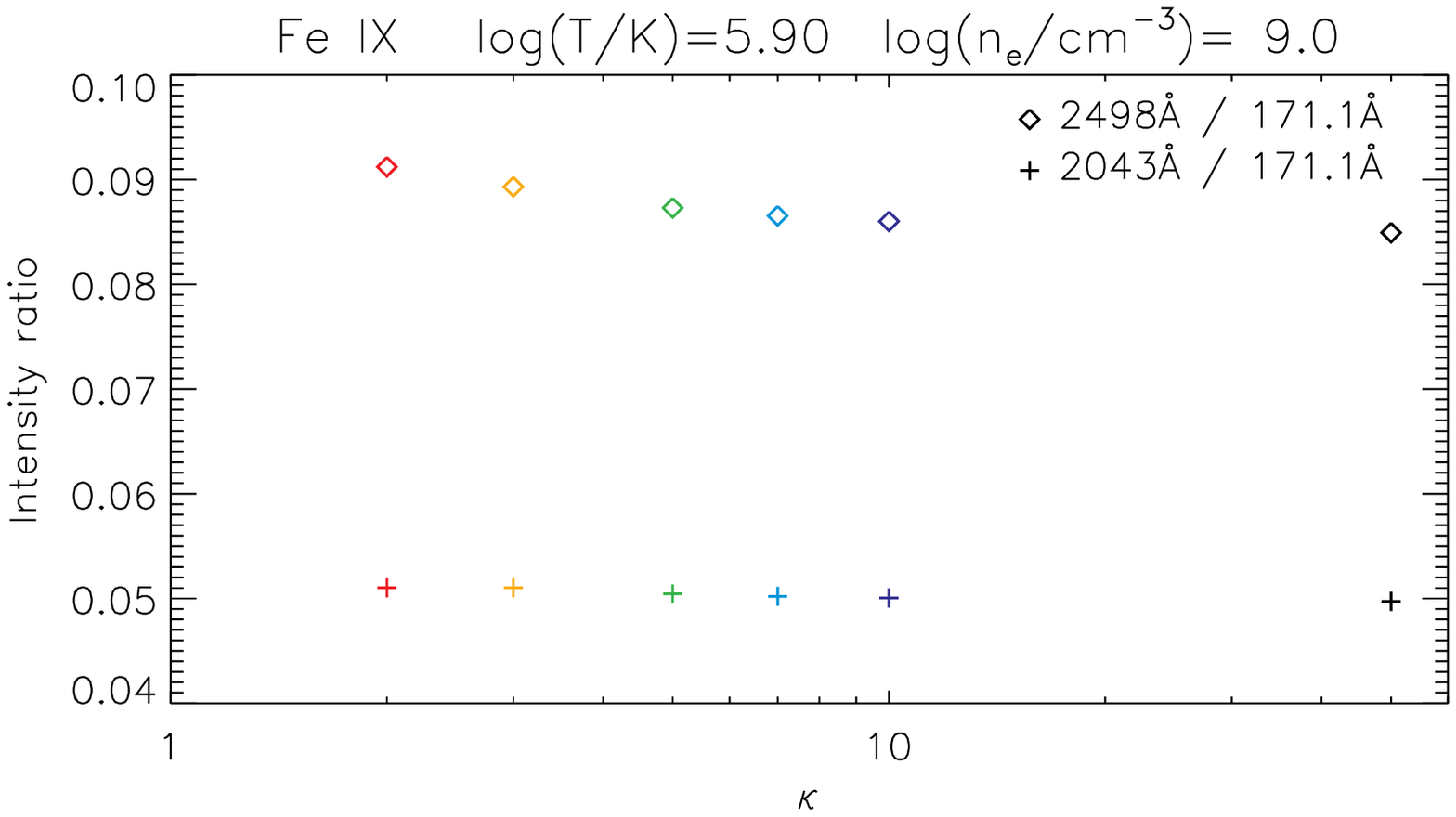}
   \includegraphics[width=8.8cm,bb= 0 48 498 283,clip]{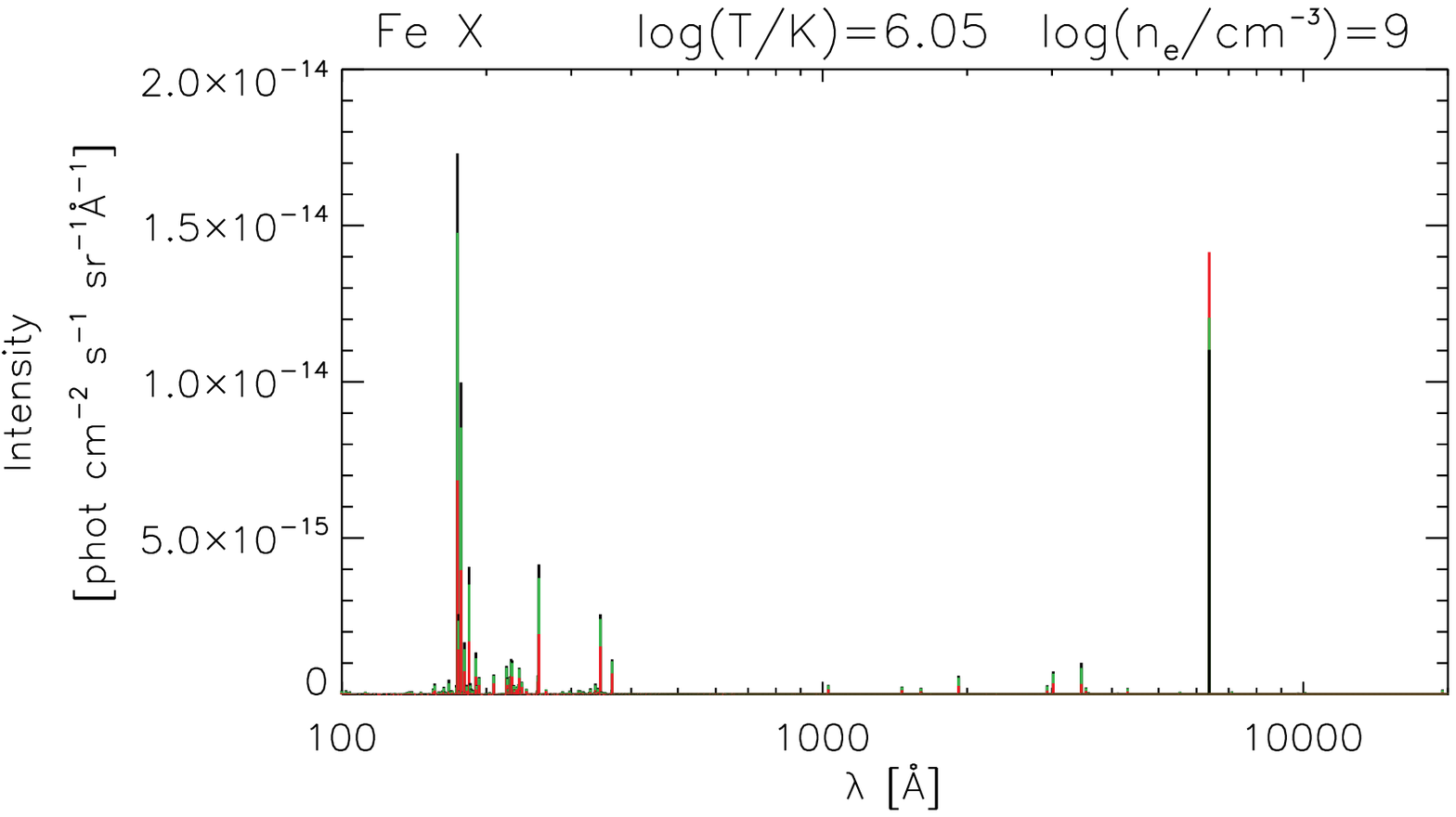}
   \includegraphics[width=8.8cm,bb= 0 48 498 292,clip]{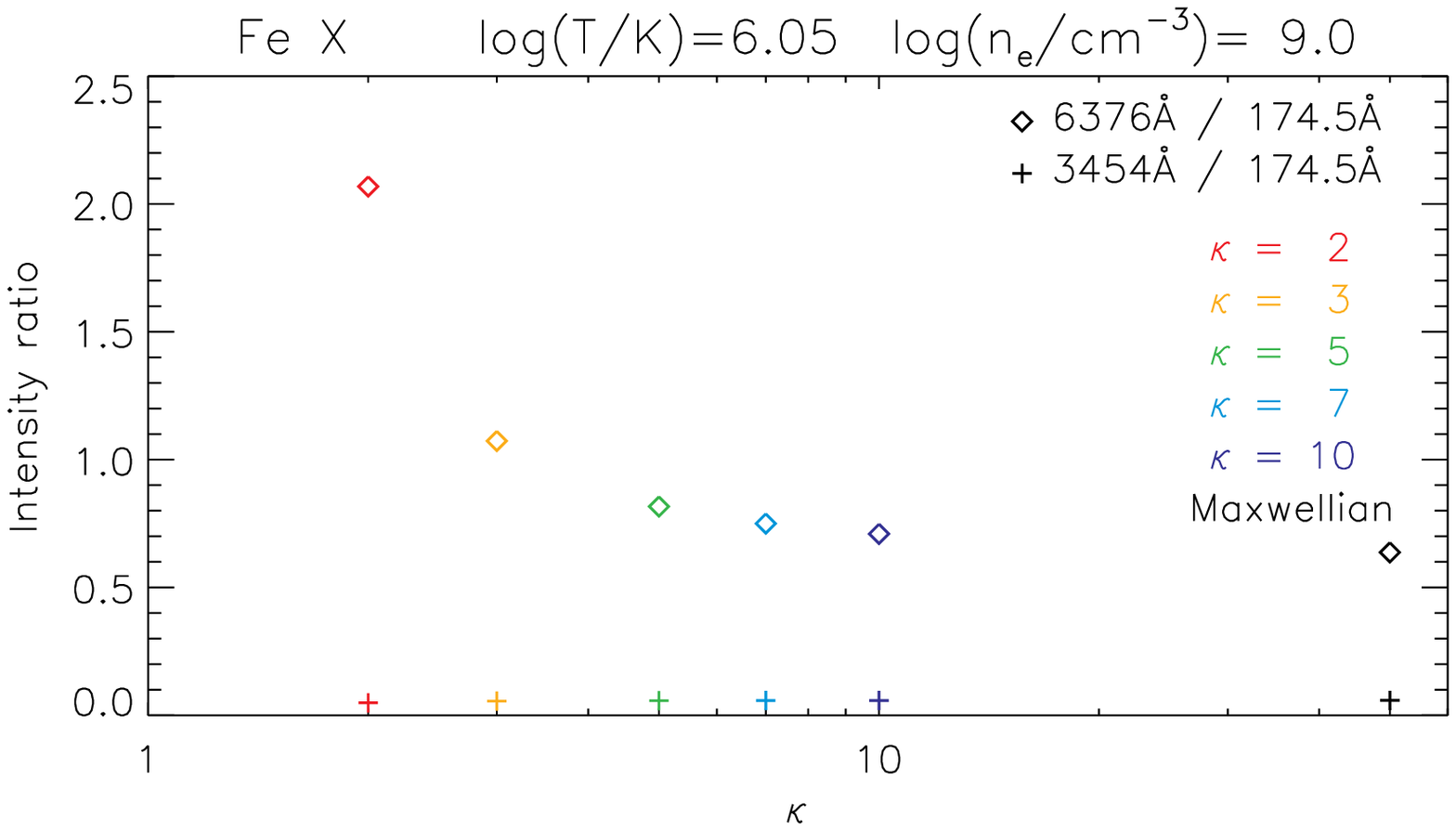}
  \includegraphics[width=8.8cm,bb= 0 48 498 283,clip]{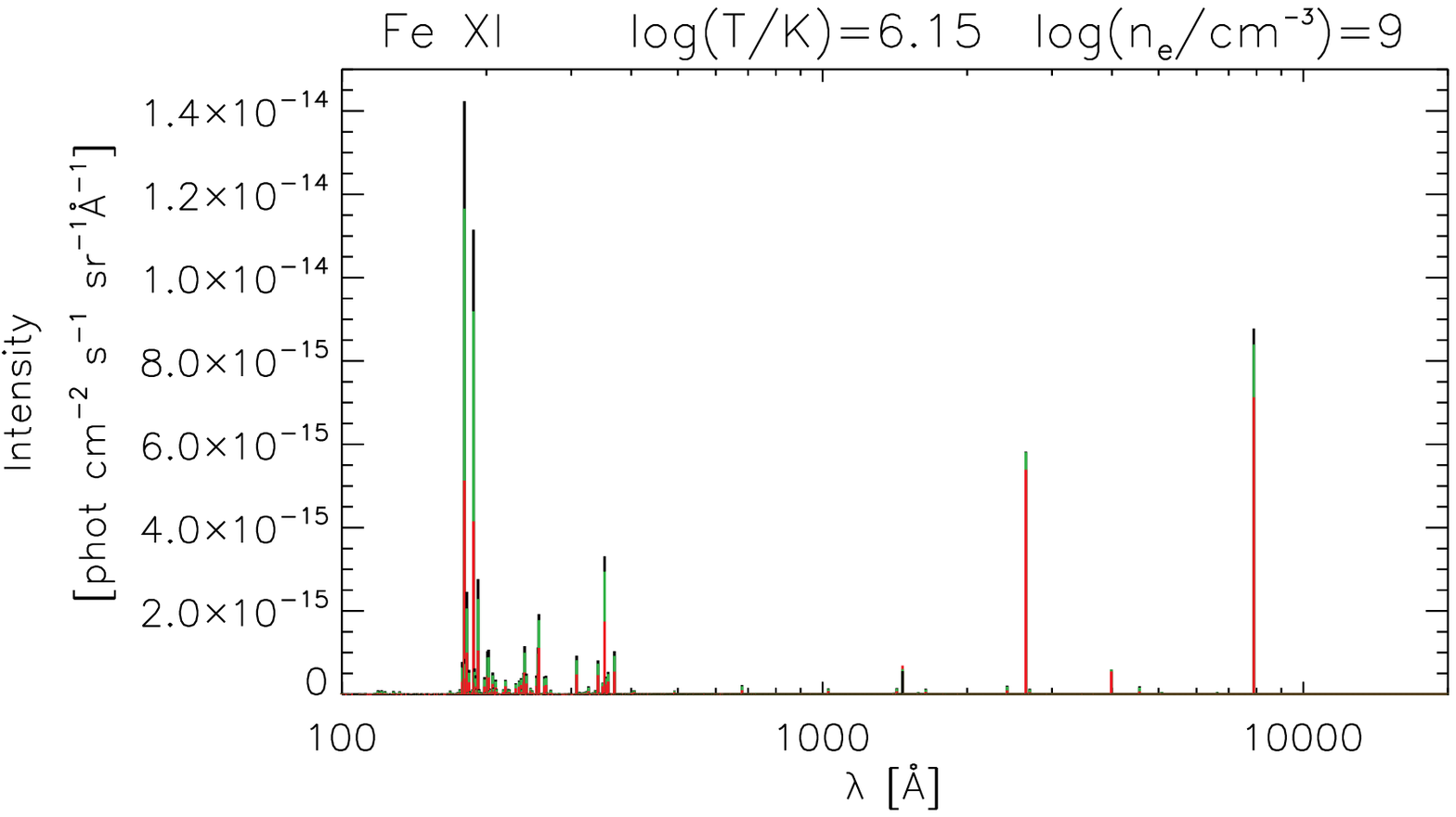}
  \includegraphics[width=8.8cm,bb= 0 48 498 292,clip]{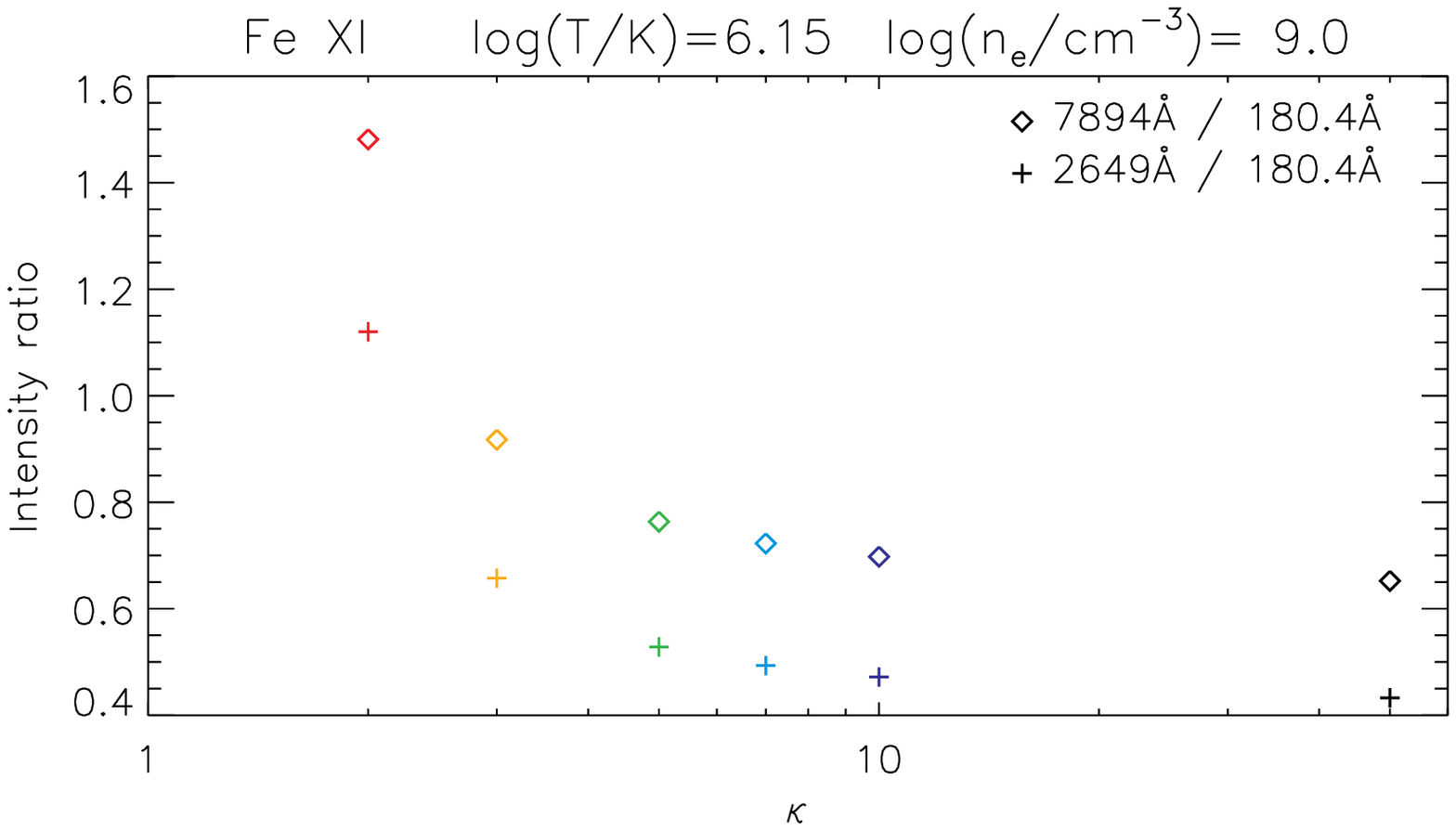}
   \includegraphics[width=8.8cm,bb= 0 48 498 283,clip]{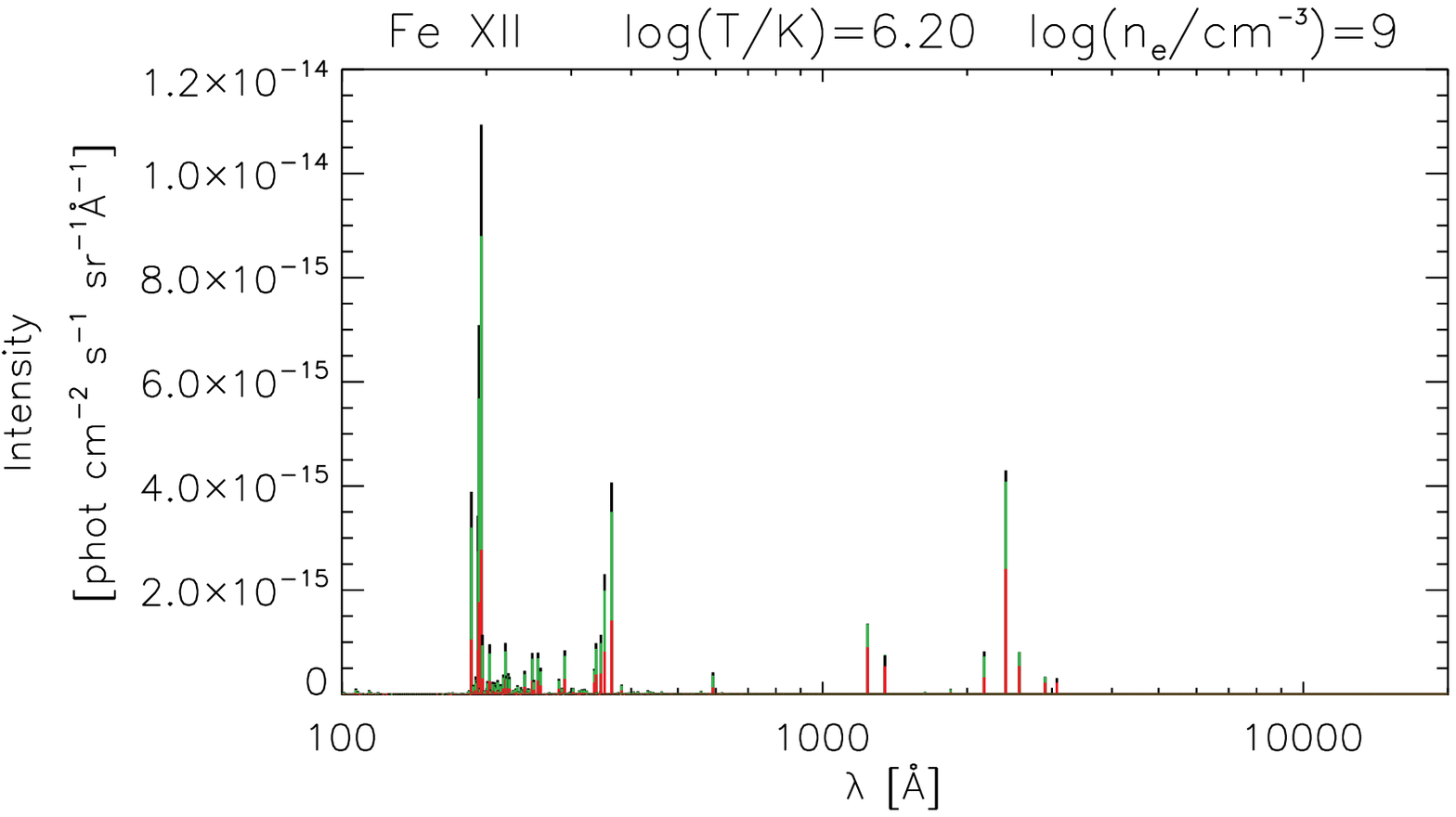}
   \includegraphics[width=8.8cm,bb= 0 48 498 292,clip]{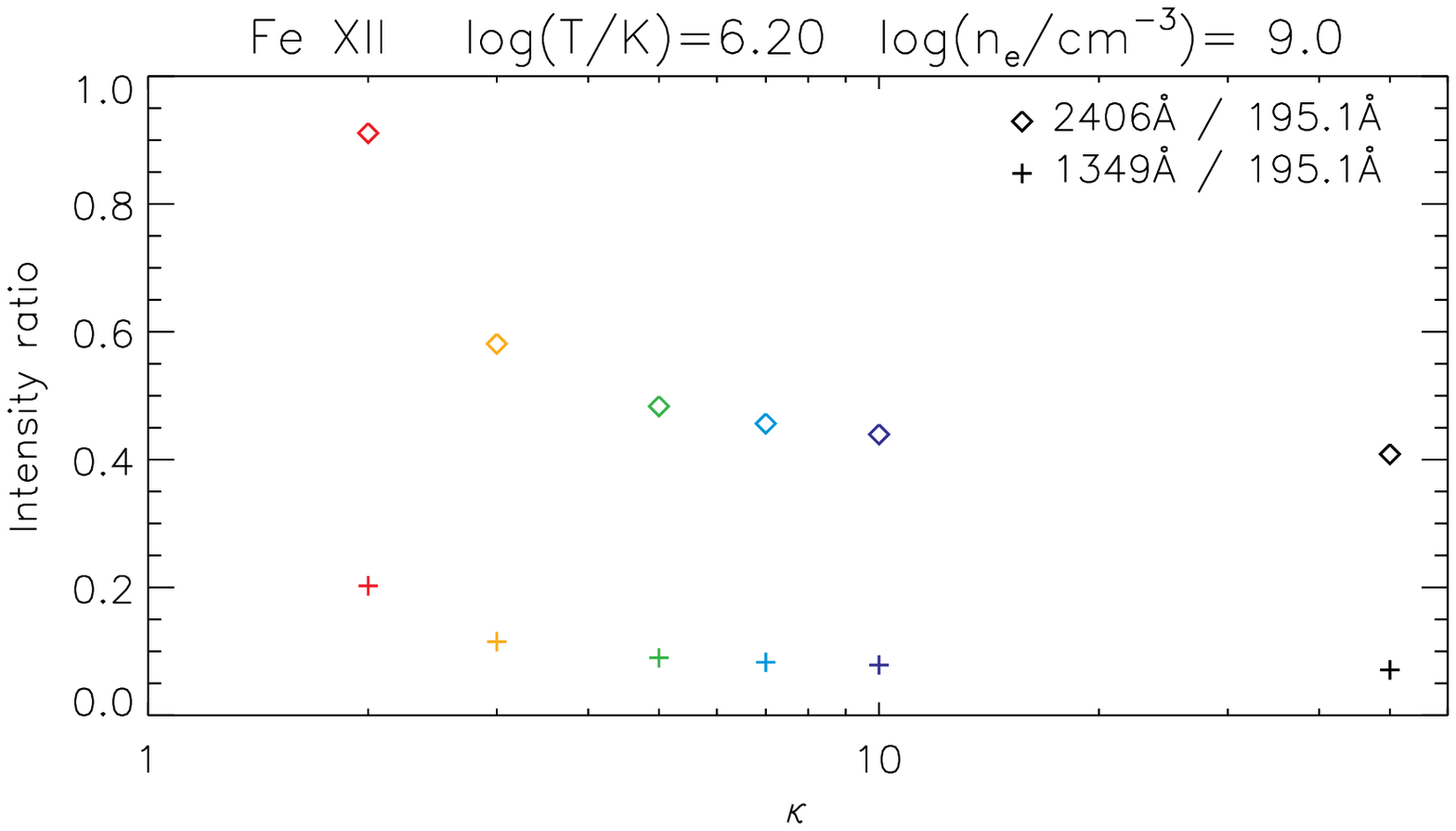}
   \includegraphics[width=8.8cm,bb= 0  0 498 283,clip]{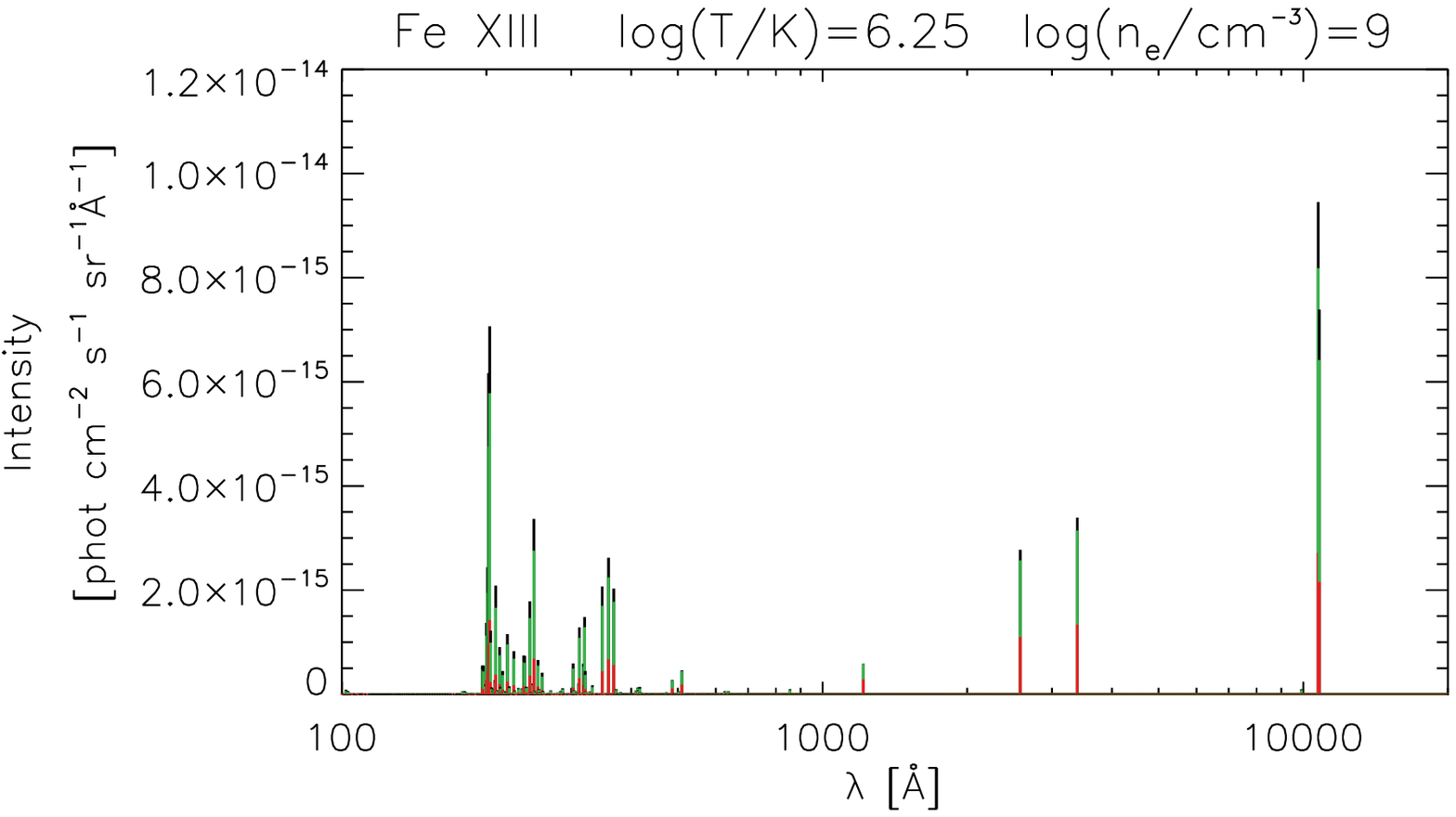}
   \includegraphics[width=8.8cm,bb= 0  0 498 292,clip]{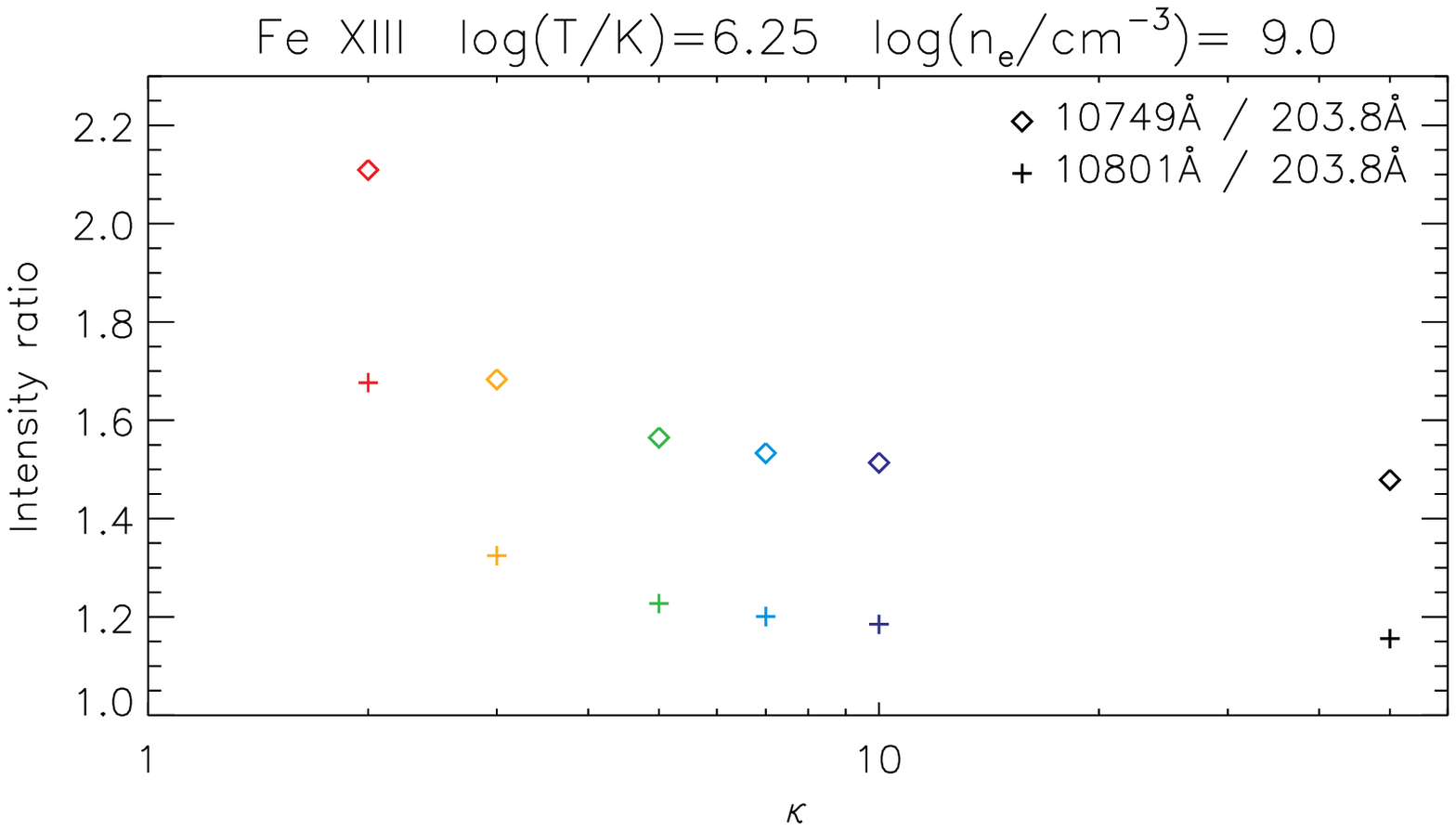}
   \caption{\ion{Fe}{IX}--\ion{Fe}{XIII} spectra between 100\AA~and 20\,000\AA. The spectra shown have a resolution of 1\AA~and are shown for temperatures $T$ corresponding to the peak of the relative ion abundance for the Maxwellian distribution. \textit{Left}: Spectra for log$(n_\mathrm{e})$\,=\,9; \textit{right}: log$(n_\mathrm{e})$\,=\,12.}
   \label{Fig:Spectra}%
\end{figure*}
%
%
\begin{figure*}
   \centering
   \includegraphics[width=8.8cm,bb= 0 28 498 283,clip]{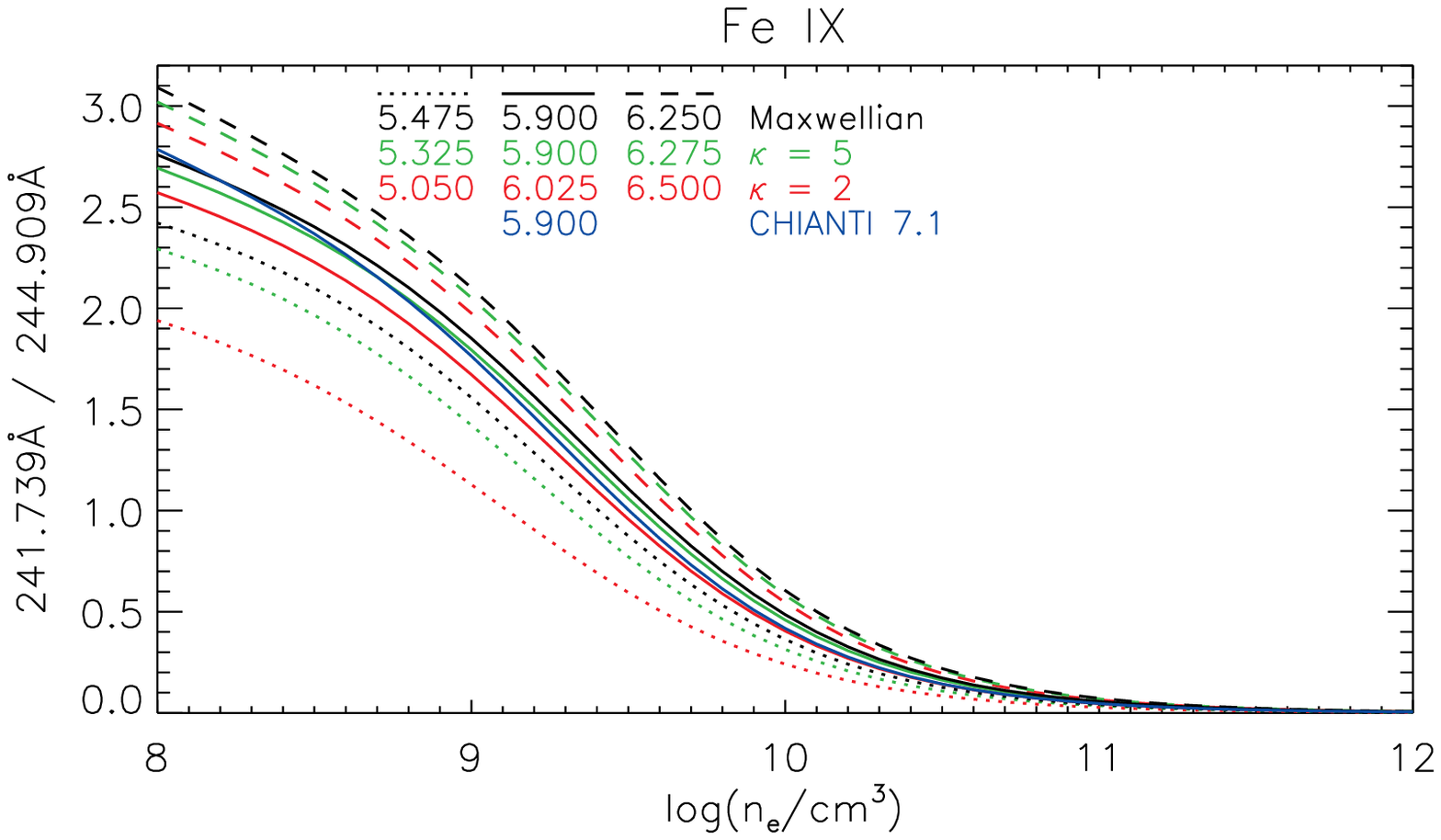}
   \includegraphics[width=8.8cm,bb= 0 28 498 283,clip]{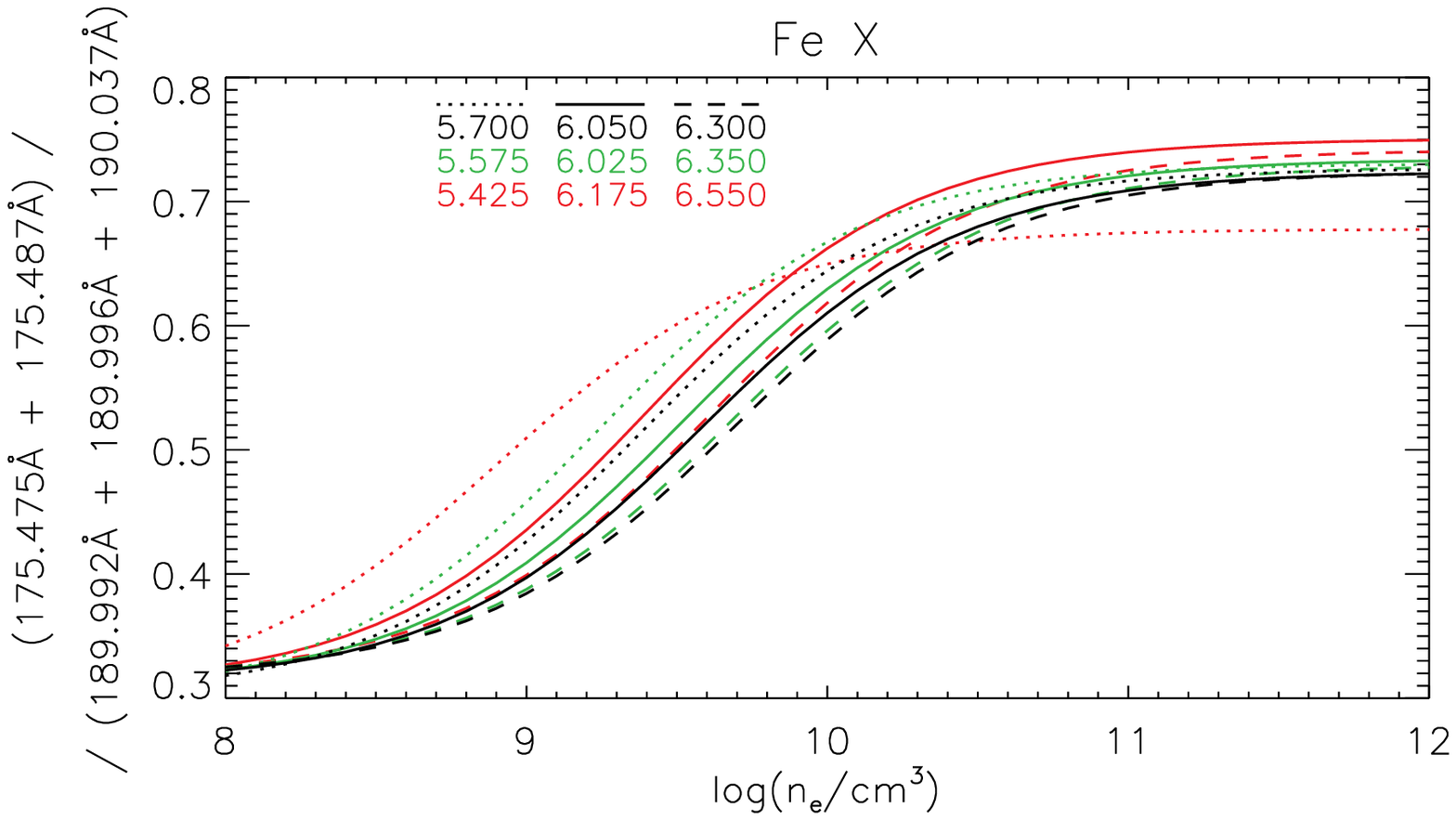}
   \includegraphics[width=8.8cm,bb= 0 28 498 283,clip]{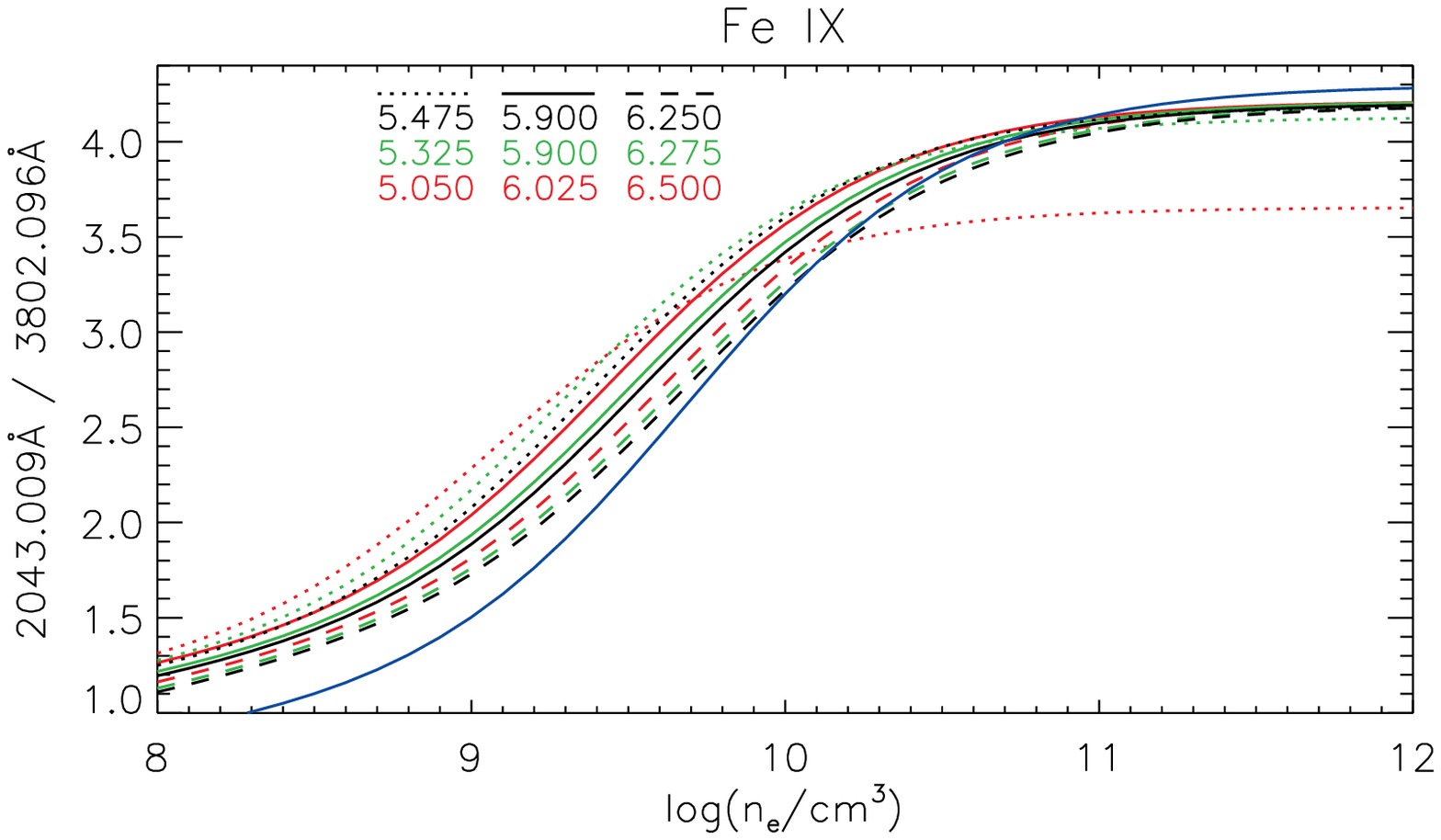}
   \includegraphics[width=8.8cm,bb= 0 28 498 283,clip]{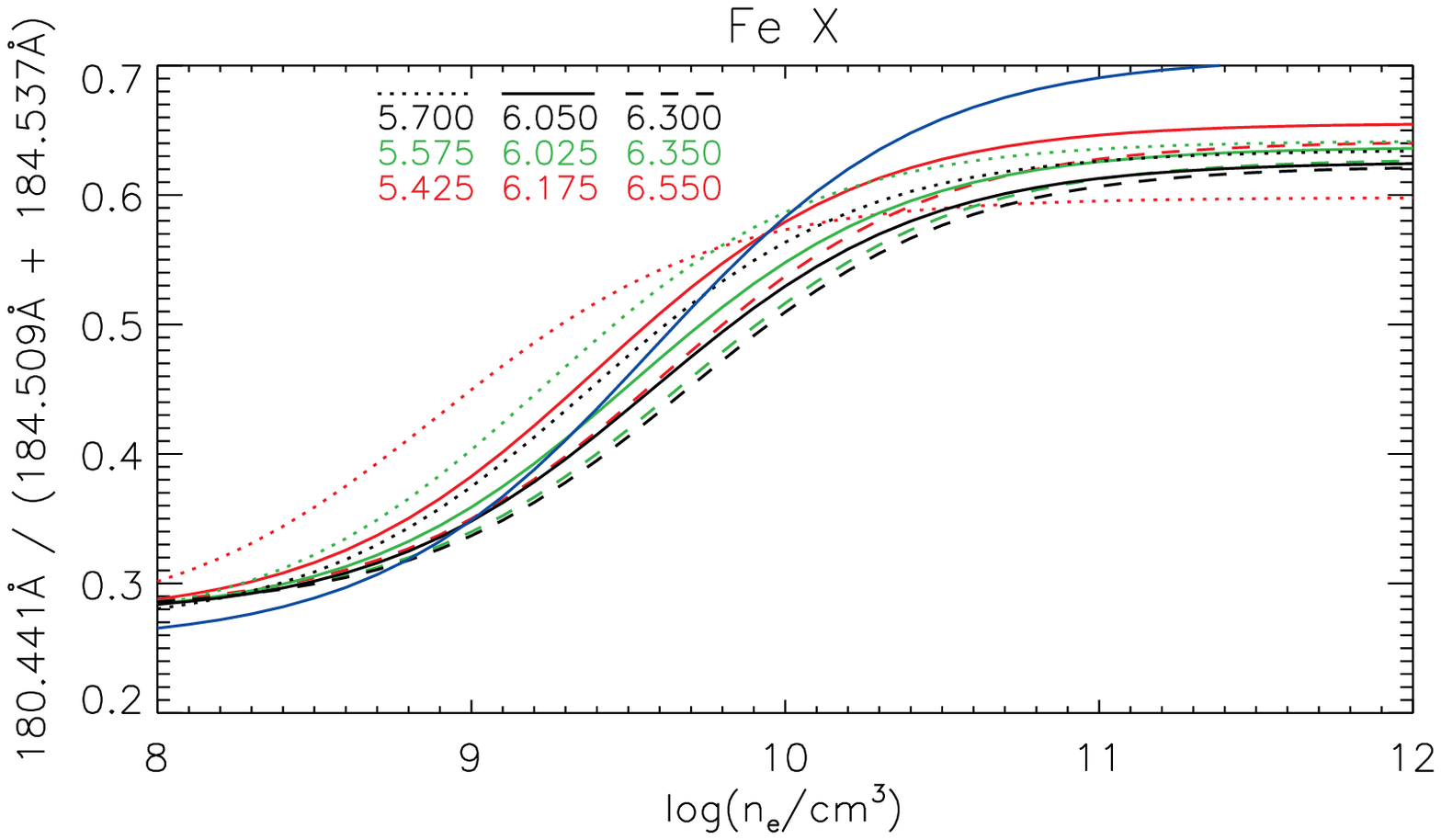}
   \includegraphics[width=8.8cm,bb= 0  0 498 283,clip]{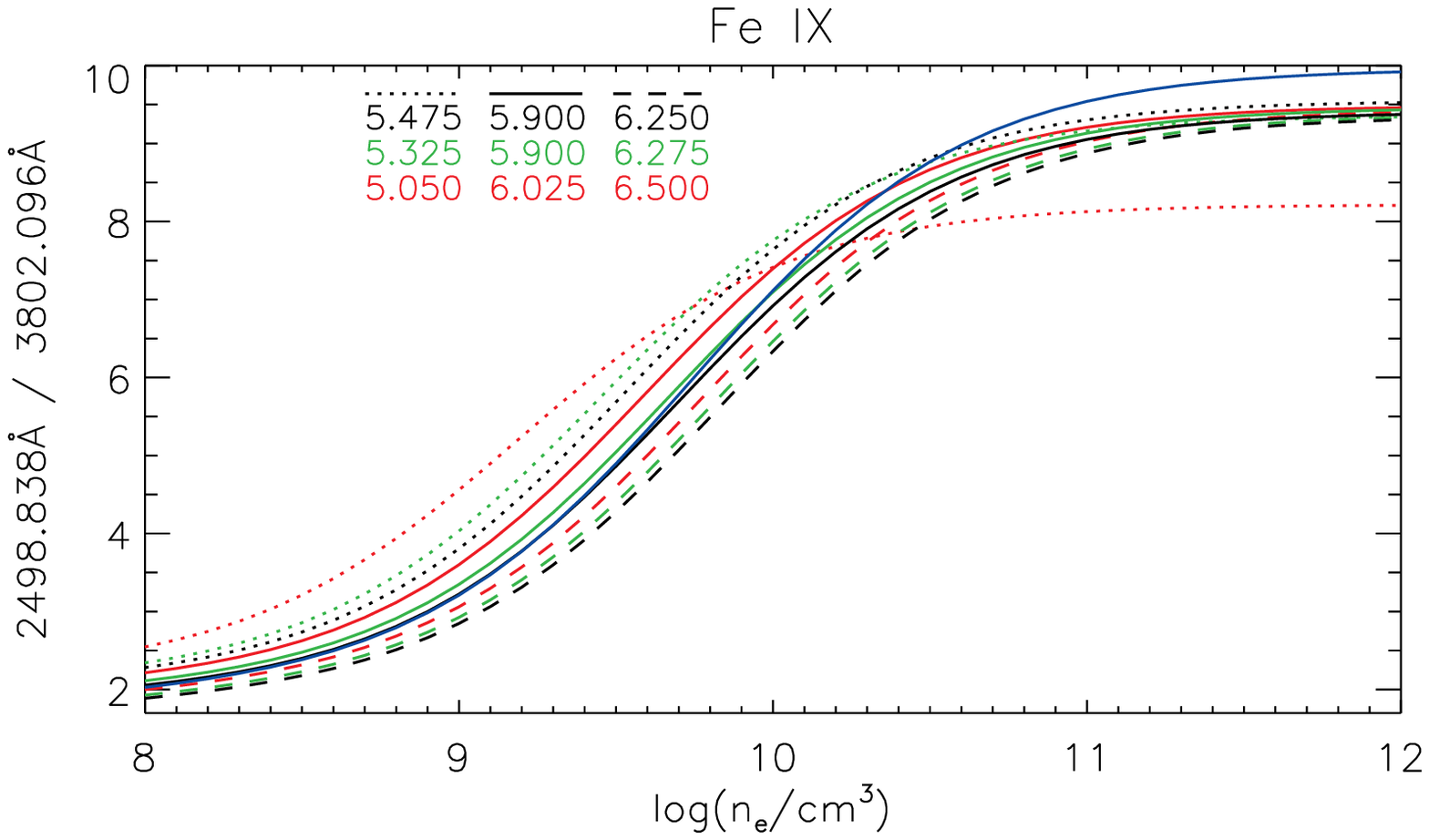}
   \includegraphics[width=8.8cm,bb= 0  0 498 283,clip]{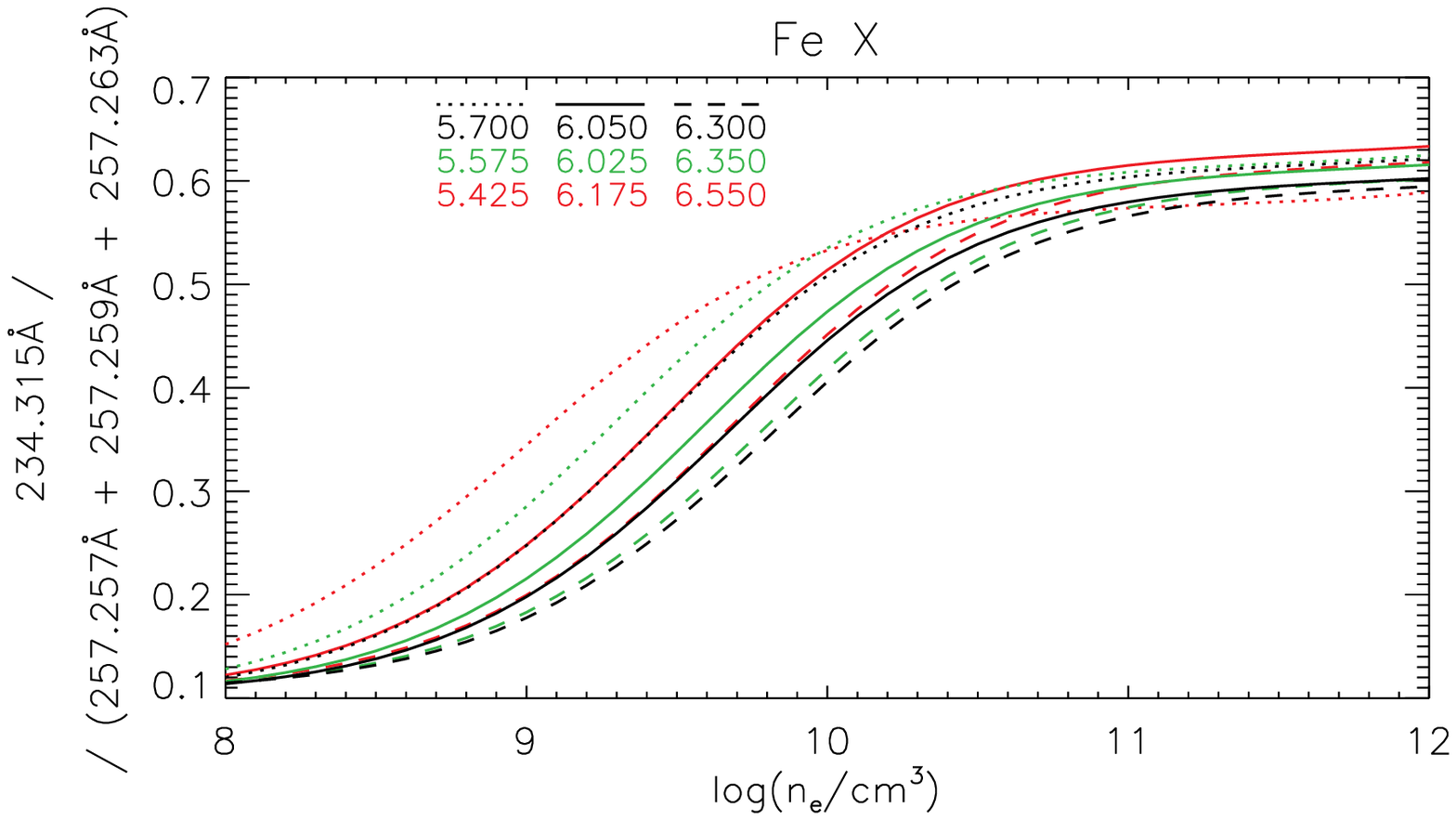}
   \caption{Electron density diagnostics from \ion{Fe}{IX} and \ion{Fe}{X}. Intensity ratios of two lines are shown. The intensity units are phot\,cm$^{-2}$\,s$^{-1}$\,sr$^{-1}$. Black lines correspond to the Maxwellian distribution, green to $\kappa$\,=\,5, and red to $\kappa$\,=\,2. Individual line styles are used to denote the $T$-dependence of the density diagnostics. Blue lines denote the same ratio according to the atomic data from CHIANTI v7.1.}
   \label{Fig:diag_ne9-10}%
\end{figure*}

\begin{figure*}
   \centering
   \includegraphics[width=8.8cm,bb= 0 28 498 283,clip]{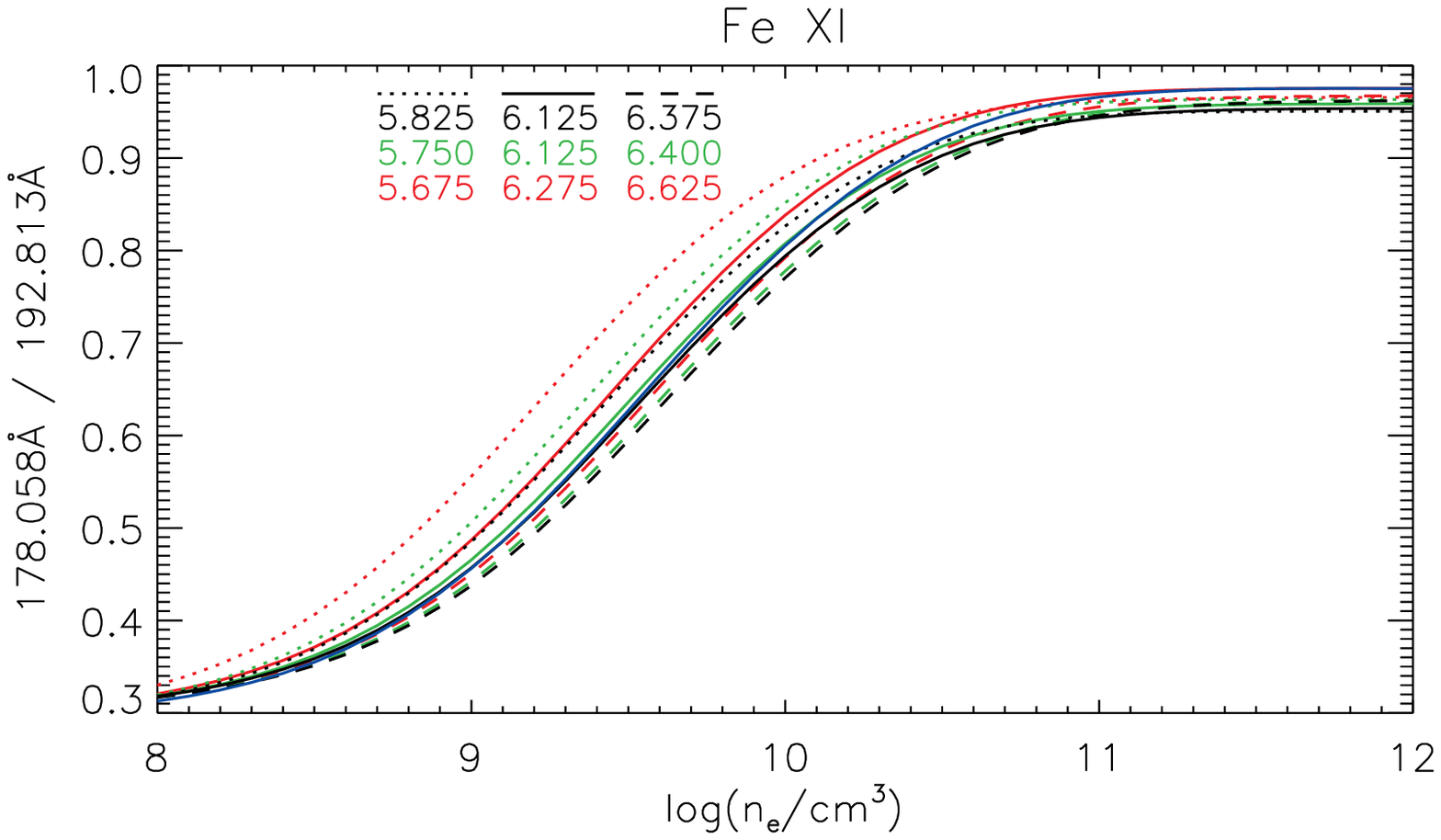}
   \includegraphics[width=8.8cm,bb= 0 28 498 283,clip]{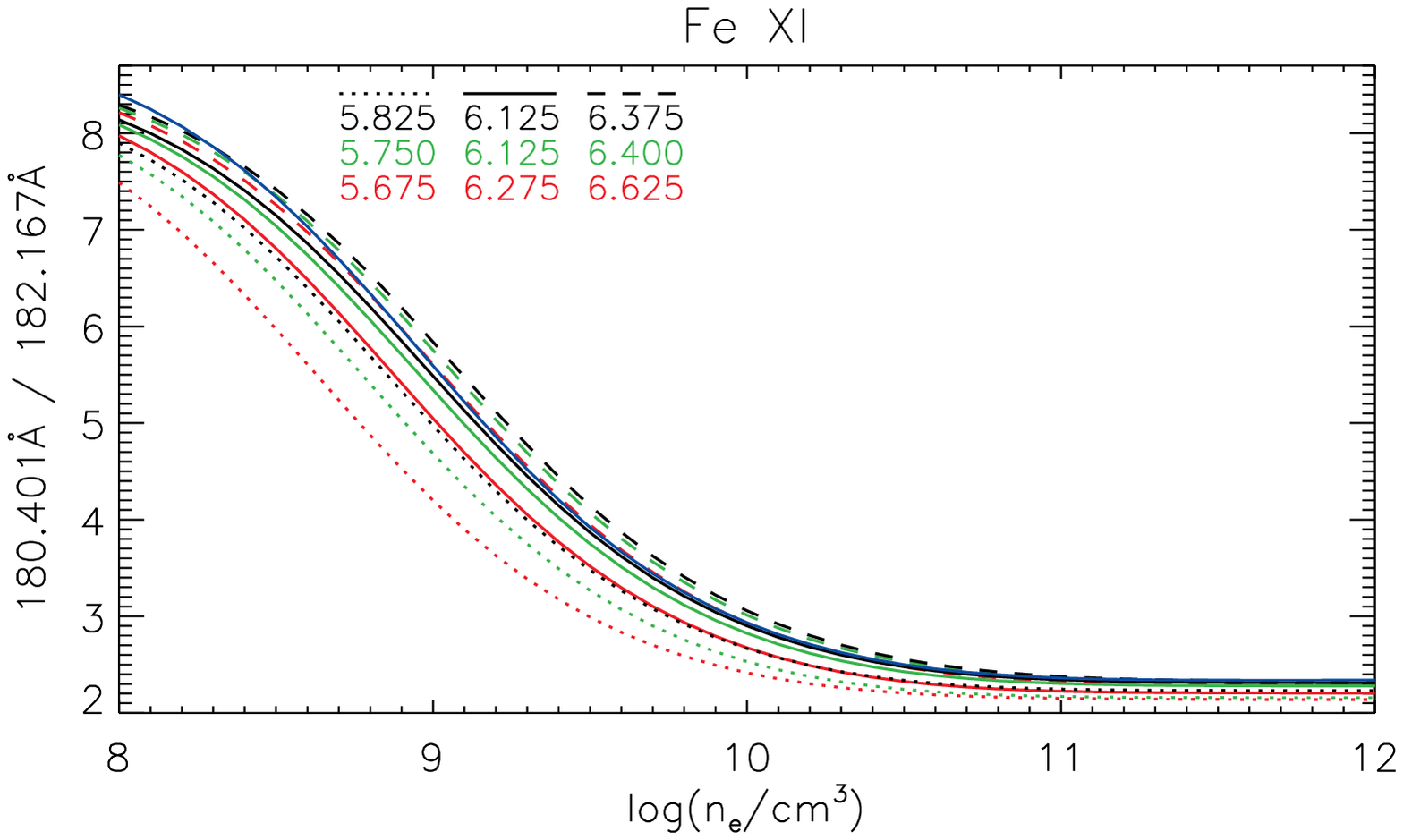}
   \includegraphics[width=8.8cm,bb= 0  0 498 283,clip]{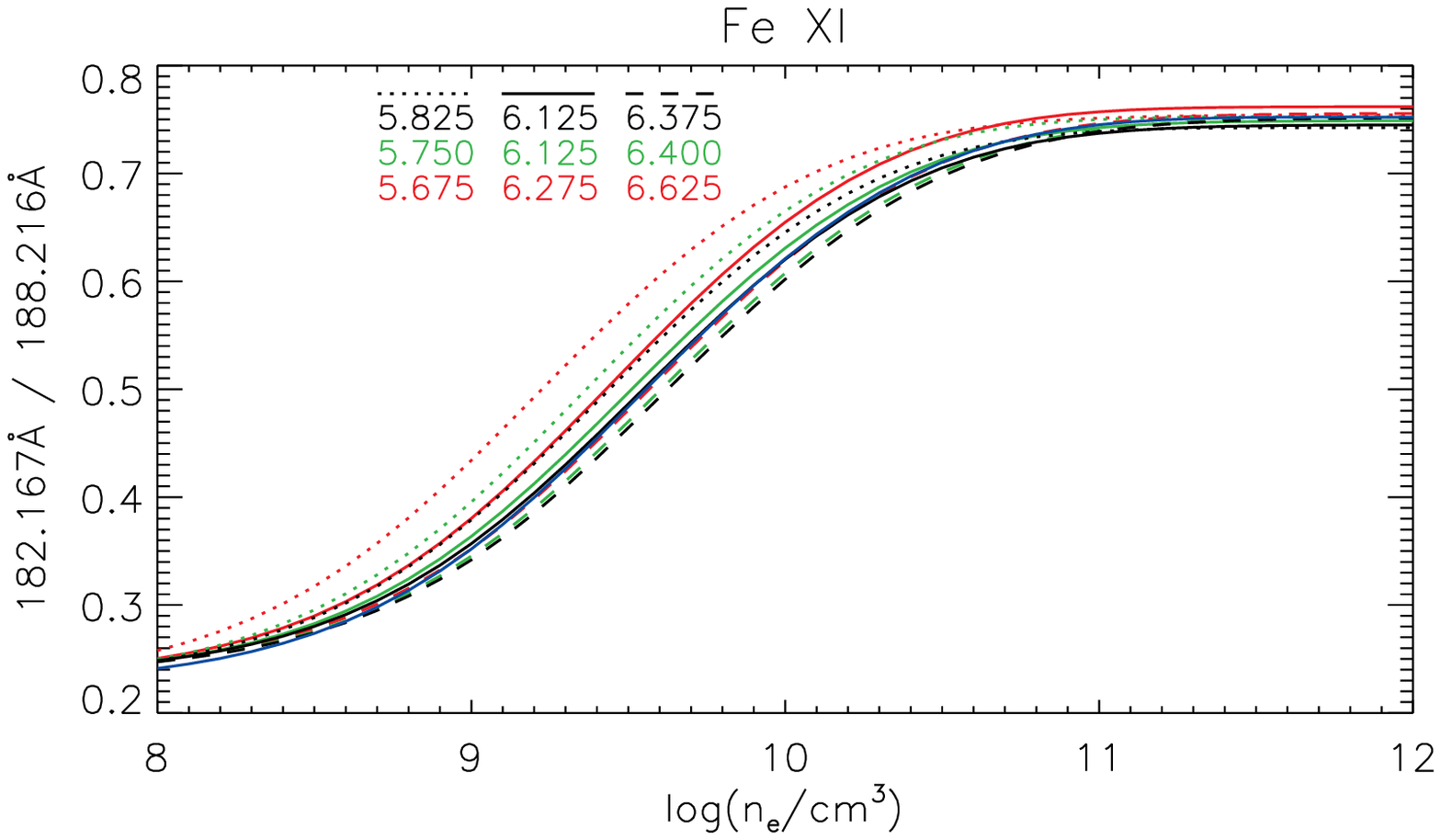}
   \includegraphics[width=8.8cm,bb= 0  0 498 283,clip]{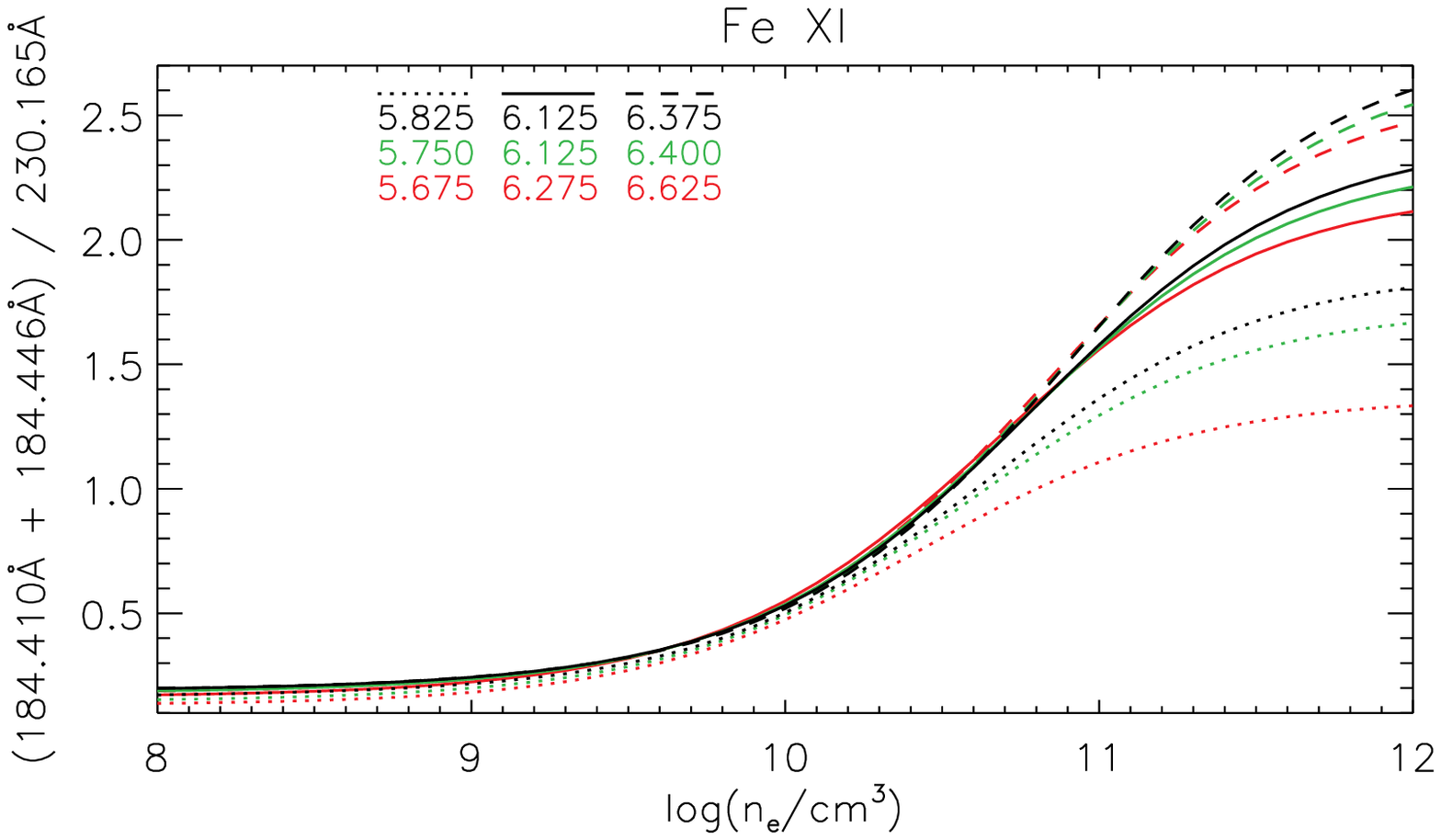}
   \caption{Electron density diagnostics from \ion{Fe}{XI}. Line styles and colors are the same as in Fig. \ref{Fig:diag_ne9-10}.}
   \label{Fig:diag_ne11}%
\end{figure*}

\begin{figure*}
   \centering
   \includegraphics[width=8.8cm,bb= 0 28 498 283,clip]{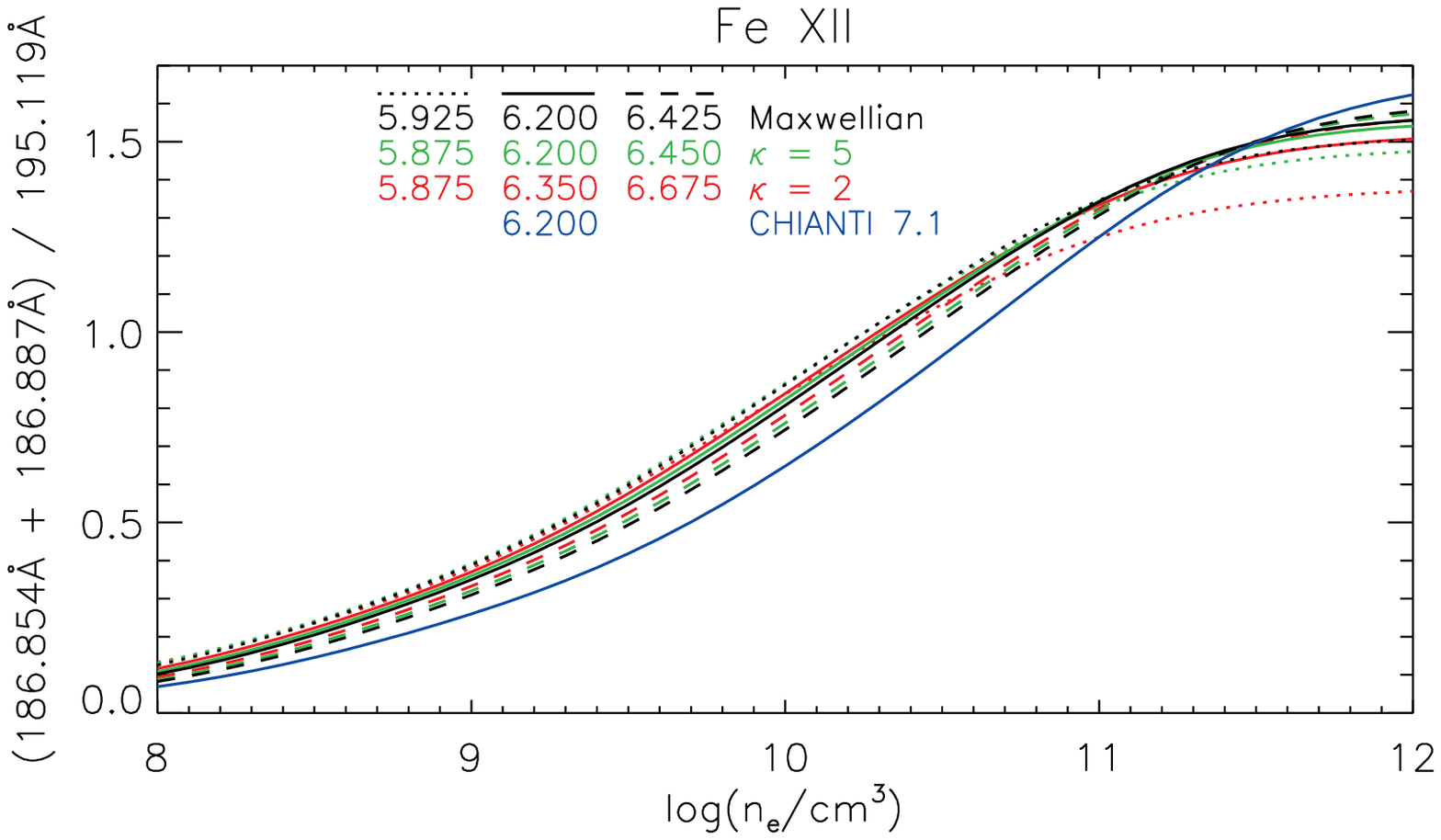}
   \includegraphics[width=8.8cm,bb= 0 28 498 283,clip]{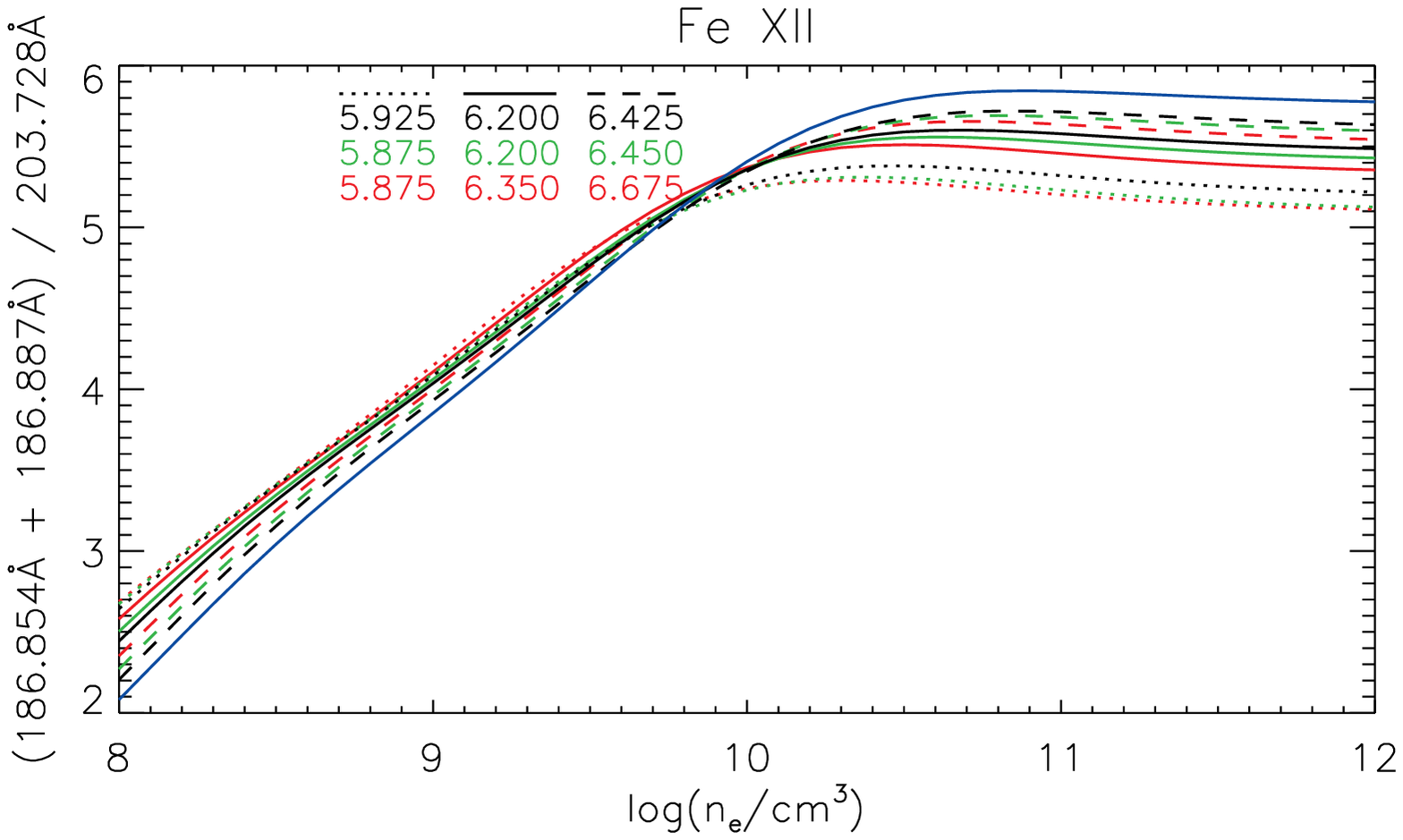}
   \includegraphics[width=8.8cm,bb= 0  0 498 283,clip]{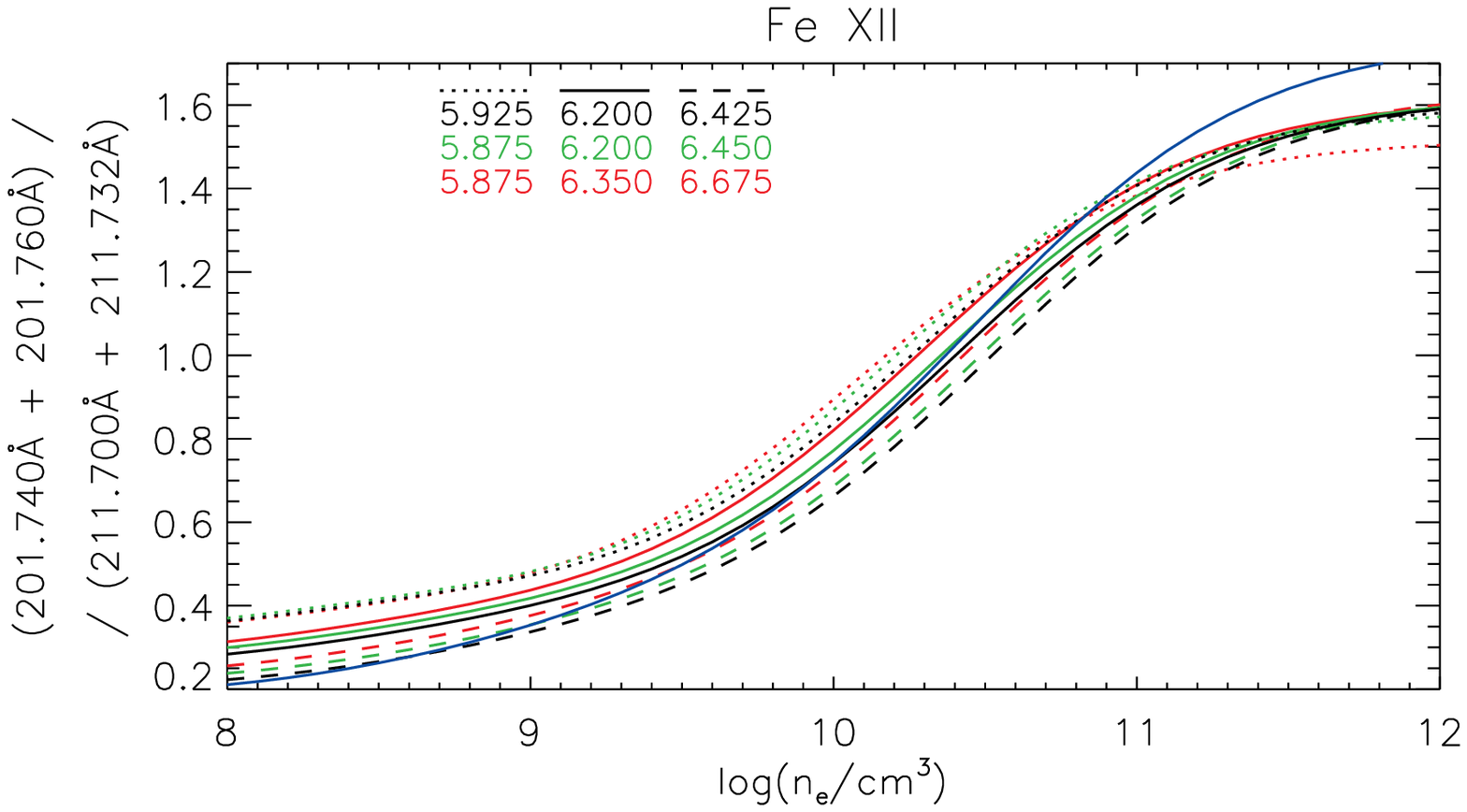}
   \includegraphics[width=8.8cm,bb= 0  0 498 283,clip]{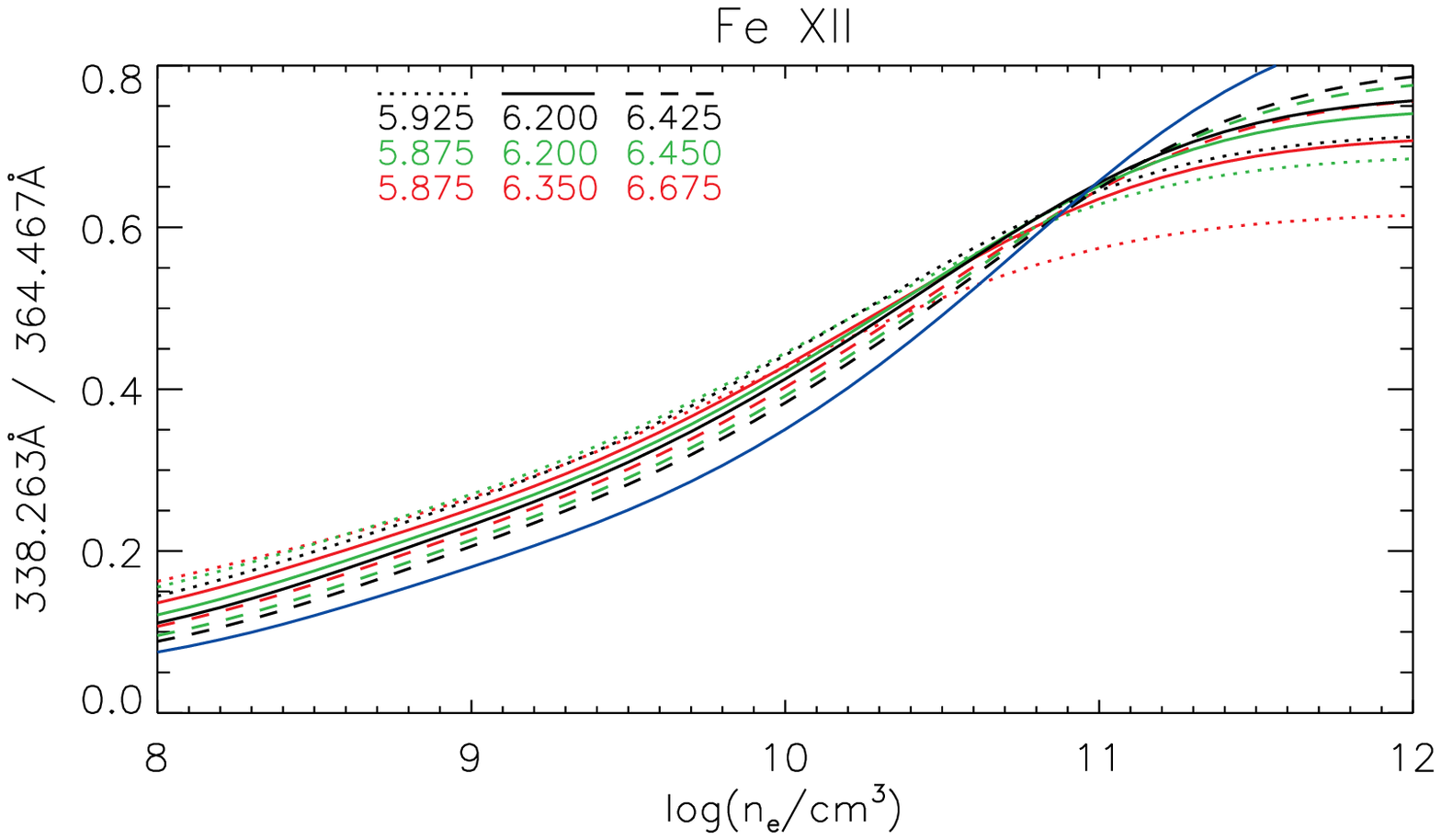}
   \caption{Electron density diagnostics from \ion{Fe}{XII}. Line styles and colors are the same as in Fig. \ref{Fig:diag_ne9-10}.}
   \label{Fig:diag_ne12}%
\end{figure*}
%
\begin{figure*}
   \centering
   \includegraphics[width=8.8cm,bb= 0 28 498 283,clip]{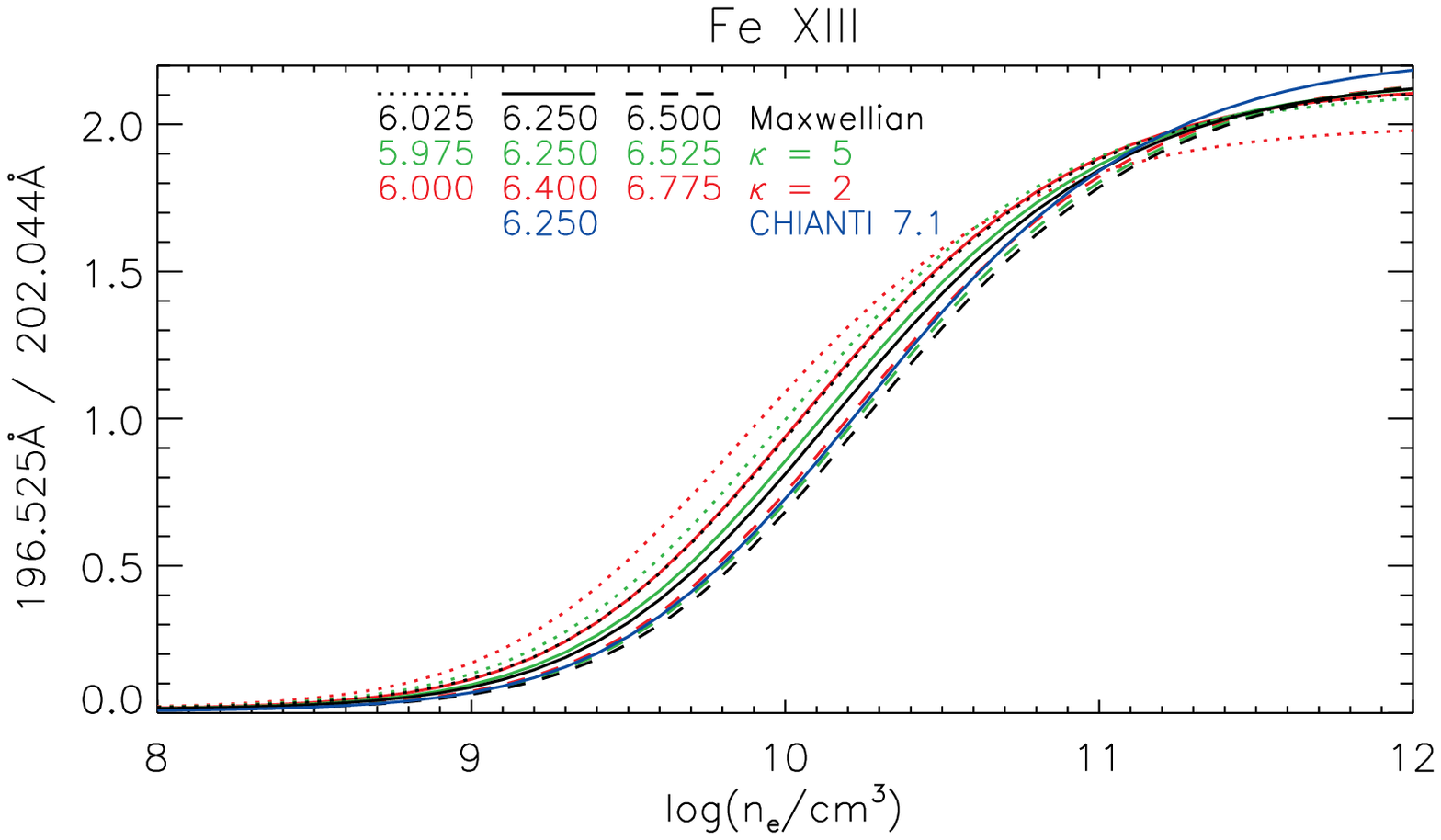}
   \includegraphics[width=8.8cm,bb= 0 28 498 283,clip]{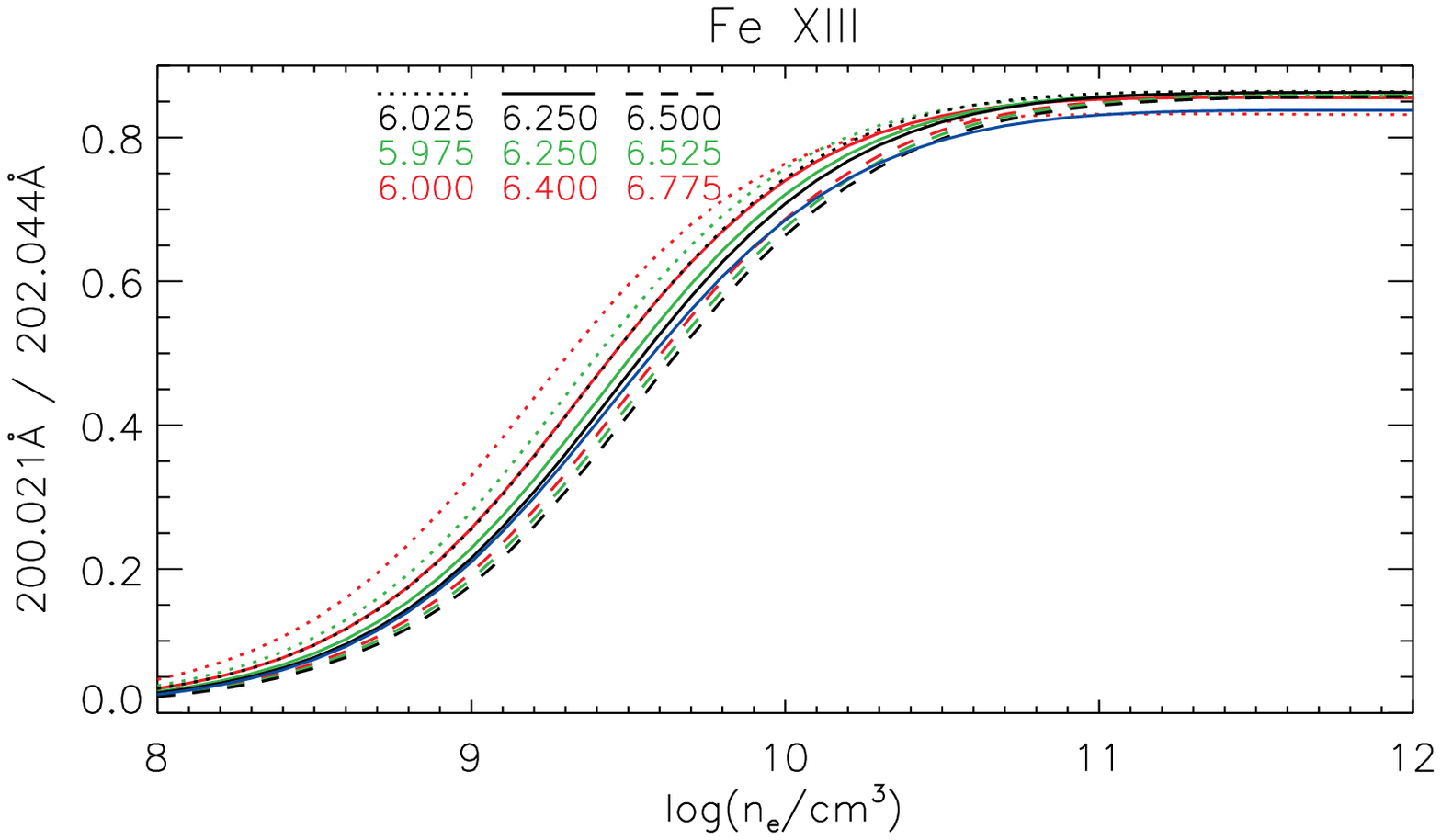}
   \includegraphics[width=8.8cm,bb= 0 28 498 283,clip]{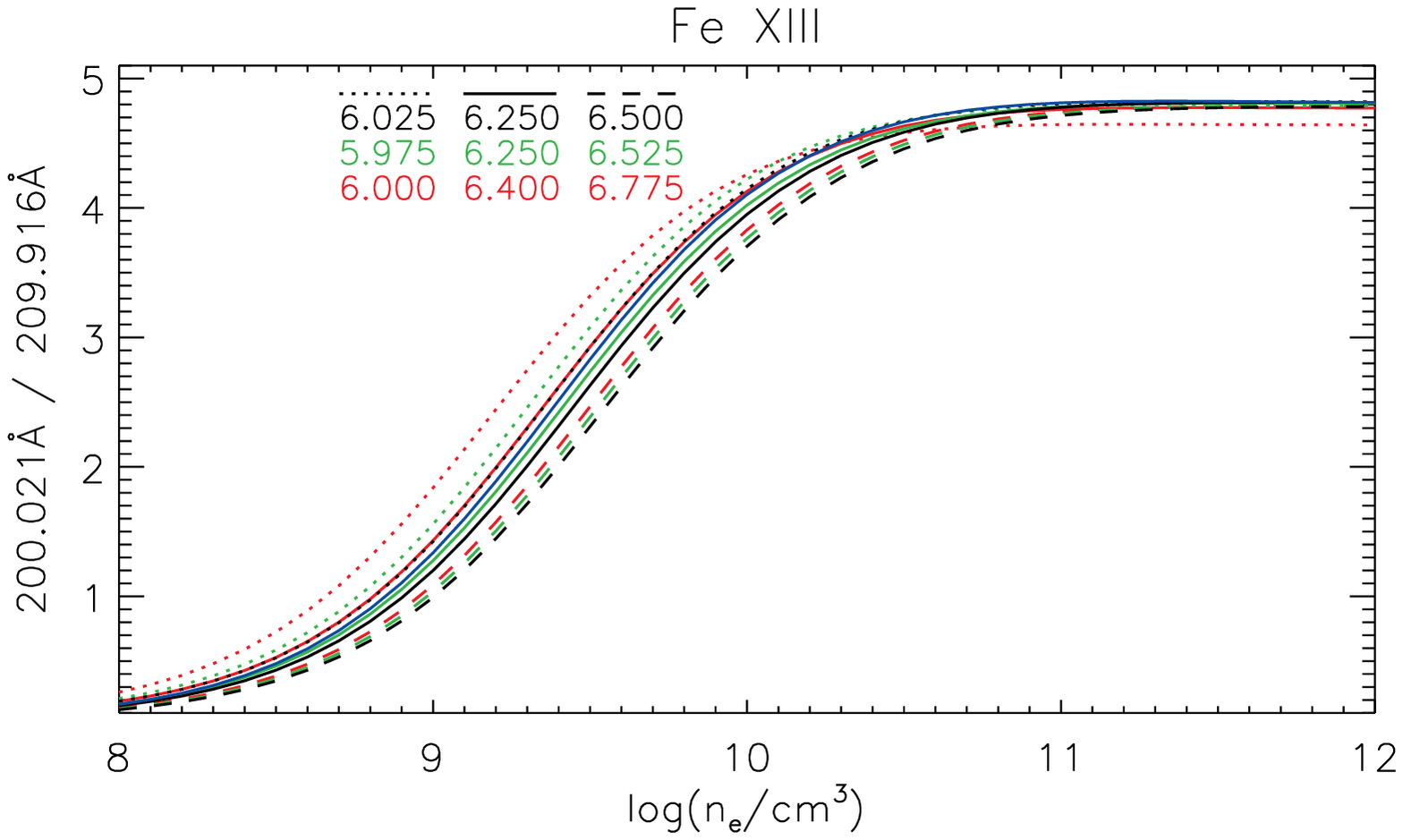}
   \includegraphics[width=8.8cm,bb= 0 28 498 283,clip]{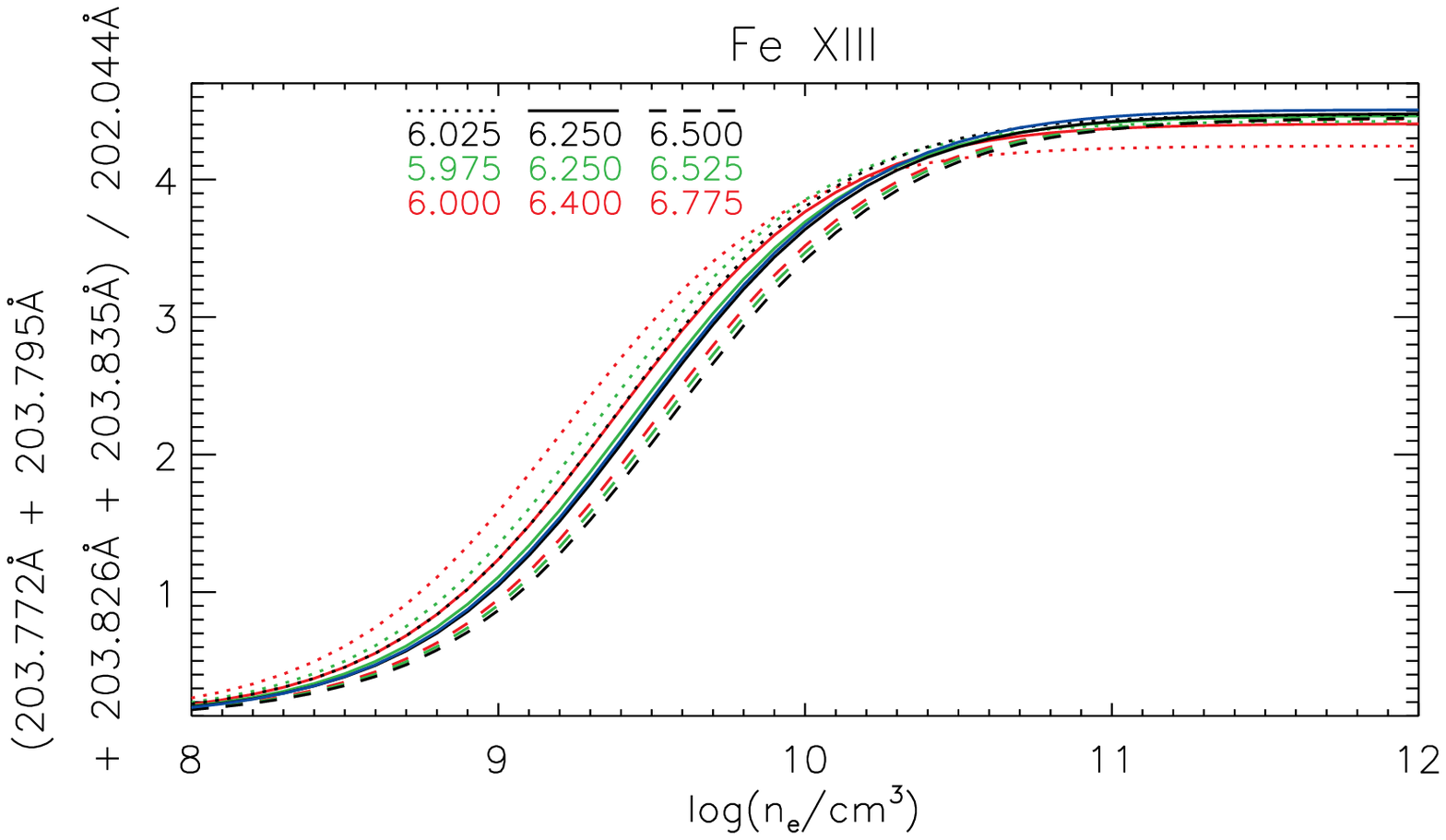}
   \includegraphics[width=8.8cm,bb= 0 28 498 283,clip]{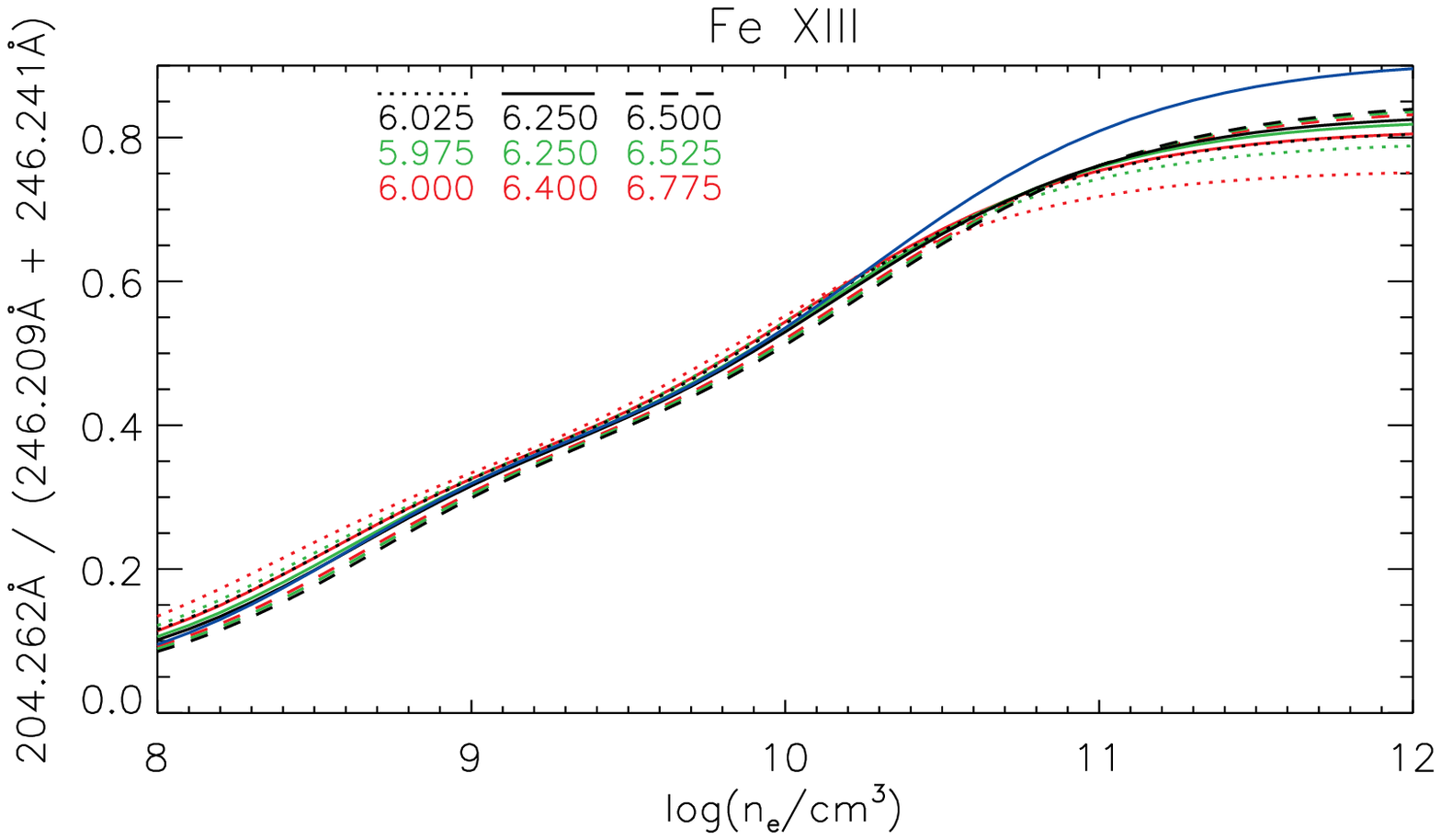}
   \includegraphics[width=8.8cm,bb= 0 28 498 283,clip]{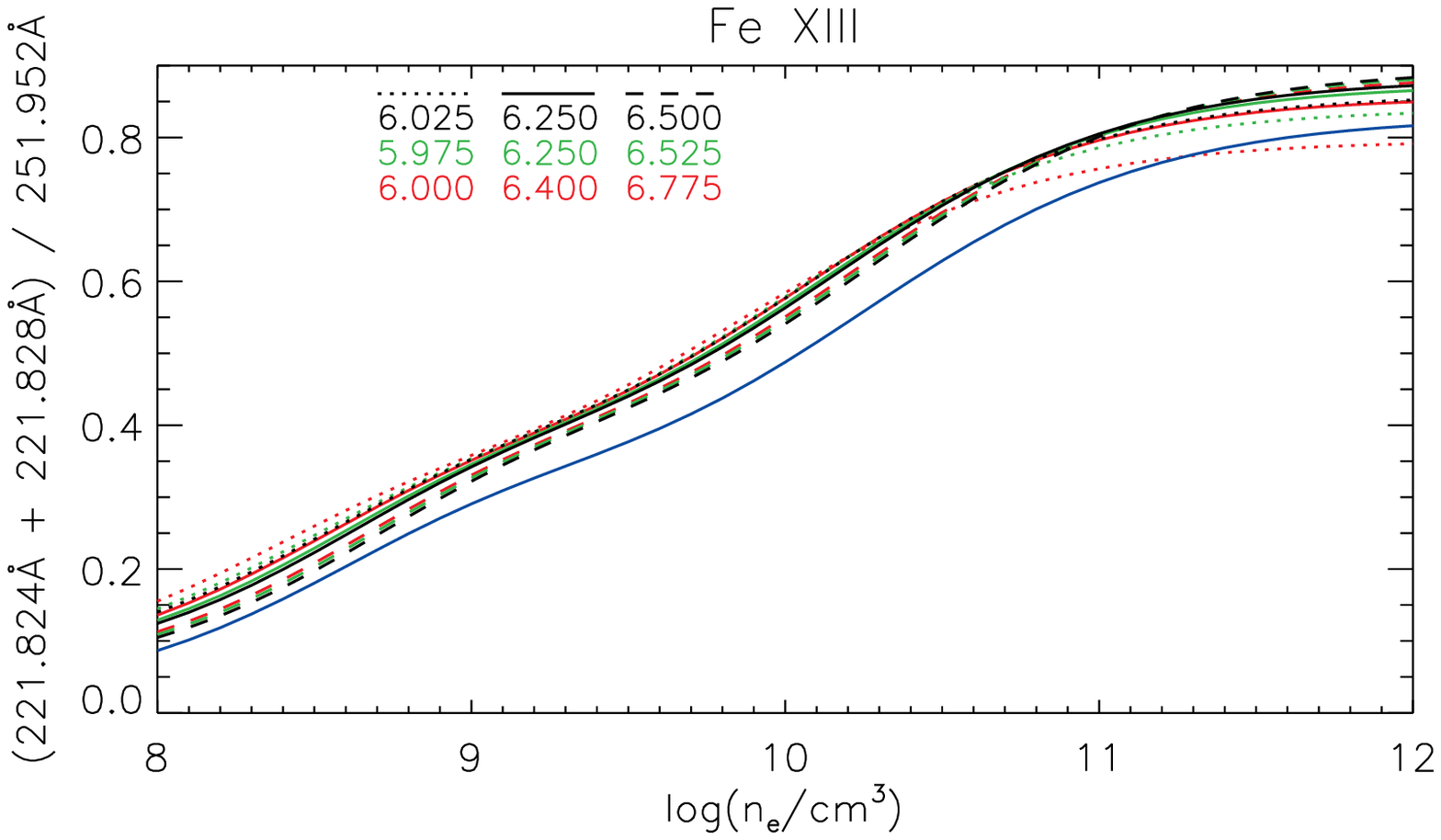}
   \includegraphics[width=8.8cm,bb= 0  0 498 283,clip]{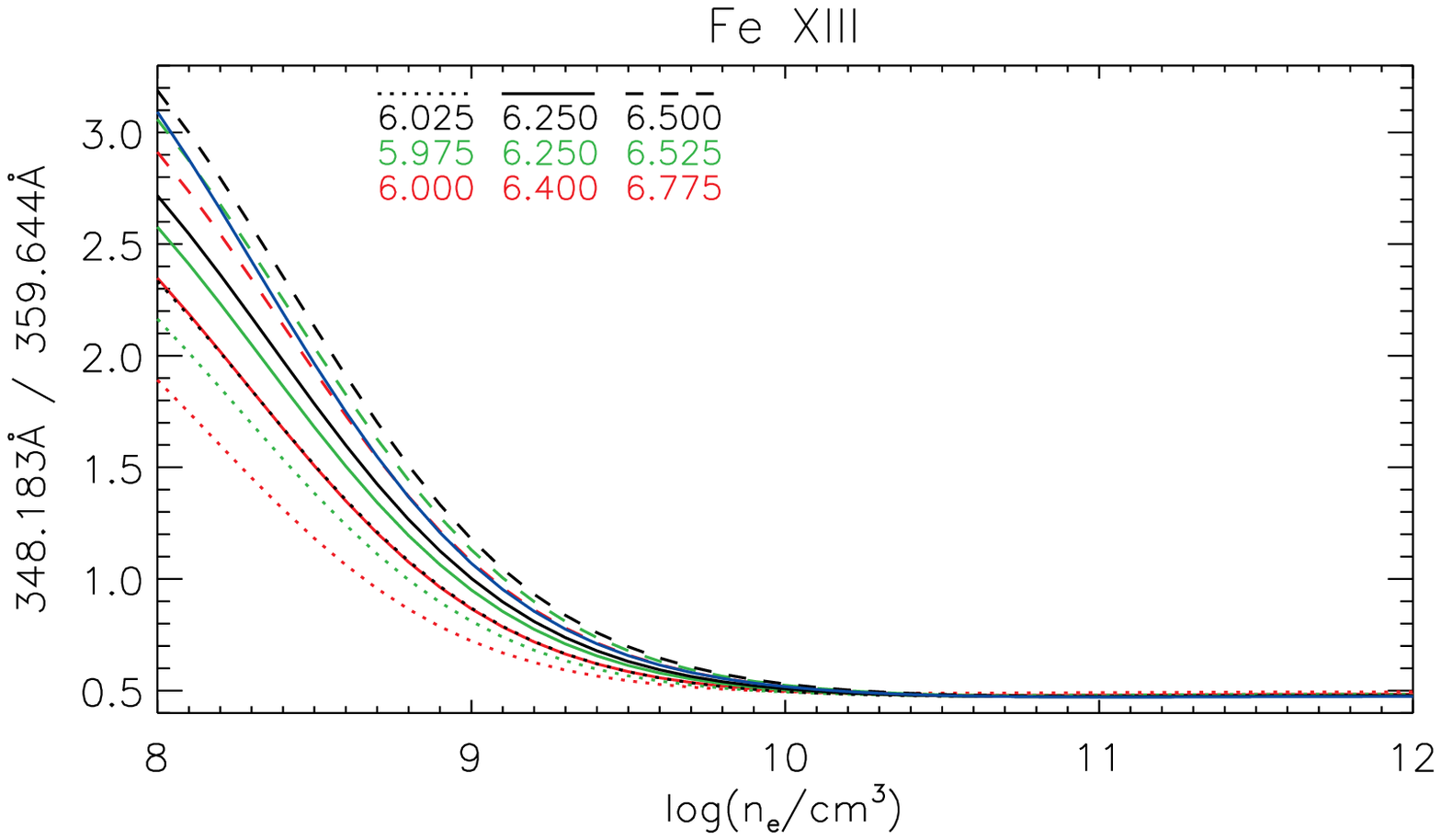}
   \includegraphics[width=8.8cm,bb= 0  0 498 283,clip]{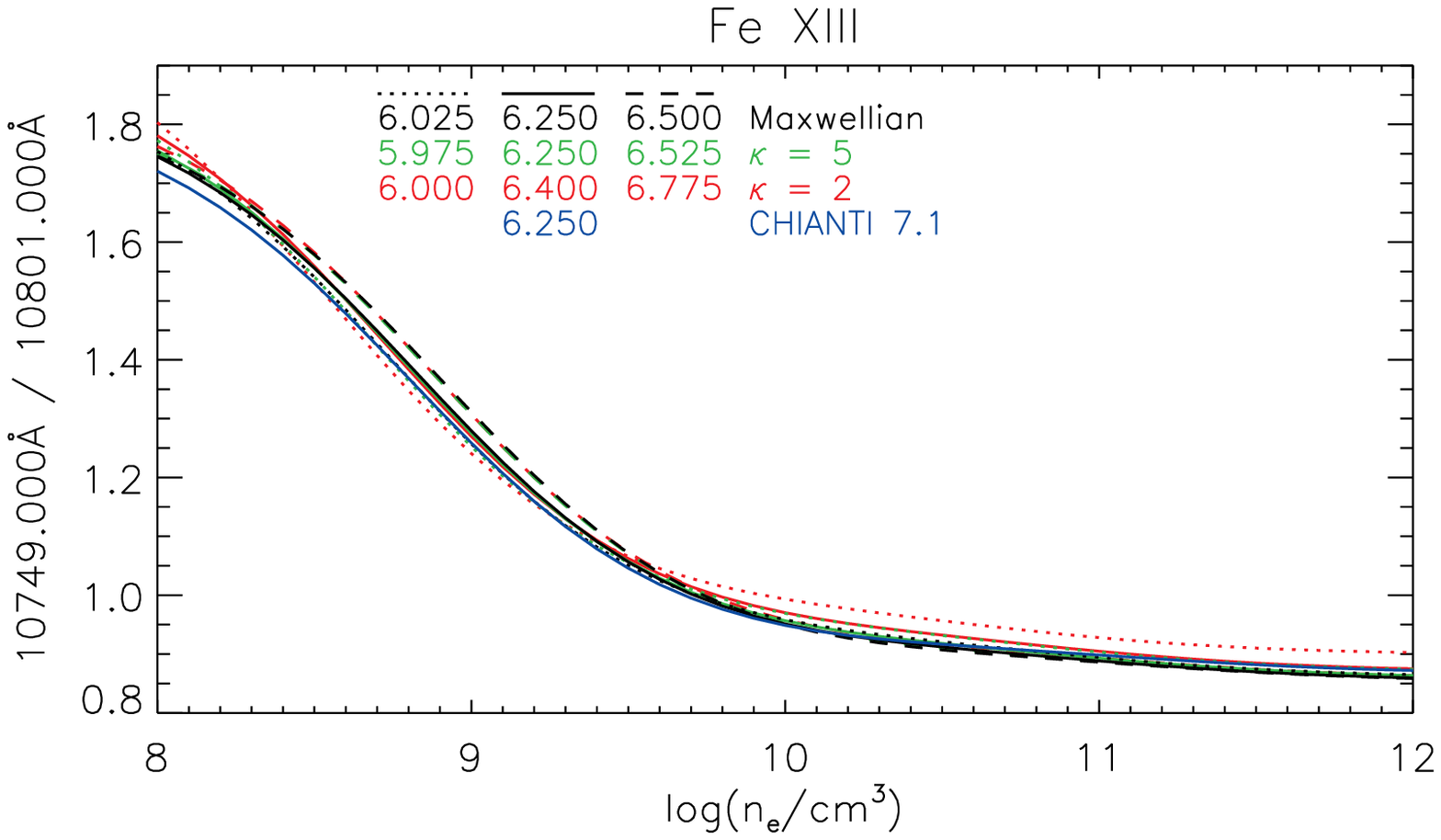}
   \caption{Electron density diagnostics from \ion{Fe}{XIII}. Line styles and colors are the same as in Fig. \ref{Fig:diag_ne9-10}.}
   \label{Fig:diag_ne13}%
\end{figure*}
%
%
%
%
%
\section{Synthetic spectra for \ion{Fe}{IX}--\ion{Fe}{XIII}}
\label{Sect:4}

Having obtained the $\Upsilon_{ij}(T,\kappa)$ and \rotatebox[origin=c]{180}{$\Upsilon$}$_{ji}(T,\kappa)$ for $\kappa$-distributions, it is straightforward to calculate the synthetic spectra for \ion{Fe}{IX}--\ion{Fe}{XIII}. In such calculations, we take into account transitions from all upper levels to only several tens of lower levels $i$. This is done in order to limit the size of calculations to several GiB, while retaining all potentially observable transitions together with their selfblends. The number of lower levels is chosen so that all of the metastable levels with relative populations greater than 10$^{-3}$ at densities of up to log$(n_\mathrm{e}$/cm$^{-3})$\,=\,12 are retained, as well as to account for contributions to excitation of metastable levels by cascading from higher levels \citep[e.g.][Sect. 4.2]{DelZanna13}. This results in calculating the transitions from all upper levels to the lower 80 levels for \ion{Fe}{IX}, lower 100 levels for \ion{Fe}{X}, 32 for \ion{Fe}{XI}, and 25 for both \ion{Fe}{XII} and \ion{Fe}{XIII}. An exception is the mestatable level 317 for \ion{Fe}{IX} (3s$^2$ 3p$^3$ 3d$^3$  $^7$K$_{8}$), which can have a relative population of $\approx 10^{-3}$ for log$(n_\mathrm{e}/\mathrm{cm}^3)$\,$>$\,11 and a Maxwellian distribution. The relative population of this level decreases with $\kappa$ and we chose not to include transitions populated from this level to limit the size of the calculations, so that they could be performed on a desktop PC. Then, the total number of resulting transitions calculated are $\approx$3.38\,$\times$10$^5$ for \ion{Fe}{IX}, 1.48\,$\times$10$^5$ for \ion{Fe}{X}, 8.2\,$\times$10$^4$ for \ion{Fe}{XI}, 4\,$\times$10$^5$ for \ion{Fe}{XII}, and 5.4\,$\times$10$^4$ for \ion{Fe}{XIII}. However, the vast majority of these transitions are extremely weak and unobservable.

The calculations of the synthetic spectra are performed for a temperature grid of 5\,$\leqq$\,log$(T/\mathrm{K})$\,$\leqq$\,7 with a step of $\Delta$log$(T/\mathrm{K})$\,=\,0.025. The range of electron densities is 8\,$\leqq$\,log$(n_\mathrm{e}$/cm$^{-3})$\,$\leqq$\,12 covering a range of astrophysical plasmas in collisional ionization equilibrium, in particular the solar and stellar coronae.

In Fig. \ref{Fig:Spectra} (\textit{left}), the \ion{Fe}{IX}--\ion{Fe}{XIII} synthetic spectra are plotted for an emission measure of unity and wavelengths $\lambda$ between 10$^2$\,\AA~and 2\,$\times$10$^4$\,\AA. The wavelength resolution of these spectra is $\Delta\lambda$\,=\,1\,\AA. The electron density chosen is log$(n_\mathrm{e}$/cm$^{-3})$\,=\,9, while the selected values of log$(T$/K) for each ion correspond to the peak of the respective ion abundance for the Maxwellian distribution. The changes in spectra with $\kappa$-distributions are illustrated by overplotting the spectra for $\kappa$\,=\,5 (green) and 2 (red color). Typically, the spectra are dominated by EUV transitions between $\approx$150\AA~and 400\AA, with few strong forbidden transitions in the UV, visible, or infrared parts of the electromagnetic spectrum. We note that these forbidden transitions disappear at higher densities because of their low $A_{ji}$ (see Eq. \ref{Eq:Excit_eq}).

The general behaviour of the EUV lines is that their intensities decrease with $\kappa$, i.e. an increasing departure from the Maxwellian distribution (Fig. \ref{Fig:Spectra}). Although the line intensities in Fig. \ref{Fig:Spectra} are plotted for log$(T$/K) corresponding to the maximum respective ion abundance for the Maxwellian distribution, this behaviour is general. A few exceptions occur, however, if the spectra are compared with spectra for e.g. $\kappa$\,=\,2 at higher log$(T/$K) corresponding to the maximum ion abundance for such $\kappa$ (see Fig. \ref{Fig:Ioneq}). In contrast, several of the forbidden lines show a reversed behaviour with $\kappa$ even at the same $T$. The best example is the \ion{Fe}{X} 6376.26\AA~the red line in the optical wavelength range (Fig. \ref{Fig:Spectra}, \textit{second row, left}). Other forbidden lines do show decreases of intensity with $\kappa$, albeit much weaker than the EUV lines. An example is the \ion{Fe}{XI} 7894\AA~infrared line or the \ion{Fe}{XII} 2406\AA~(Fig. \ref{Fig:Spectra}, \textit{left, rows 3--4}). Figure \ref{Fig:Spectra}, \textit{right}, shows the intensity ratios (in photon units) of some of the forbidden transitions to the strongest EUV transition as a function of $\kappa$. The line intensities for the Maxwellian, $\kappa$\,=\,5 and 2 and temperatures corresponding to the maximum of the ionization peak for the Maxwellian distribution are listed in Tables \ref{Table:fe9}--\ref{Table:fe13} in Appendix \ref{Appendix:Transitions}.

This behaviour of forbidden lines with $\kappa$ shows that the $\kappa$-distributions can contribute to the observed enhanced intensities of these lines compared to the EUV ones, reported from eclipse observations \citep[][Figs. 2--4]{Habbal13}. The forbidden lines are observed up to $\approx$2\,$R_\odot$. Although the enhanced excitation of the forbidden lines may be to some degree a result of photoexcitation by photospheric radiation \citep{Habbal13}, the enhancement of the \ion{Fe}{X} 6376.26\AA~line occurs especially in less dense regions of the corona, such as in a coronal hole or at the boundary of a helmet streamer. Both these regions contain predominantly open magnetic field lines, and thus are source regions of the fast and slow solar wind, respectively. We note that especially the fast solar wind exhibits $\kappa$-distributions \citep[e.g.][]{Maksimovic97a,Maksimovic97b,LeChat11}. 

%
\section{Electron density diagnostics independent of $\kappa$}
\label{Sect:5}

Rhe line contribution function $G_{X,ji}(T,n_\mathrm{e},\kappa)$  (see Eq. \ref{Eq:line_emissivity}) is a function of all three plasma parameters. Determining electron density from observations prior to and independently of $T$ and $\kappa$ would greatly simplify the plasma diagnostics \citep[][]{Dzifcakova10,Mackovjak13}. Therefore, we searched preferentially for density-sensitive line ratios that are not strongly sensitive to $T$ and $\kappa$. Such line ratios have to contain lines belonging to the same ion in order to avoid strong sensitivity to $T$ and $\kappa$ coming from the relative ion abundance term $n(X^{+k})/n(X)$ in Eq. (\ref{Eq:line_emissivity}).

%
\subsection{Line selection procedure and selfblends}
\label{Sect:5.1}

Since the calculated synthetic spectra contain tens of thousands of transitions (Sect. \ref{Sect:5}), the vast majority of which are very weak and unobservable, we performed the search only on the strong observable lines. We consider a line as observable if its intensity is at least 0.05$I_\mathrm{max}$, where $I_\mathrm{max}$ is the intensity of the strongest line produced by the same ion. In order to take into account changes in the spectra produced by electron density (Fig. \ref{Fig:Spectra}, Sect. \ref{Sect:5}), we include lines having $I$\,$\geqq$\,0.05$I_\mathrm{max}$ at either log$(n_\mathrm{e}$/cm$^{-3})$\,=\,8 or 12, or both. For \ion{Fe}{IX}, a value of 0.02$I_\mathrm{max}$ is chosen instead of the 0.05$I_\mathrm{max}$. This is due to the very strong intensity of the 171.073\AA~bright line, which in turn would lead to dismissing most of the \ion{Fe}{IX} lines observed by EIS, notably the 197.862\AA~\citep[first identified by][]{Young09b}.

The selected observable lines also include selfblending transitions originating within a given ion. A weaker transition is considered to be a selfblend if its wavelength $\lambda$ is located within 50\,m\AA~of the stronger transition, and if its intensity is at least 5\,$\times$10$^{-4}I_\mathrm{max}$. Although these values are arbitrary, they have been chosen to correspond to the typical wavelength resolution of EUV spectrometers \citep[e.g. 47 m\AA~at 185\,\AA~for Hinode/EIS,][]{Culhane07}, as well as to limit the number of selfblending transitions only to the ones with a relevant intensity contribution.

The selection procedure including the selfblends is run automatically on synthetic spectra calculated for the Maxwellian distribution. Subsequently, the same lines are picked from the synthetic spectra calculated for the $\kappa$-distributions. This can be done, since there are no lines which are strong for a $\kappa$-distribution but unobservable for the Maxwellian distribution (Fig. \ref{Fig:Spectra}). After the selection procedure, we are left with 30 observable lines of \ion{Fe}{IX}, 20 for \ion{Fe}{X}, 40 for \ion{Fe}{XI}, and 31 for \ion{Fe}{XII} as well as \ion{Fe}{XIII}. The lines together with their selfblends are listed in Tables \ref{Table:fe9}--\ref{Table:fe13}. There, possible presence of blends from transitions in other ions are indicated as well.

%
\subsection{Line ratios recommended for density-diagnostics independently of $\kappa$}
\label{Sect:5.2}

We identify the best density diagnostics as line ratios that are both strongly dependent on $n_\mathrm{e}$ and relatively independent of $T$ and $\kappa$. The recommended ratios for density diagnostics are shown in Figs. \ref{Fig:diag_ne9-10}--\ref{Fig:diag_ne13}. There, density-sensitive ratios of two lines are plotted as a function $n_\mathrm{e}$ for the Maxwellian distribution (black lines) together with $\kappa$\,=\,5 (green) and 2 (red lines). Blue lines correspond to the intensity ratios according to the CHIANTI v7.1 \citep{Landi13}.

For the present calculations, the ratios are plotted at three different temperatures. These are the temperature of the peak of the relative ion abundance for the respective distribution (full lines), as well as temperatures corresponding to the $\approx$1\% of the maximum of the relative ion abundance. We note that 1\% of the relative ion abundance is an extreme value; usually, it is expected that the ion is formed at temperatures much closer to the peak of the ion abundance. However, this allows us to capture the sensitivity of the ratios to $T$ in the entire temperature range corresponding to the formation of each ion. It also provides an estimate of the maximum uncertainty of the diagnosed log$(n_\mathrm{e}$/cm$^{-3})$ if the simultaneous diagnostics of $T$ and $\kappa$, described in Sect. \ref{Sect:6} are not performed.

In the following, we report on density-sensitive line ratios throughout the electromagnetic spectrum. However, the discussion will be more focused on the line ratios observed by the Hinode/EIS, since this is a recent spectroscopic instrument with a large observed dataset.

%
\subsubsection{\ion{Fe}{IX}}
\label{Sect:5.2.1}

The ion \ion{Fe}{IX} has only a few UV and EUV lines whose ratios can be used to determine $n_\mathrm{e}$. For line identifications, see e.g. \citet{DelZanna14} and references therein. The \ion{Fe}{IX} 241.739\AA\,/\,244.909\AA~\citep{Storey02,DelZanna14} is density-sensitive for log$(n_\mathrm{e}/\mathrm{cm}^3)$\,$\lessapprox$\,10; however it has a non-negligible dependence on $T$ that increases for low $\kappa$ (Fig. \ref{Fig:diag_ne9-10}, \textit{top left}). We note that these lines are just outside of the EIS long-wavelength channel.

\subsubsection{\ion{Fe}{IX} lines longward of 1000\AA}
\label{Sect:5.2.2}

The \ion{Fe}{IX} ion also offers several density-sensitive ratios involving forbidden lines \citep{DelZanna14}. Two examples, the 2043\AA\,/\,3802\AA~and 3644\AA\,/\,3802\AA~are shown in Fig. \ref{Fig:diag_ne9-10}. The latter ratio shows a good agreement with the CHIANTI v7.1 atomic data (Fig. \ref{Fig:diag_ne9-10}, \textit{left, bottom}). Nevertheless, there are differences in present calculations for the Maxwellian distribution and the CHIANTI v7.1 atomic data, mainly for log$(n_\mathrm{e}/\mathrm{cm}^3)$\,$\gtrapprox$\,10. The 2043\AA~line has different intensities for all densities, resulting in disagreement with the CHIANTI v7.1 calculations (Fig. \ref{Fig:diag_ne9-10}, \textit{left, middle}). We note that these line ratios also have non-negligible sensitivity to $T$.

%
\subsubsection{\ion{Fe}{X}}
\label{Sect:5.2.3}

For line identifications, see \citet{DelZanna04}, \citet{DelZanna12e}, and references therein. The \ion{Fe}{X} ion has only a few density-sensitive line ratios that also show non-negligible sensitivity to $T$.  All of these ratios are usable only for log$(n_\mathrm{e}/\mathrm{cm}^3)$\,$\lessapprox$\,10. The three best ones are plotted in Fig. \ref{Fig:diag_ne9-10}. For the 175.475\AA\,/\,190.037\AA~ratio, the sensitivity to $T$ increases with decreasing $\kappa$ (Fig. \ref{Fig:diag_ne9-10}, \textit{top right}). Both the lines involved are selfblended by transitions contributing $\approx$0.4--1.4\%, with the exception of 189.996\AA~transition, which contributes up to 2.7\% at log$(n_\mathrm{e}/\mathrm{cm}^3)$\,=\,12.  We note that we do not show the corresponding ratio for the CHIANTI 7.1 atomic data due to differences in atomic data and numerous difficulties in identifying individual selfblending transitions.

The 180.441\AA\,/\,(184.509\AA\,+\,184.537\AA) is also usable; however, the density sensitivity is rather weak. The 234.315\AA\,/\,(257.257\AA\,+\,257.259\AA\,+\,257.263\AA) shows a strong density-sensitivity, but involves the 234.315\AA~line, which is unobservable by the EIS instrument. The selfblend at 257.3\AA~can be used to diagnose $n_\mathrm{e}$ also in conjuction with other lines, e.g. 207.449\AA, 220.247\AA, 224.800\AA, 225.856\AA, or 226.998\AA. We note that the selfblend at 257.3\AA~is dominated by the 257.263\AA~line at log$(n_\mathrm{e}/\mathrm{cm}^3)$\,$\lessapprox$\,8, but at higher densities, the 257.259\AA~line is the dominant one. Finally, the 234.315\AA~line is not included in CHIANTI v7.1.

%
\subsubsection{\ion{Fe}{XI}}
\label{Sect:5.2.4}

For line identifications, see \citet{DelZanna10c}, \citet{DelZanna12d}, and references therein. The \ion{Fe}{XI} ion offers several density-sensitive ratios (Fig. \ref{Fig:diag_ne11}). Here we report only on those that are not strongly sensitive to $\kappa$. The 178.058\AA\,/\,192.813\AA, 180.401\AA\,/\,182.167\AA, and 182.167\AA\,/\,188.216\AA~are all usable at log$(n_\mathrm{e}/\mathrm{cm}^3)$\,$\lessapprox$\,10. They show a weak dependence on $\kappa$, with $\kappa$\,=\,2 providing systematically lower log$(n_\mathrm{e})$ by about 0.2\,dex \citep{Mackovjak13}. At higher densities, the 184.410\AA\,/\,230.165\AA~ratio can be used. This ratio shows very low dependence on $T$. However, it involves a 230.165\AA~line not observable by Hinode/EIS. The 184.410\AA~line is blended by a 184.446\AA~transition contributing of up to 7\% of the intensity, especially at low densities. We note that both the 184.446\AA~and 230.165\AA~lines are not included in CHIANTI 7.1.

%
\subsubsection{\ion{Fe}{XII}}
\label{Sect:5.2.5}

For line identifications, see \citet{DelZanna05}, \citet{DelZanna12d}, and references therein. The \ion{Fe}{XII} also offers several density-sensitive ratios that are insensitive to $\kappa$ (Fig. \ref{Fig:diag_ne12}). Two of these are the (186.854\AA\,+\,186.887\AA)\,/\,195.119\AA~and (186.854\AA\,+\,186.887\AA)  /\,203.728\AA~ratios. The 186.887\AA~is the stronger line, with the 186.854\AA~having 20--60\% of the 186.887\AA~intensity at log$(n_\mathrm{e}/\mathrm{cm}^3)$\,=\,8--12, respectively. We note that these ratios are shifted to lower densities compared to the CHIANTI 7.1 results \citep{DelZanna12a}. We also note that the 203.728\AA~line is actually blending the stronger \ion{Fe}{XIII} selfblend at 203.8\AA~\citep[e.g.][Fig. 9 therein]{Young09}. Careful deblending should be performed if this line is to be used for density diagnostics. Other available \ion{Fe}{XII} ratios include the 338.263\AA\,/\,364.467\AA~and the (201.740\AA\,+\,201.760\AA)\,/\,(211.700\AA\,+\,211.732\AA). The 201.760\AA~has about 60--90\% intensity of the 201.740\AA~line. On the other hand, the 211.700\AA~has only about 2--3\% intensity compared to the 211.732\AA~line.

%
\subsubsection{\ion{Fe}{XIII}}
\label{Sect:5.2.6}

For line identifications, see \citet{DelZanna11} and references therein. The \ion{Fe}{XIII} has many density-sensitive line ratios, several of which are observable by Hinode/EIS \citep{Young09,Watanabe09}. In Fig. \ref{Fig:diag_ne13} we show the best ones. The 196.525\AA\,/\,202.044\AA, 200.021\AA\,/\,202.044\AA, and 200.021\AA\,/\,209.916\AA~are all strongly density-sensitive and all contain unblended lines. The usually used 203.8\AA\,/\,202.044\AA~ratio involves a complicated selfblend at 203.8\AA~that contains four transitions, 203.772\AA, 203.765\AA, 203.826\AA, and 203.835\AA~\citep{DelZanna11}. The main contributor is the 203.835\AA~line, with the contribution of other lines at log$(n_\mathrm{e}/\mathrm{cm}^3)$\,=\,8--12 ranging from $\approx$\,23 to 1\%, 38--2\%, and 47--36\% of the 203.835\AA~intensity, respectively. The 203.8\AA\,/\,202.044\AA~ratio saturates at $\approx$\,4.4 for log$(n_\mathrm{e}/\mathrm{cm}^3)$\,=\,10.5.

The 204.262\AA\,/\,(246.209\AA\,+\,246.241\AA) and the (221.824\AA\,+\,221.828\AA)\,/\,251.952\AA~are also excellent density diagnostics, both having negligible dependence on both $T$ and $\kappa$. The intensity of the 246.241\AA~line is about 0.1--0.8\% of the 246.209\AA~line, its contribution to the selfblend increasing with density. Similarly, the intensity of the 221.824\AA~is about 19--2\% of the 221.828\AA~line, again strongly decreasing with density. The 249.241\AA~and 221.824\AA~lines are however absent from CHIANTI 7.1. There are also differences in our atomic data compared to the CHIANTI 7.1 for the 221.828\AA~line.

\subsubsection{\ion{Fe}{XIII} lines longward of 1000\AA}
\label{Sect:5.2.7}

The pair of infrared lines at 10749.0\AA~and 10801.0\AA~also provide a good density diagnostic \citep[e.g.][]{Pineau73,Wiik94,Singh02} (Fig. \ref{Fig:diag_ne13}, \textit{bottom right}). Here, the ratio is usable up to higher densities of up to log$(n_\mathrm{e}/\mathrm{cm}^3)$\,$\approx$\,9.5, compared to the original paper of \citet{Pineau73}. These lines can be measured by spectrographs mounted on the ground-based coronagraphs, such as the Coronal Multichannel Polarimeter (COMP-S) instrument being installed at the Lomnicky Peak observatory \citep{Tomczyk07,Tomczyk09,Schwartz12,Schwartz14}.

\begin{figure*}[!ht]
   \centering
   \includegraphics[width=8.8cm]{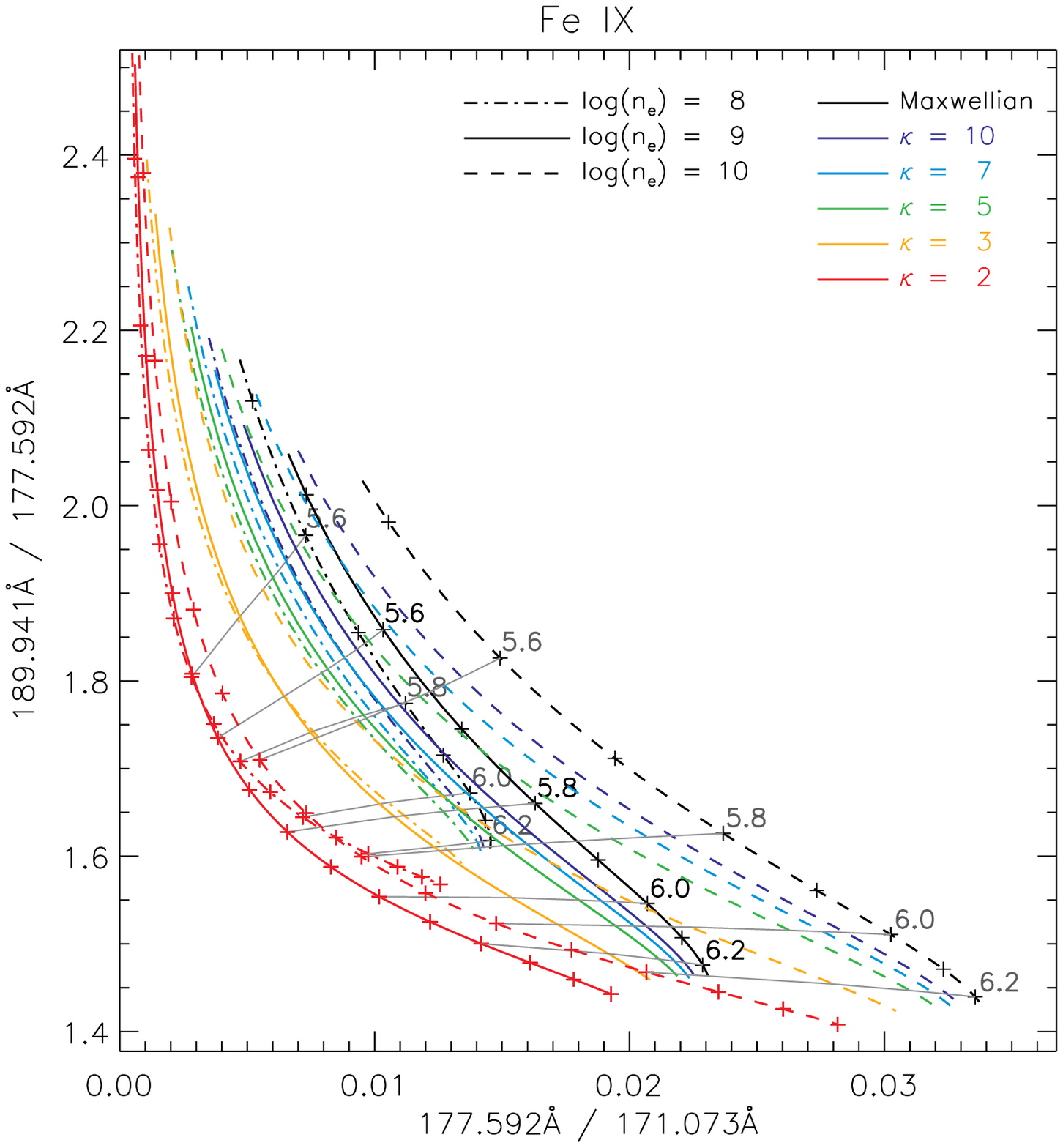}
   \includegraphics[width=8.8cm]{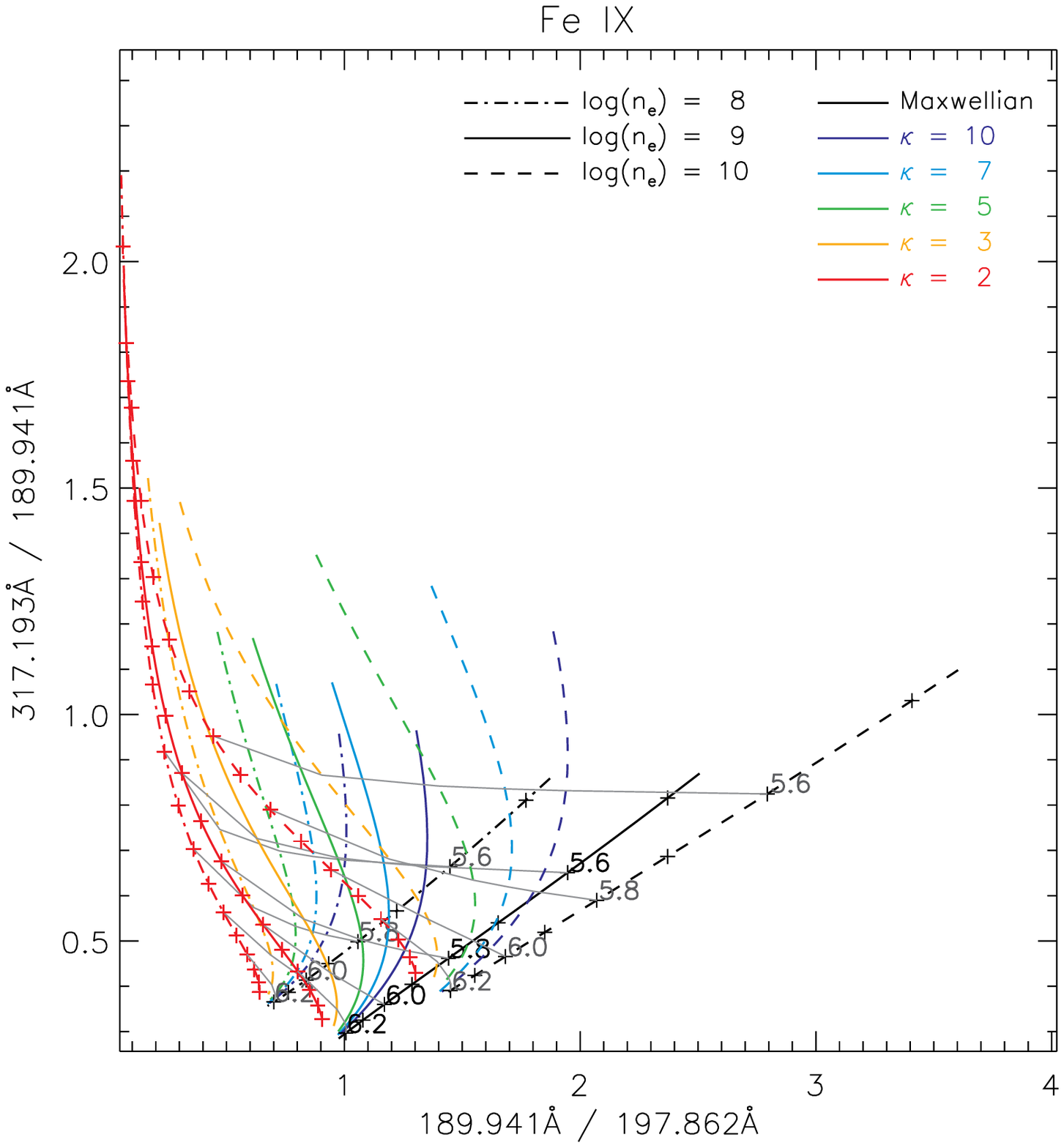}
   \includegraphics[width=8.8cm]{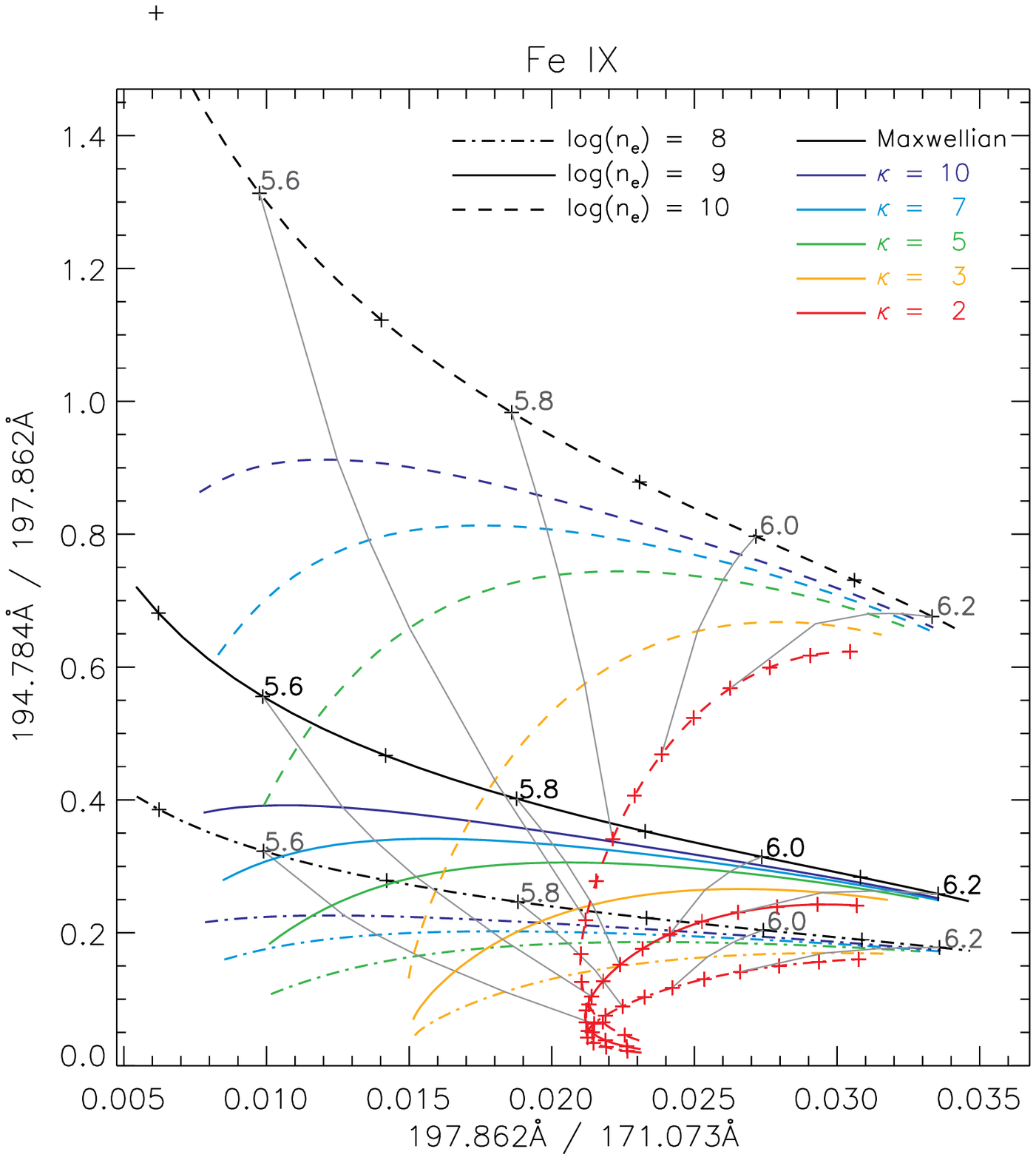}
   \includegraphics[width=8.8cm]{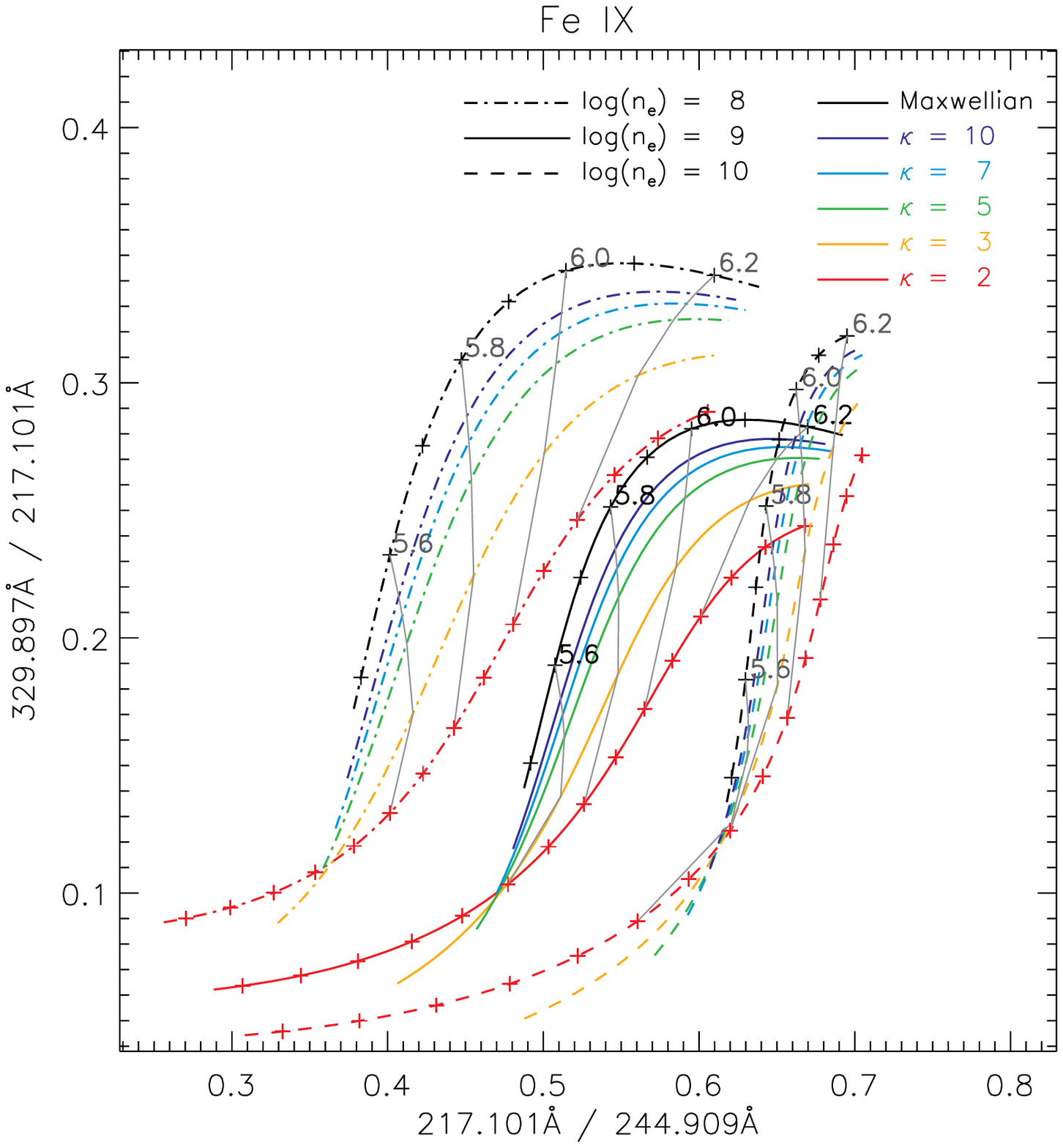}
   \caption{Examples of the theoretical plots for simultaneous diagnostics of $T$ and $\kappa$ from \ion{Fe}{IX}. Both axes show intensity ratios of two lines, with the line intensities in the units of phot\,cm$^{-2}$\,s$^{-1}$\,sr$^{-1}$. Individual line styles correspond to different log$(n_\mathrm{e}/\mathrm{cm}^3)$, colors to individual values of $\kappa$, and gray lines denote isotherms connecting points having the same log$(T/$K).}
   \label{Fig:diag_tk9}%
\end{figure*}
%
\begin{figure*}
   \centering
   \includegraphics[width=8.8cm]{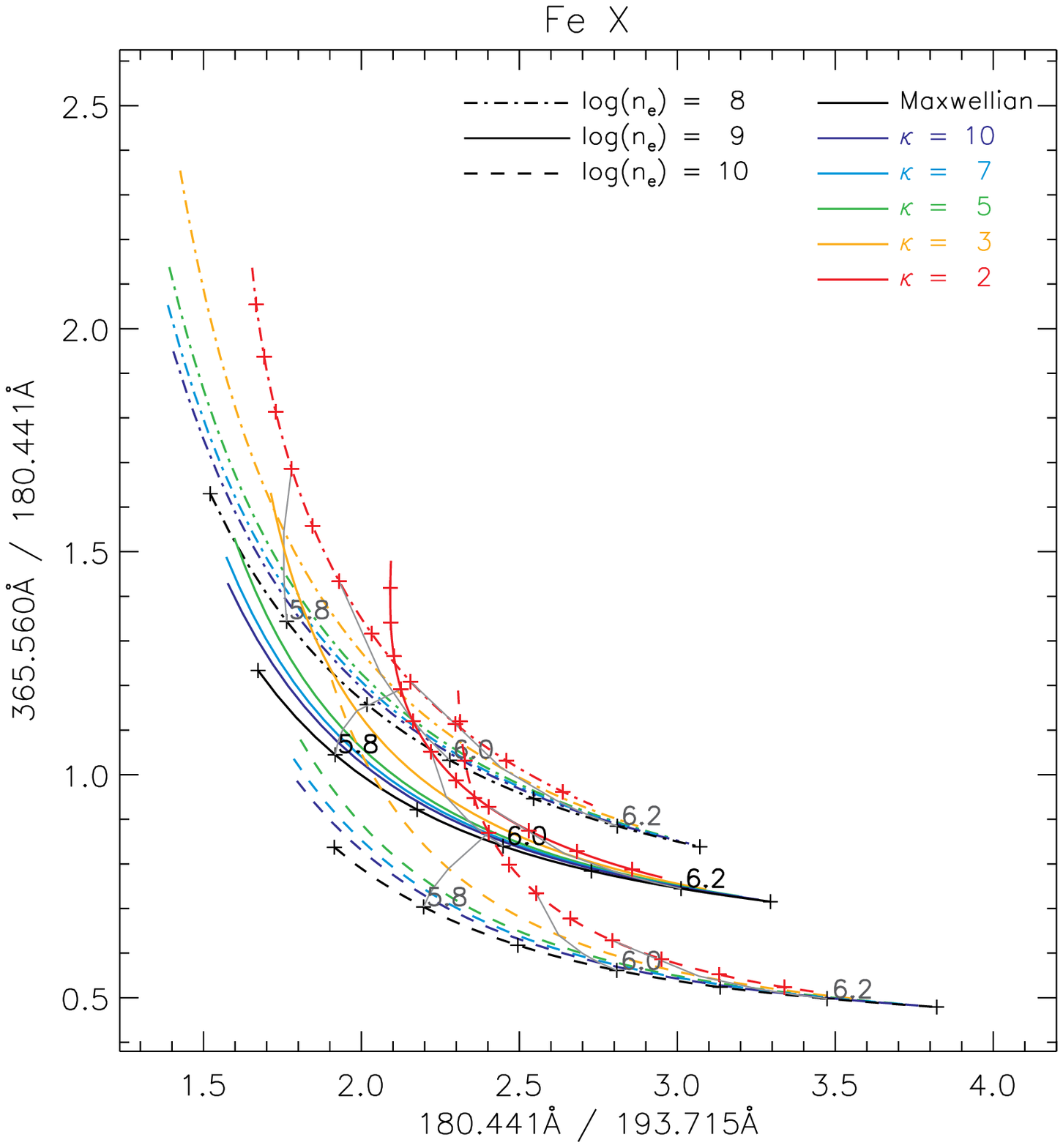}
   \includegraphics[width=8.8cm]{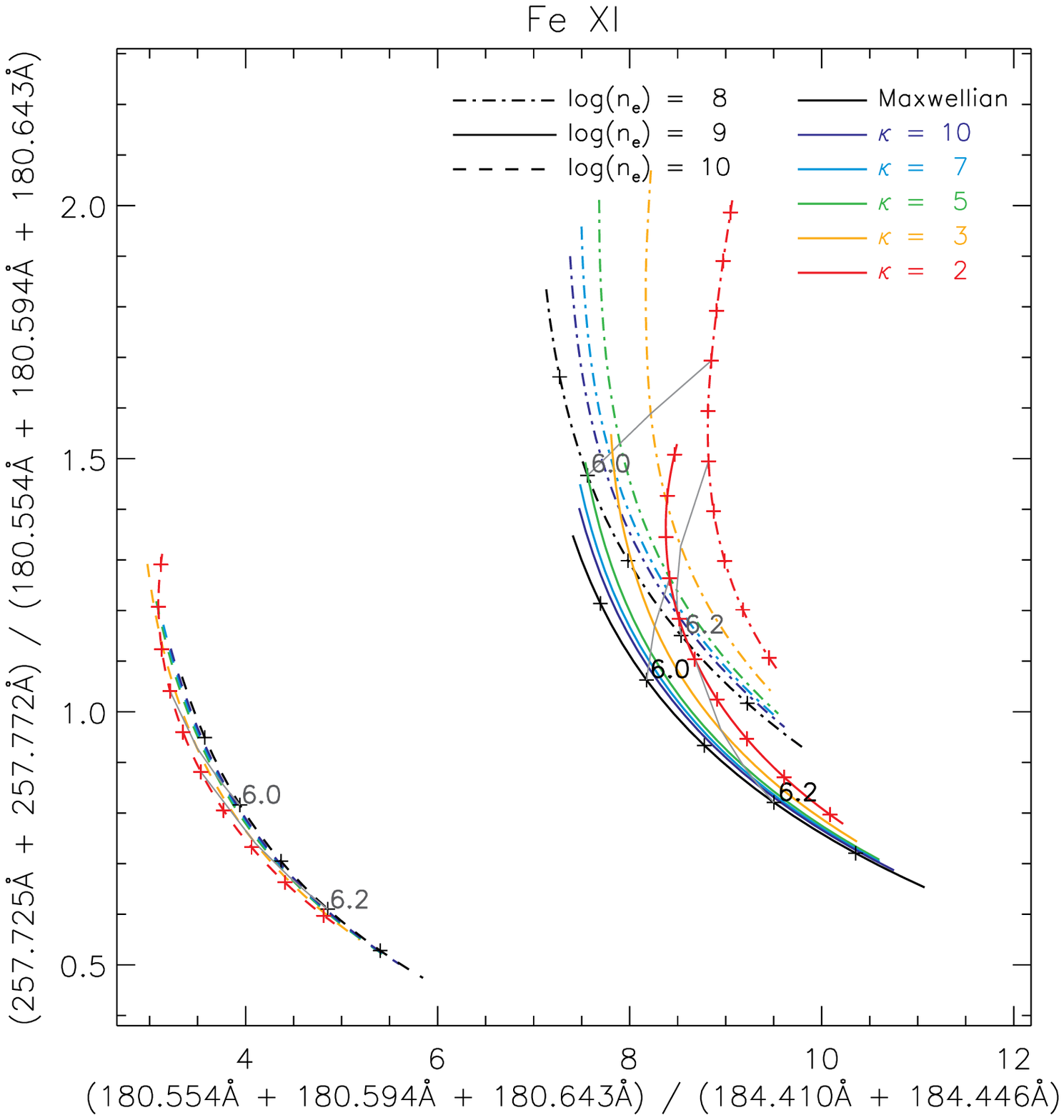}
   \includegraphics[width=8.8cm]{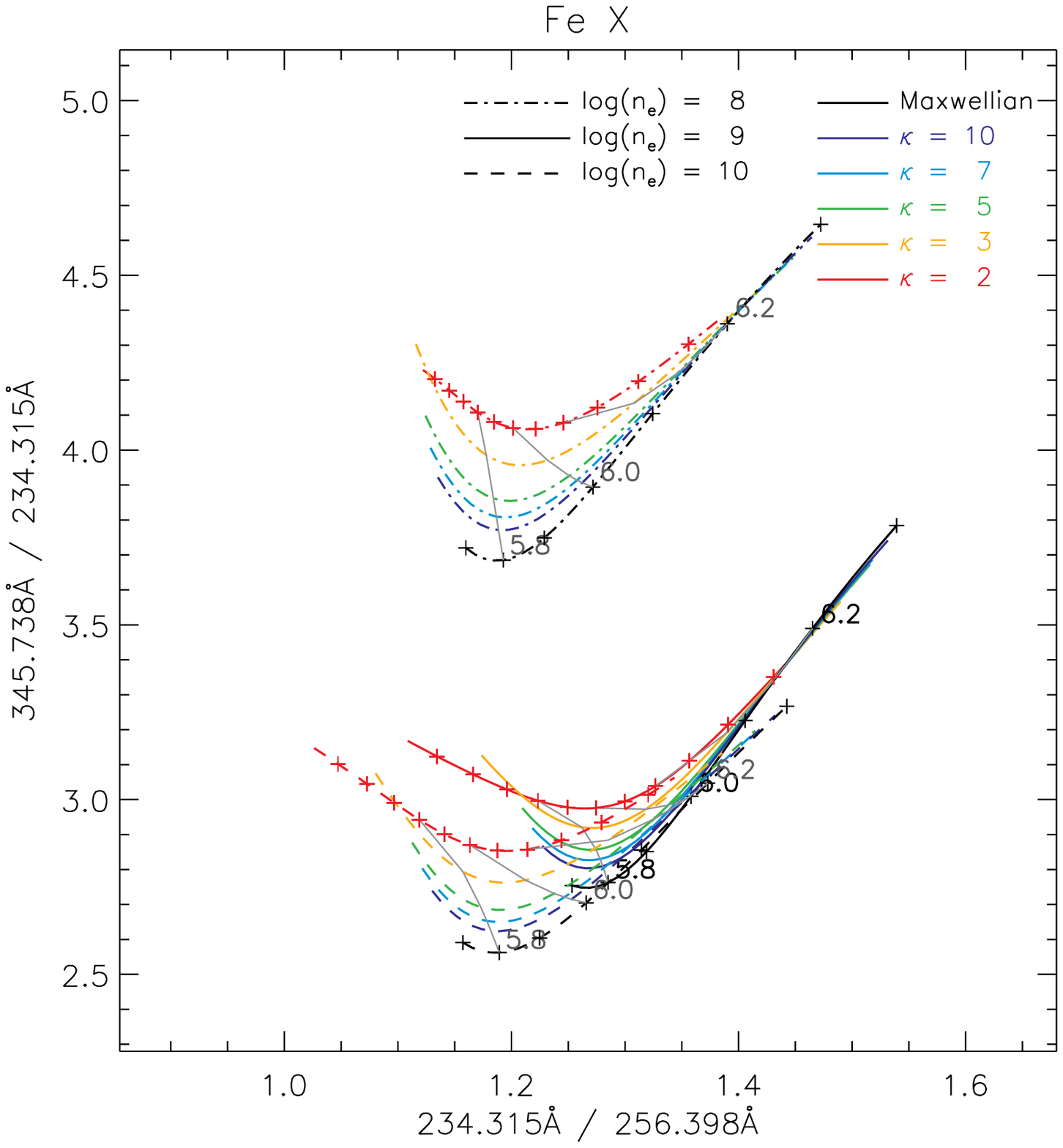}
   \includegraphics[width=8.8cm]{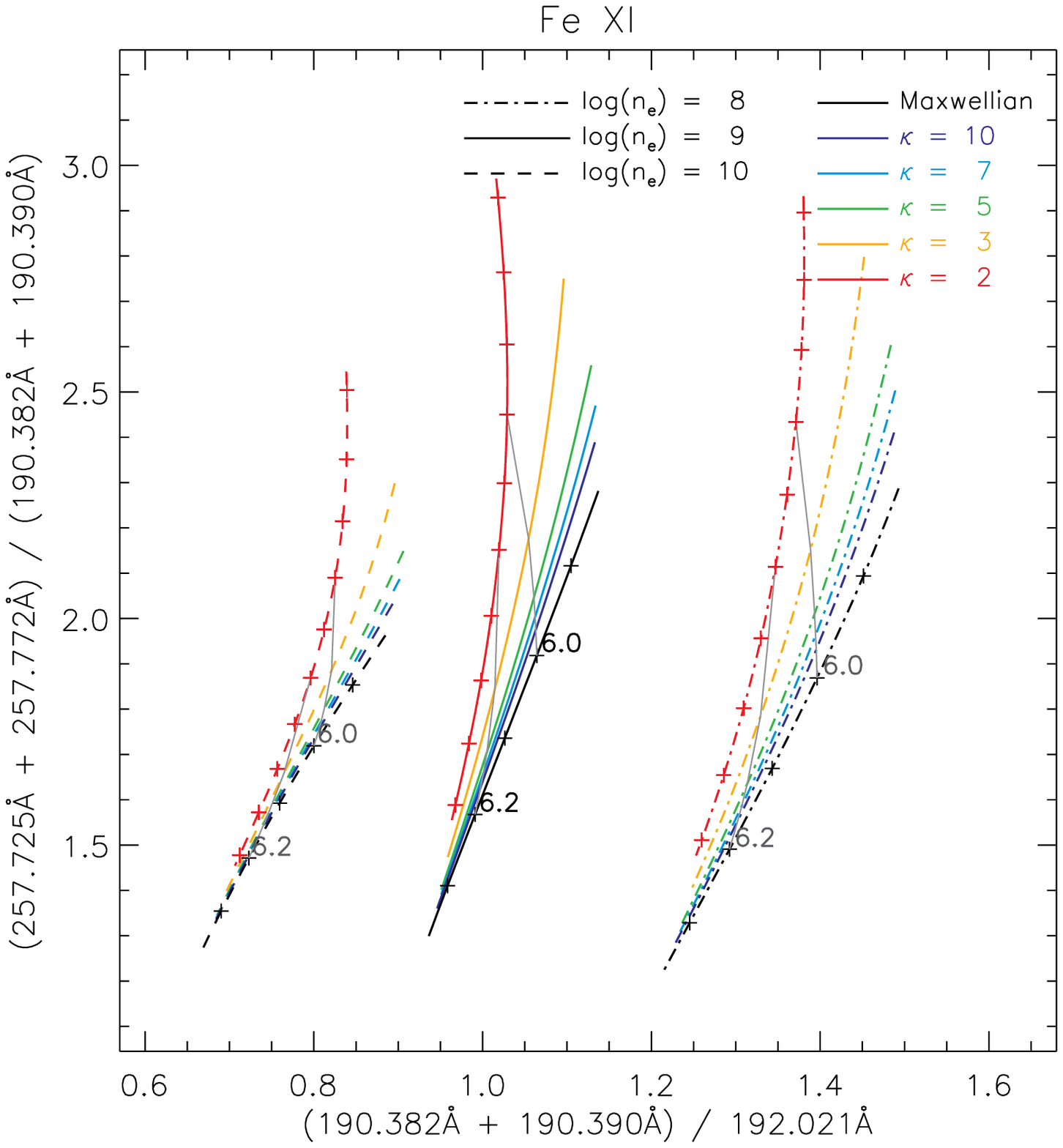}
   \includegraphics[width=8.8cm]{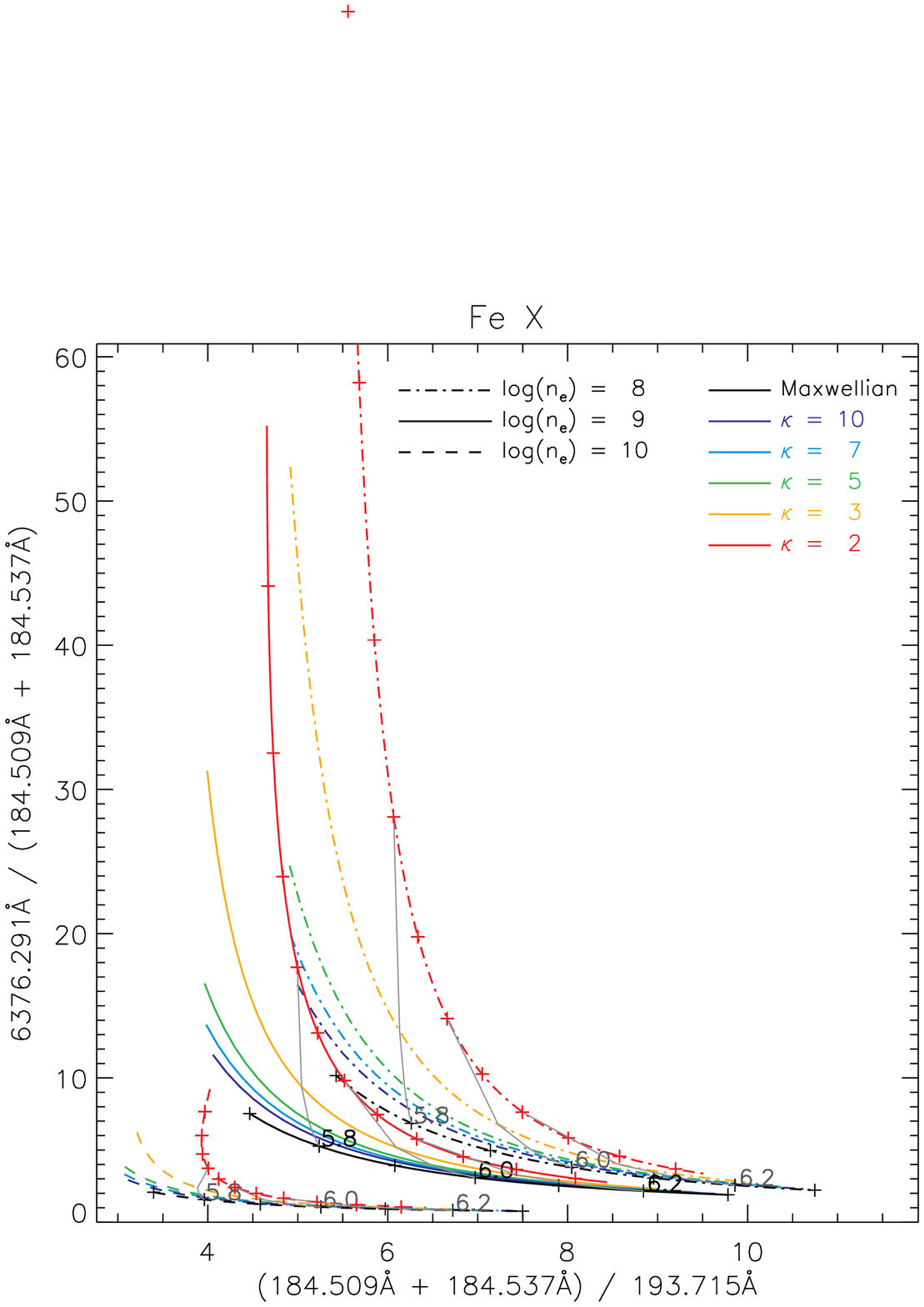}
   \includegraphics[width=8.8cm]{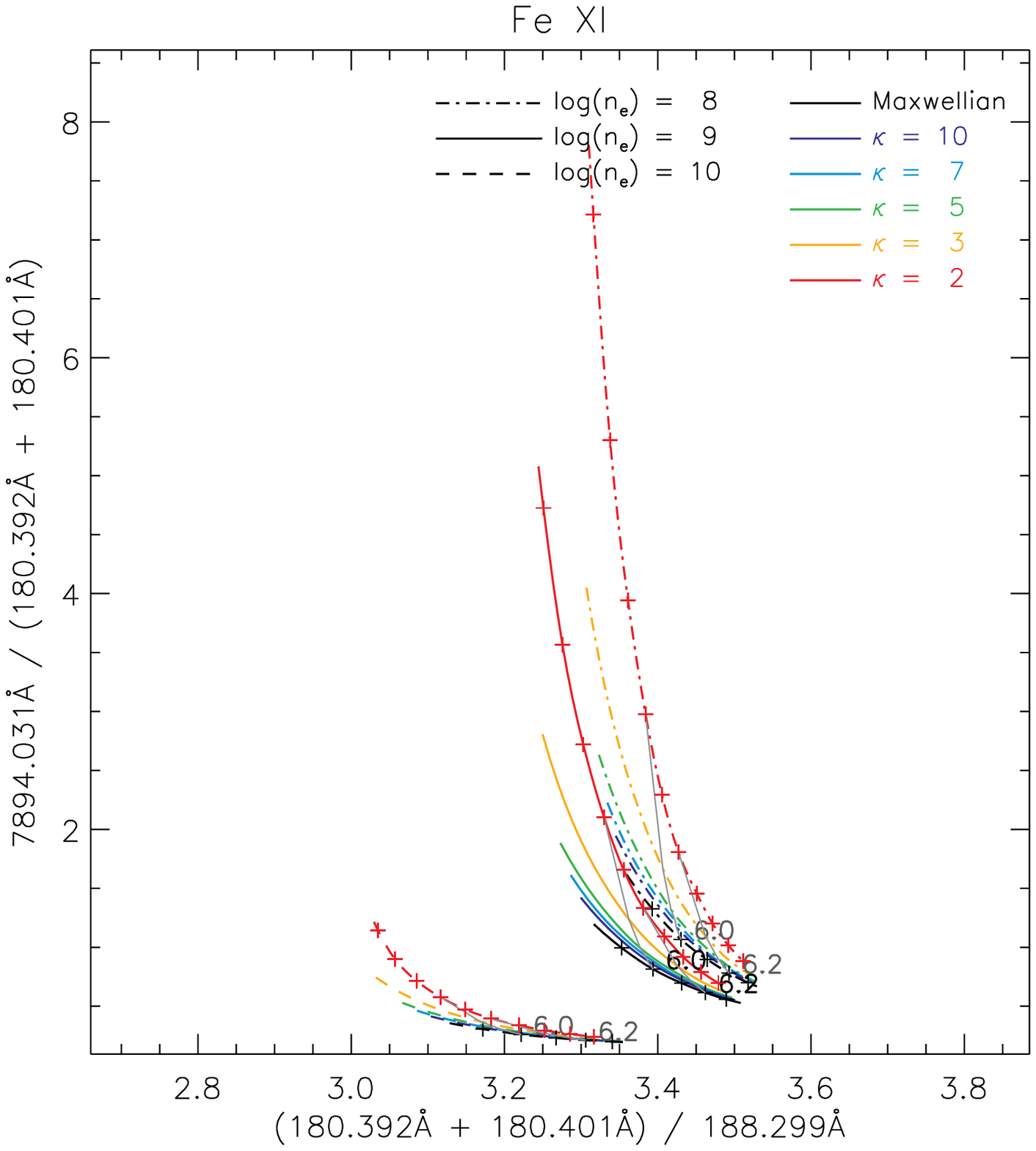}
   \caption{Examples of the theoretical plots for simultaneous diagnostics of $T$ and $\kappa$ from \ion{Fe}{X} (\textit{left}) and \ion{Fe}{XI} (\textit{right}). Line styles and colors are the same as in Fig. \ref{Fig:diag_tk9}.}
   \label{Fig:diag_tk10-11}%
\end{figure*}
%
%
%
%
\section{Theoretical diagnostics of $T$ and $\kappa$}
\label{Sect:6}

Once the electron density $n_\mathrm{e}$ is determined, $T$ and $\kappa$ can be diagnosed simultaneously using the ratio-ratio technique involving line ratios sensitive to both $\kappa$ and $T$ \citep{Dzifcakova10,Dzifcakova11Si,Mackovjak13}. We note that these papers investigated specific Fe lines observable by EIS \citep{Dzifcakova10}, non-Fe EIS lines \citep{Mackovjak13}, and \ion{Si}{III} observed by SUMER \citep{Dzifcakova11Si}. Other wavelength ranges have not yet been investigated.

Here, we investigate the lines selected using the procedure described in Sect. \ref{Sect:5.1} for sensitivity to $\kappa$-distributions using the ratio-ratio technique. The ratio-ratio diagrams are constructed using intensities of three lines, with $R_X$\,=\,$I_1/I_2$ and $R_Y$\,=\,$I_3/I_1$, where the wavelengths obey the relation $\lambda_1 < \lambda_2 < \lambda_3$. If the number of lines selected (Sect. \ref{Sect:5.1}) is ${\cal N}$, then the number of possible ratio-ratio diagrams, i.e. the $R_Y(R_X)$ combinations is given by ${\cal N}!/(3! ({\cal N}-3)!)$. We note that in principle, four different lines can be used, with $R_Y$\,=\,$I_3/I_4$. The number of possible ratio-ratio diagrams would then be increased by a factor of $({\cal N}-3)/4$. For the \ion{Fe}{XI} this would increase the possible number of ratio-ratio diagrams from 9\,880 by a factor of 9.25 (see Sect. \ref{Sect:5.1}), which is impractical. We thus limited our investigation to the ratio-ratio diagrams using only intensities of three lines, with the additional restriction that the ratios of line intensities either larger than 20 or lower than 1/20 are not investigated except for \ion{Fe}{IX} (Sect. \ref{Sect:5.1}), because with such different line intensities, the weaker line would have considerably more photon noise, limiting the usefulness of the diagnostics. Furthermore, ratio-ratio diagrams that exhibit sensitivity to $\kappa$ in only a very limited range of electron densities (less than an order of magnitude) are not considered here.

The ratio-ratio diagrams here are also limited to the lines emitted by the same ion \citep{Dzifcakova10,Mackovjak13}. This is done to avoid an additional source of errors coming from uncertainties in the ionization and recombination rates. Therefore, any sensitivity of the line ratios to $\kappa$ has to originate in the excitation processes alone.

In the following subsections, we report only on the ratios that show the highest sensitivity to $\kappa$. The vast majority of the possible $R_Y(R_X)$ combinations either do not show sensitivity to $\kappa$, or only a very low one. This is important, since the \textit{presence of many insensitive lines can be misleading} when interpreting solar spectra, as the spectra can then be fitted simply with some values of $T$ and emission measure under the assumption of the Maxwellian distribution, without providing any clue to the nature of the distribution function of the emitting plasma.

%
\subsection{\ion{Fe}{IX}}
\label{Sect:6.1}

This ion offers the best options for diagnosing $\kappa$ and $T$, since unlike the \ion{Fe}{X}--\ion{Fe}{XIII}, the transitions selected for \ion{Fe}{IX}  contain several that do not involve the first few energy levels (Tables \ref{Table:fe9}--\ref{Table:fe13}). The most conspicuous example is the 13--148 transition constituting a line at 197.862\AA.

Examples of the best diagnostics options are shown in Fig. \ref{Fig:diag_tk9}. The 177.592\AA\,/\,171.073\AA~-- 189.941\AA\,/\,177.592\AA~has the lowest sensitivity to density, while having a high sensitivity to $\kappa$, by about a factor of 2 difference between $\kappa$\,=\,2 and the Maxwellian distribution. Strong sensitivity to $T$ is also present, mainly due to the 189.941\AA\,/\,177.592\AA~ratio. The 189.941\AA\,/\,197.862\AA~-- 317.193\AA\,/\,189.941\AA~and the 197.862\AA\,/\,171.073\AA~-- 189.941\AA\,/\,197.862\AA~ratio-\-ratio diagrams also have a strong sensitivity to $\kappa$, but in these cases the sensitivity decreases strongly with increasing log$(T/$K). The sensitivity of the 217.101\AA\,/\,244.909\AA~-- 329.897\AA\,/\,217.101\AA~to $\kappa$ decreases with increasing $n_\mathrm{e}$ and at low log$(T/$K) is further complicated by the overlap of various $\kappa$. Nevertheless, this diagram still can be used for log$(T/$K)\,$\approx$\,5.8--6.1.

The ratio-ratio diagrams presented here emphasize the need for independent diagnostics of the electron density (Sect. \ref{Sect:5}), which would in most cases complicate the diagnostics of $\kappa$. 

%
\subsection{\ion{Fe}{X}}
\label{Sect:6.2}

The \ion{Fe}{X} ion offers only a few possibilities of diagnosing $T$ and $\kappa$. Figure \ref{Fig:diag_tk10-11} (\textit{left}) shows three examples. The first one is 180.441\AA\,/\,193.715\AA~--~365.560\AA\,/\,180.441\AA. This ratio-ratio diagram exhibits sensitivity to density. The sensitivity to $\kappa$ comes mainly from the 365.560\AA\,/\,180.441\AA~ratio involving lines with different excitation thresholds. The sensitivity to $\kappa$ is better at lower log$(T/$K) and decreases towards higher values of log$(T/$K)\,$\approx$\,6.2. The sensitivity to $T$ is given mainly by the 180.441\AA\,/\,193.715\AA~ratio. Other alternative ratios with similar sensitivities are, for example, the 180.441\AA\,/\,207.449\AA~--~345.738\AA\,/\,180.441\AA, with the 345.738\AA~and the 365.560\AA~lines being formed from the same upper level (Table \ref{Table:fe10}), as well as the (184.509\AA\,+\,184.537\AA)\,/\,193.715\AA~--~365.560\AA\,/\,(184.509\AA\,+ +\,184.537\AA).

The second example shown in Fig. \ref{Fig:diag_tk10-11}, \textit{left, middle} is the 234.315\AA\,/\,256.398\AA~--~345.738\AA\,/\,234.316\AA. We note that the sen\-sitivity of these \ion{Fe}{X} ratios to $\kappa$ is again not much greater than a few tens of per cent. Such sensitivity occur for low $\kappa$\,=\,2 and decrease strongly with increasing log$(T/$K). We also note that this sensitivity is comparable to the typical calibration uncertainty of the EUV spectrometers \citep[e.g.][]{Culhane07,Wang11,DelZanna13a}, which makes the diagnostics using these \ion{Fe}{X} lines unrealistic at present. 

\subsubsection{\ion{Fe}{X} diagrams involving the 6376\AA~line}
\label{Sect:6.2.1}

The last example involves the red forbidden line at 6376.3\AA~(Fig. \ref{Fig:diag_tk10-11}, \textit{left, bottom}). The sensitivity to $\kappa$ comes from the 6376.3\AA\,/\,(184.509\AA\,+\,185.537\AA), while the sensitivity to $T$ arises from the (184.509\AA\,+\,185.537\AA)\,/\,193.715\AA~ratio. The overall shape is similar to Fig. \ref{Fig:diag_tk10-11}, \textit{left, top}. The sensitivity to $\kappa$ is greatest at low log$(T/$K) and increases with decreasing density. This density-dependence of this ratios could make this diagram especially useful for low-density regions higher in the atmosphere, where the red line can still be strong.

\begin{figure}
   \centering
   \includegraphics[width=7.8cm]{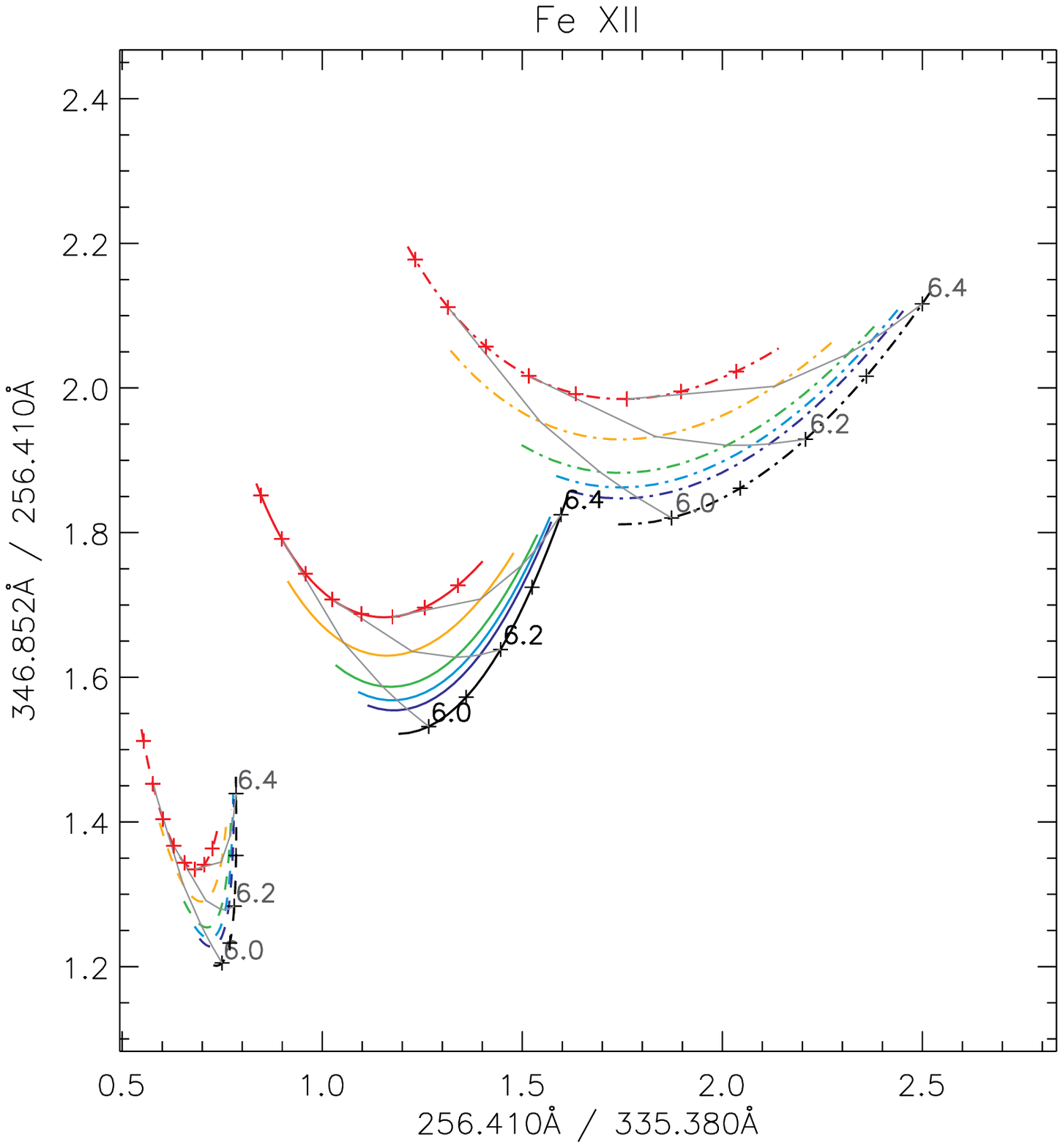}
   \includegraphics[width=7.8cm]{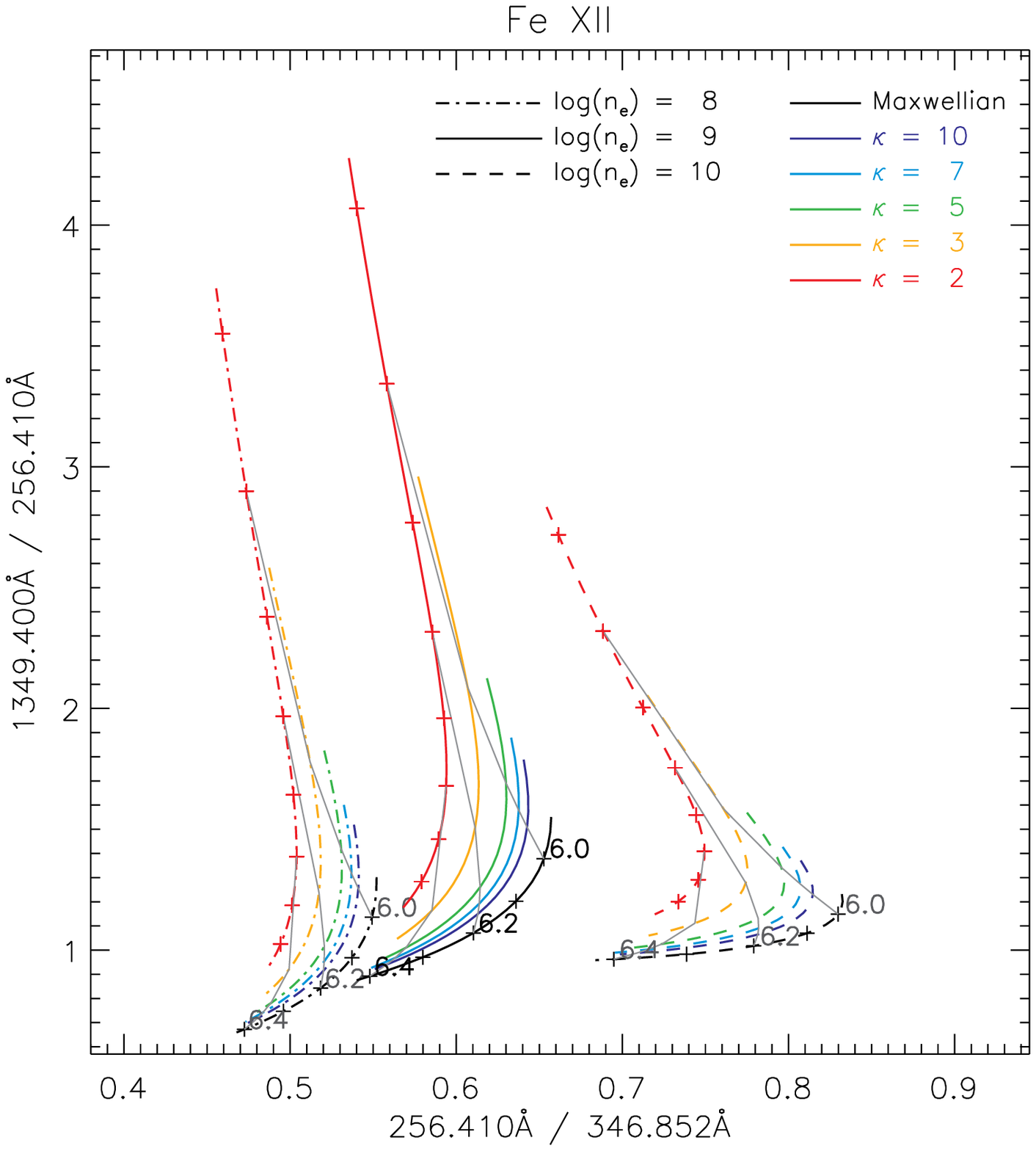}
   \includegraphics[width=7.8cm]{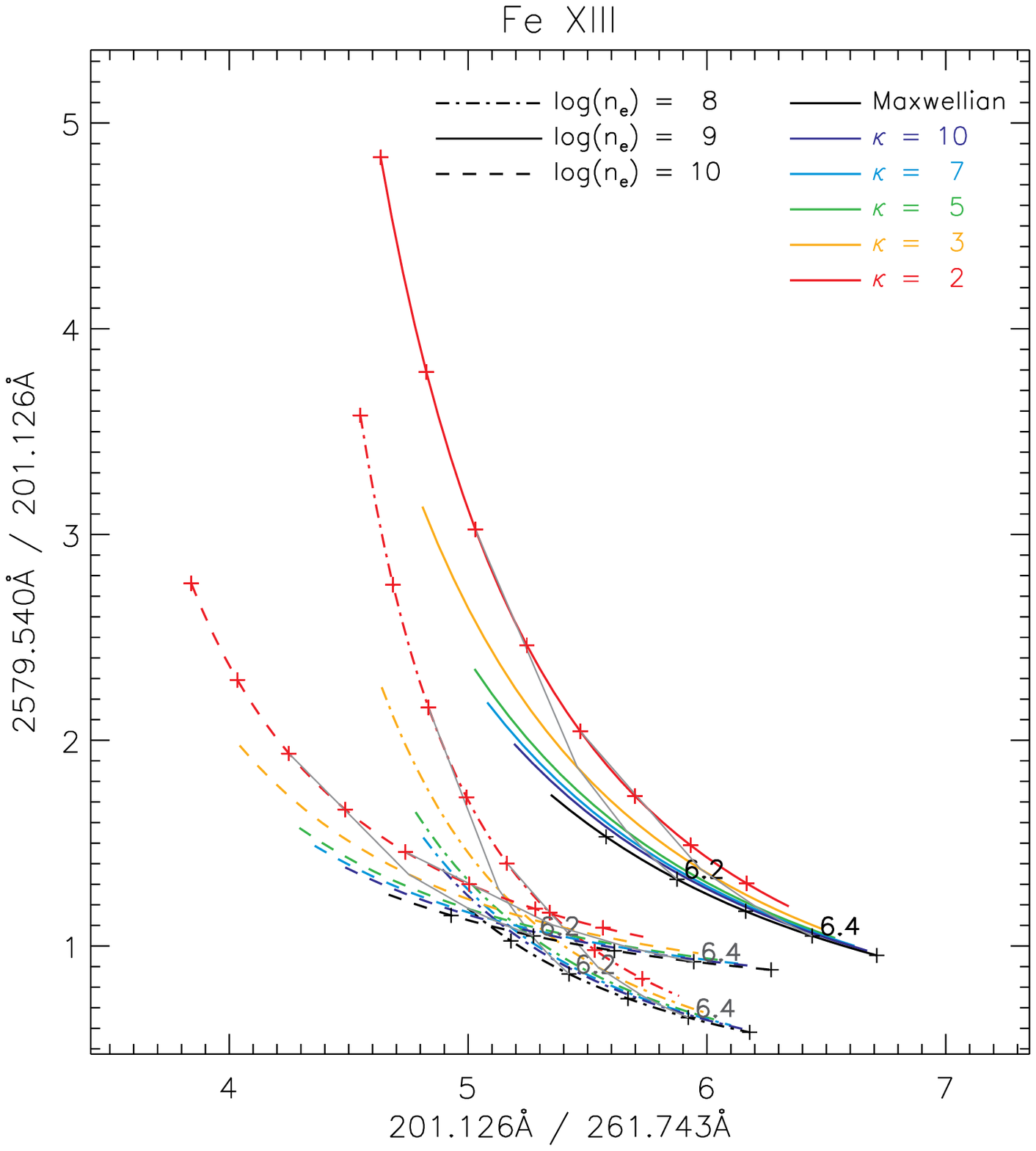}
   \caption{Examples of the theoretical plots for simultaneous diagnostics of $T$ and $\kappa$ from \ion{Fe}{XII} (\textit{top} and \textit{middle}), as well as \ion{Fe}{XIII} (\textit{bottom}). Line styles and colors are the same as in Fig. \ref{Fig:diag_tk9}.}
   \label{Fig:diag_tk12-13}%
\end{figure}
%
%
\subsection{\ion{Fe}{XI}}
\label{Sect:6.3}

The \ion{Fe}{XI} ion has the largest number of observable lines (Sect. \ref{Sect:5.1}). It offers several options for simultaneous diagnostics of $T$ and $\kappa$. Here, we present several typical examples (Fig. \ref{Fig:diag_tk10-11}, \textit{right}. These ratio-ratio diagrams show that the sensitivity of individual line ratios to $\kappa$ is again typically only a few tens of per cent. In most instances, the magnitude of sensitivity depends on log$(T)$, with greater sensitivities at lower temperatures as in the case of \ion{Fe}{X}. 

The sensitivity to $T$ is obtained by using lines such as the selfblends at 257\AA~\citep{DelZanna10a}, i.e. the 257.538\AA\,+\,257.547\AA\,+\,257.554\AA\, +\,257.558\AA, dominated by the 257.547\AA~and 257.554\AA~lines; or the 257.725\AA\,+\,257.772\AA, dominated by the 257.772\AA~line. The theoretical diagrams presented here confirm that the ratios of these lines to lines at 178--211\AA~(i.e. short-wavelength channel of the EIS instrument) are strongly sensitive to temperature. We note that other lines, such as the 234.730\AA, 236.494\AA, 239.780\AA\,+\,239.787\AA, 308.554\AA, 349.046\AA, 356.519\AA, or 358.613\AA~can be used instead, depending on the particular combination with other lines in a given ratio-ratio diagram.

We note that some ratio-ratio diagrams can be used in a limited range of densities, for example, (180.554\AA\,+\,180.594\AA\,+\,180.643\AA) /\,(184.410\AA\,+\,180.446\AA) -- (257.725\AA\, +\,257.772\AA)\,/\,(180.554\AA\, +\,180.594\AA\, +\,180.643\AA) (Fig. \ref{Fig:diag_tk10-11}, \textit{right, top}) can be used for log$(n_\mathrm{e}/\mathrm{cm}^3)$\,$\approx$\,8--9, i.e. at quiet Sun densities. Other ratios have similar sensitivity to $\kappa$ independently of density. An example is the (190.382\AA\,+\,190.390\AA)\,/\,192.021\AA~-- (257.725\AA\,+\,257.772\AA)\,/\,(190.382\AA\,+\,190.390\AA), see Fig. \ref{Fig:diag_tk10-11}, \textit{right, middle}. In this diagram, changes in $\kappa$ can be masked by relatively small changes in the electron density. This emphasizes the need for diagnosing $n_\mathrm{e}$  together with $T$ and $\kappa$, since some of the density-sensitive ratios presented in Sect. \ref{Sect:5} have non-negiligible dependence on $T$.

\subsubsection{\ion{Fe}{XI} diagrams involving the 7894\AA~line}
\label{Sect:6.3.1}

The forbidden 7894.03\AA~line can be a strong indicator of departures from the Maxwellian distribution (see also Sect. \ref{Sect:4} and Fig. \ref{Fig:Spectra}). Its intensity can be enhanced several times relative to the EUV lines, e.g. 180.392\AA\,\-+\,180.401\AA~(Fig. \ref{Fig:diag_tk10-11}, \textit{bottom, right}), which is the strongest EUV line of \ion{Fe}{XI}. However, the enhancement decreases with electron density, and at log$(n_\mathrm{e}/\mathrm{cm}^3)$\,$\approx$\,10 is only a negligible one.

%
\subsection{\ion{Fe}{XII}}
\label{Sect:6.4}

The \ion{Fe}{XII} ion offers only a few opportunities for determining $\kappa$. The ratio-ratio diagrams are found to always be density-dependent, with only a very weak sensitivity to $\kappa$, of the order of several tens of per cent. Two typical examples, both involving forbidden lines, are shown in Fig. \ref{Fig:diag_tk12-13}, \textit{top} and \textit{middle}. In these examples, the sensitivity to $\kappa$ originates from the 256.410\AA\,/\,346.852\AA~ratio. The 256.410\AA\,/\,346.852\AA~-- 1349.4\AA\, /\,256.410\AA~involves the 1349.4\AA~\ion{Fe}{XII} line observable by the IRIS spectrometer \citep{DePontieu14}. This is a forbidden transition that shows temperature sensitivity when combined with EUV lines. The temperature sensitivity increases with $\kappa$\,$\to$\,2 by more than a factor of two. Therefore, any anomalous intensities of this line could be caused by the presence of $\kappa$-distributions. We note that other forbidden lines, such as the 1242.01\AA~or the 2566.8\AA~can be used instead.

%
\subsection{\ion{Fe}{XIII}}
\label{Sect:6.5}

The \ion{Fe}{XIII} ion has very few line ratios sensitive to $\kappa$. The sensitivity is generally very weak and the ratio-ratio diagrams involve the 201.126\AA~line together with a forbidden transition in the near-ultraviolet. Examples include e.g. the 201.126\AA\,/\,261.743\AA~-- 2579.540\AA\,/\,201.126\AA~(Fig. \ref{Fig:diag_tk12-13}, \textit{bottom}). A similar alternative is the 201.126\AA\,/\,239.030\AA~-- 3388.9\AA\,/\,201.126\AA, except that the the 201.126\AA\,/\,239.030\AA~ratio increases monotonically with decreasing density. Similarly to the \ion{Fe}{XII} case (Sect. \ref{Sect:6.4}), anomalous intensities of the forbidden lines can be a signature of the non-Maxwellian $\kappa$ distributions.

%
\section{Summary and discussion}
\label{Sect:7}

We performed a calculation of the distribution-averaged collision strengths for \ion{Fe}{IX}--\ion{Fe}{XIII} and subsequently the spectral synthesis for all wavelengths for the non-Maxwellian $\kappa$-distributions. We used the state-of-the art atomic data and searched for line ratios sensitive to electron density, temperature, and $\kappa$. In doing so, the previous exploratory work of \citet{Dzifcakova10} was extended and superseded. We also investigated various collision strength approximations and their accuracy. The most important conclusions can be summarized as follows:

   \begin{enumerate}
      \item The calculated \ion{Fe}{IX}--\ion{Fe}{XIII} synthetic spectra show  consistent behaviour with $\kappa$ for most of the EUV lines. Typically, the line intensities decrease with $\kappa$ at temperatures corresponding to the peak of the relative ion abundance for the Maxwellian distribution.
      \item It is possible to perform a reliable diagnostics of the electron density without diagnostics of $\kappa$. \ion{Fe}{IX} and \ion{Fe}{X} offer only a few opportunities. The best density-diagnostics that are not strongly sensitive to $T$ or $\kappa$ come from \ion{Fe}{XI}--\ion{Fe}{XIII}. Some of the line ratios, such as the \ion{Fe}{XII} (186.854\AA\,+\,186.887\AA)\,/\,195.119\AA, \ion{Fe}{XIII} 204.262\AA\,/\,(246.209\AA\,+\,246.241\AA), or the \ion{Fe}{XIII} 10749.0\AA\,/\,10801.0\AA~show only a very weak dependence on $T$ and $\kappa$, and are excellent density diagnostics. 
      \item The consistent behaviour of EUV line intensities with $\kappa$ makes it is very difficult to perform diagnostics of $\kappa$ using only lines of one ion. The vast majority of line ratios have no sensitivity to $\kappa$ at all. A small number of line ratios exhibit sensitivity to $\kappa$. Typically, this sensitivity is of the order of only several tens of per cent, which is comparable to the calibration uncertainties of the present EUV instruments, such as the Hinode/EIS. Only very few line ratios, in particular for \ion{Fe}{IX}, exhibit larger sensitivity. The best diagnostics options presented here often include lines unobservable by present-day EUV spectroscopic instrumentation.
      \item Because of the above, signatures of the departures from the Maxwellian distribution will be inconspicuous in most of the observed EUV solar coronal spectra. In most instances, small changes in electron density can obscure the changes in the spectra due to $\kappa$. 
      \item Several forbidden lines, such as the \ion{Fe}{X} 6378.26\AA~line show reversed behaviour with $\kappa$. That is, the intensity of this line increases with decreasing $\kappa$. Other forbidden lines show decrease of intensity with $\kappa$; however, the decrease is not as strong as for the EUV lines. An example is the \ion{Fe}{XI} 7894.03\AA~line whose intensity changes only weakly with $\kappa$. Therefore, these forbidden lines are a good indicator of the departures from the Maxwellian distribution.
      \item Averaging the collision strengths $\Omega_{ji}(E_i)$ over the regularly-spaced grid in log$(E_i/\mathrm{Ryd})$ with a step of 0.01 gives the best approximation of the resulting $\Upsilon_{ij}(T,\kappa)$. The error for the strongest transitions is typically very small, less than 0.5\%. However, for weak transitions, low $\kappa$\,$\approx$\,2, and ions formed at transition-region temperatures the error can reach $\approx$20--30\%. There are two reasons for this: the shift of the relative ion abundance to low log$(T/\mathrm{K})$ for low $\kappa$, as well as the strongly decreasing $\Omega_{ji}(E_i)$ with $E_i$ for such weak transitions. In these cases, the $\Upsilon_{ij}(T,\kappa)$ calculated from the averaged $\left<\Omega_{ji}\right>_{\Delta \mathrm{log}(E_i/\mathrm{Ryd}) = 0.01}$ are dominated by the errors in the approximation near its first energy point. However, given that the first energy point always has a large uncertainty in the scattering calculations, we still consider the error of $\approx$20--30\% for weak transitions an acceptable one. We recommend using this method to decrease the size of the $\Omega_{ji}(E_i)$ datasets.
   \end{enumerate}

We have investigated all strong lines throughout the wavelength range unrestricted by the constraints of a given instrument. We have found only a few diagnostic options for $\kappa$ using lines of a single ion. Therefore, the diagnostic options including lines from neighbouring ionization stages need to be investigated in the future. Direct modelling of the entire observed spectrum including a few lines sensitive to $\kappa$ would be the best possibility. However, this is beyond the scope of this paper. Increase of sensitivity to $\kappa$ is expected if lines from neighbouring ionization stages are included \citep{Dzifcakova10,Mackovjak13} because of the changes in the ionization equilibrium \citep[][Fig. 6 therein]{Dzifcakova13}. The ionization equilibrium is, however, an additional source of uncertainty because of the individual ionization and recombination rates, as well as possible departures from the ionization equilibrium \citep[e.g.][]{Bradshaw04,Bradshaw09,deAvillez12,Reale12,Doyle12,Doyle13,Olluri13} at low electron densities.

The results presented here highlight the need for spectroscopic observations over a wide wavelength range, especially  in the 170\AA\,--\,370\AA~range, which include many strongly temperature-sensitive line pairs, such as the \ion{Fe}{IX} 244.909\AA\,/\,171.073\AA~or 244.909\AA\,/\,188.497\AA, or analogous combinations from other ions investigated here. Such a large wavelength range is necessary to keep the best options for diagnosing $\kappa$ from EUV lines. The proposed LEMUR instrument \citep{Teriaca12} will, however, observe the 170\AA\,--\,210\AA~wavelength range together with many other spectral windows longward of 482\AA. These will include the \ion{Fe}{XII} 1242.01\AA~line together with the strongest \ion{Fe}{XII} EUV lines, which may lead to a successful diagnostics. Inclusion of the forbidden lines in the near-UV, visible, or the infrared parts of the spectrum could, in principle, be done by multi-instrument observations involving space-borne spectrometers together with eclipse observations or ground-based coronagraphs, such as the COMP-S instrument being installed at the Lomnicky Peak observatory.

Finally, we stress the particular importance and necessity of high-quality intensity calibration. Given the sensitivities to departures from the Maxwellian distribution presented here, the $\approx$20\% uncertainties may no longer be sufficient enough.

\begin{acknowledgements}
The authors thank Peter Cargill for stimulating discussions. JD acknowledges support from the Royal Society via the Newton Fellowships Programme. GDZ and HEM acknowledge STFC funding through the DAMTP astrophysics grant, as well as the University of Strathclyde UK APAP network grant ST/J000892/1. EDz acknowledges Grant 209/12/1652 of the Grant Agency of the Czech Republic. The authors also acknowledge the support from the International Space Science Institute through its International Teams program. Hinode is a Japanese mission developed and launched by ISAS/JAXA, with NAOJ as domestic partner and NASA and STFC (UK) as international partners. It is operated by these agencies in cooperation with ESA and NSC (Norway). CHIANTI is a collaborative project involving the NRL (USA), the University of Cambridge (UK), and George Mason University (USA).
\end{acknowledgements}


\bibliographystyle{aa}         
\bibliography{Omega}   

\begin{appendix}

%
\section{Collision strength approximations and their accuracy}
\label{Appendix:Omega}

\begin{figure*}
   \centering
   \includegraphics[width=8.8cm]{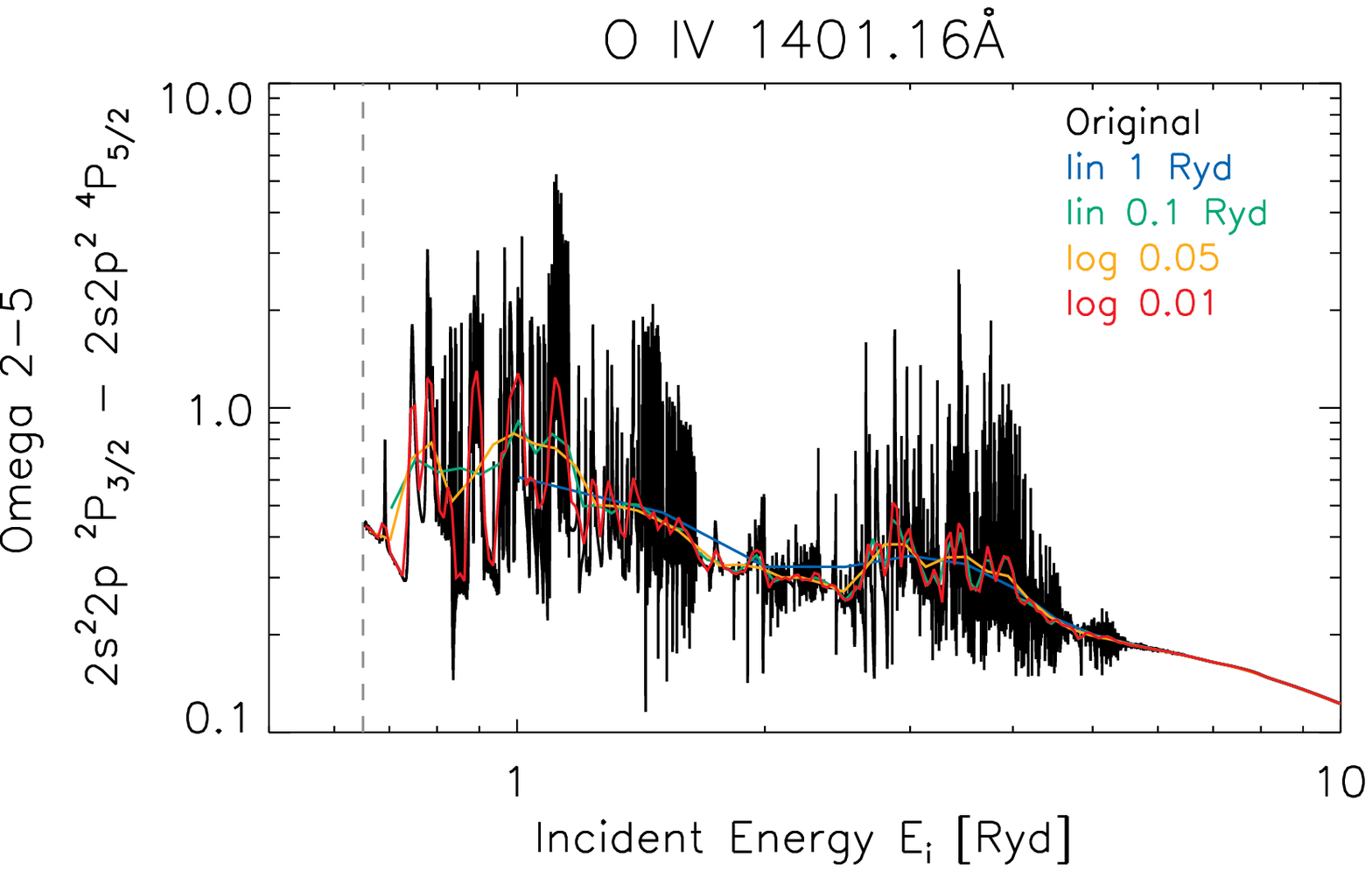}
   \includegraphics[width=8.8cm]{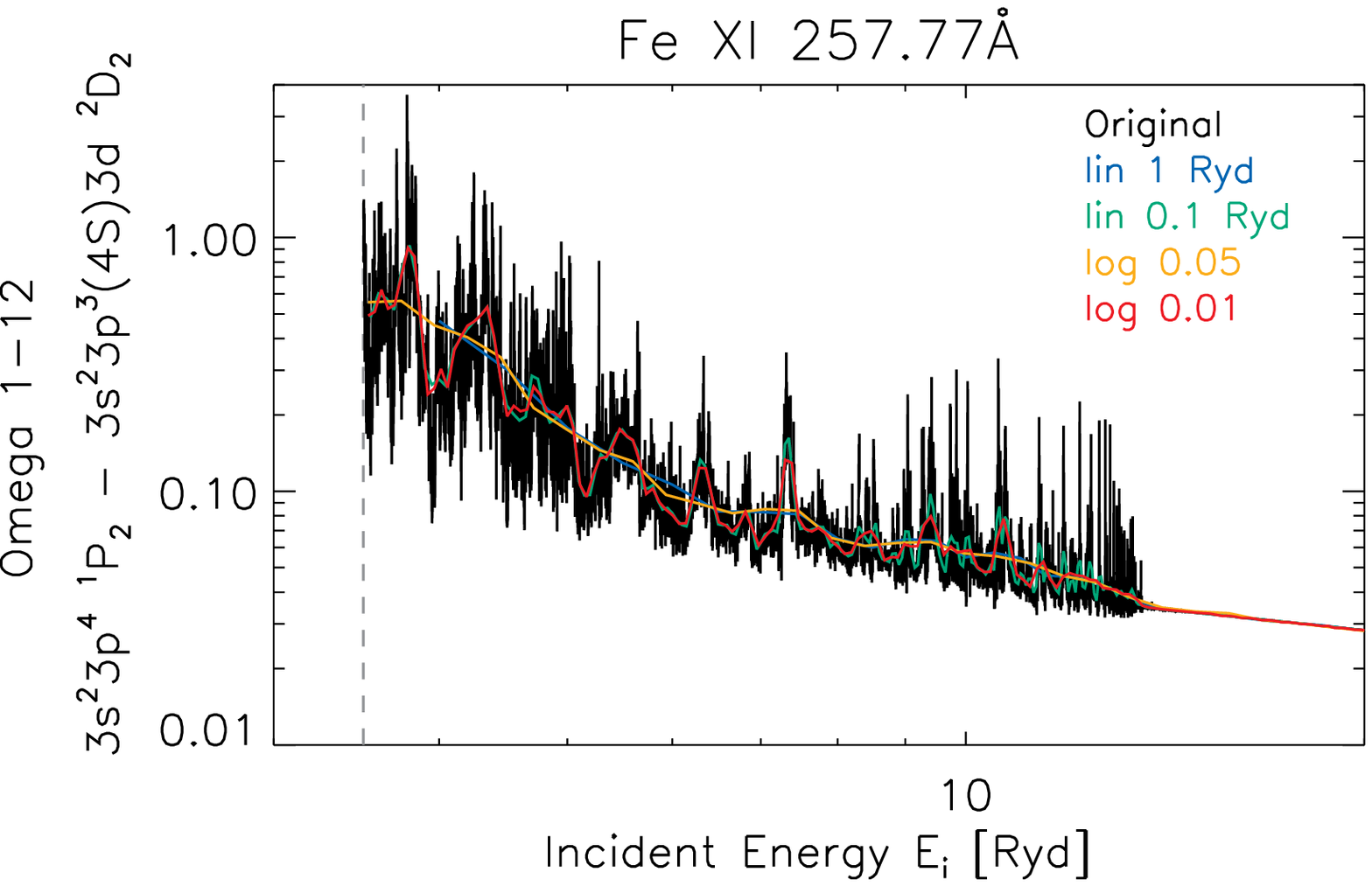}
   \includegraphics[width=8.8cm]{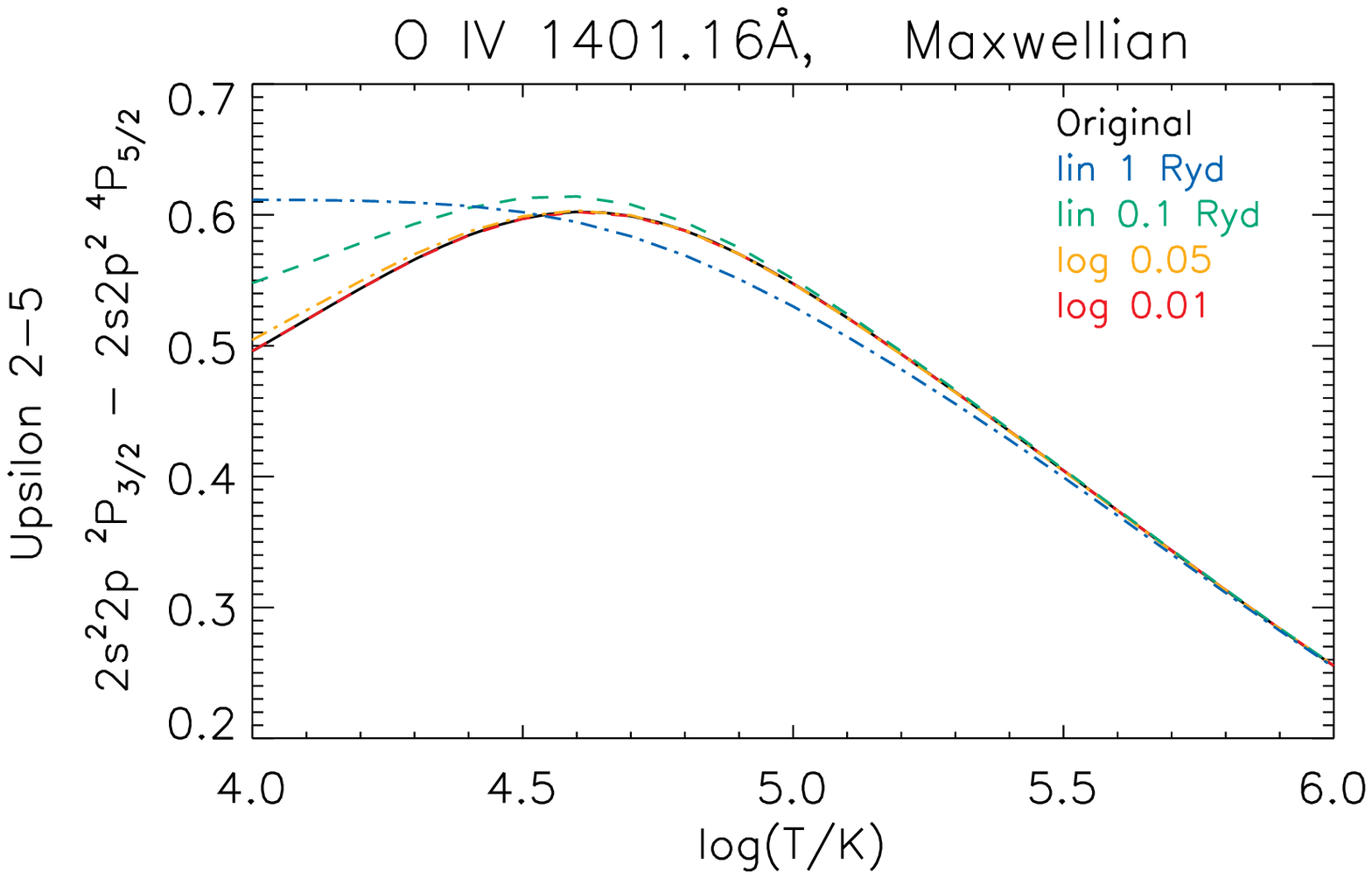}
   \includegraphics[width=8.8cm]{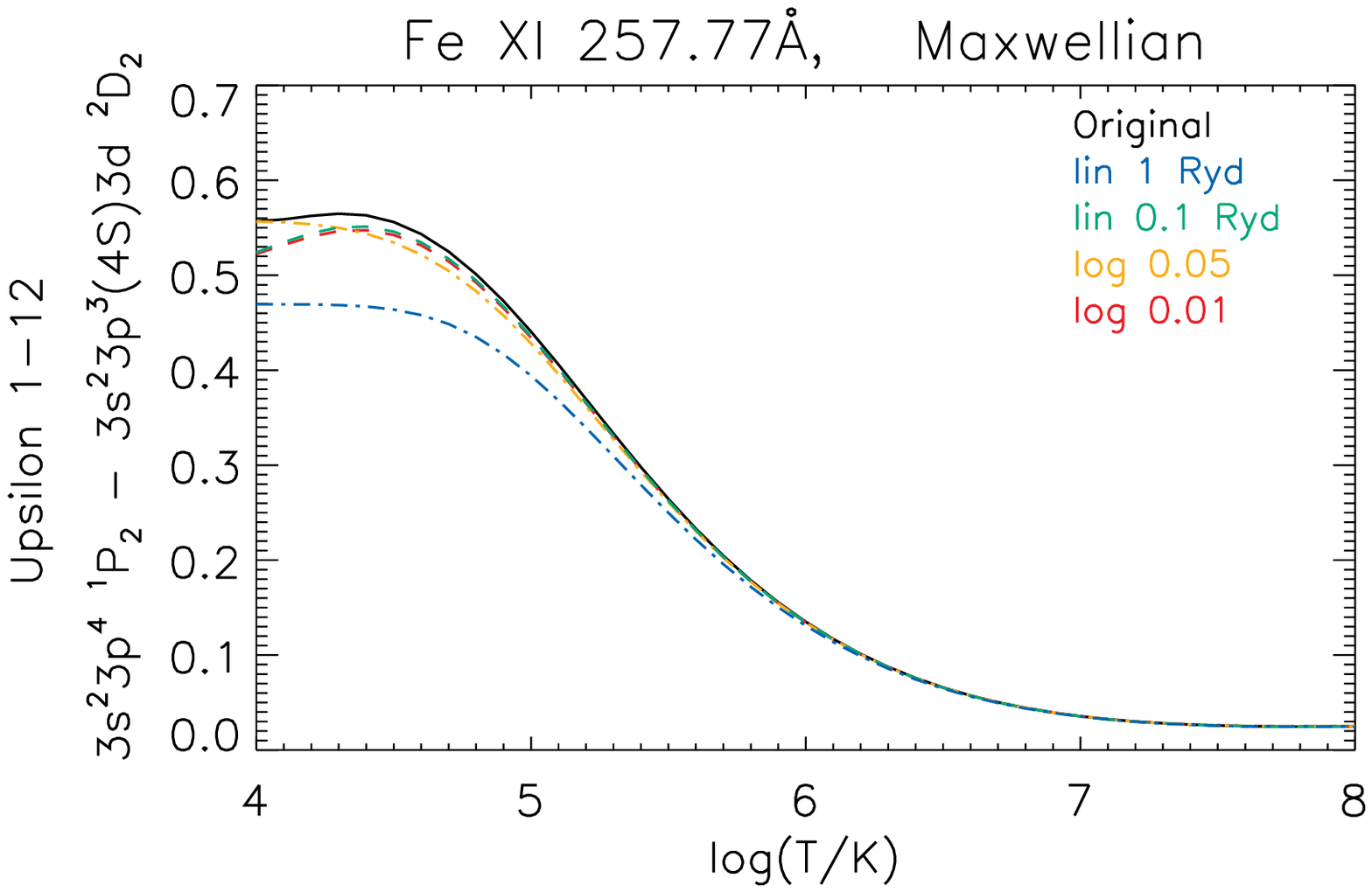}
   \includegraphics[width=8.8cm]{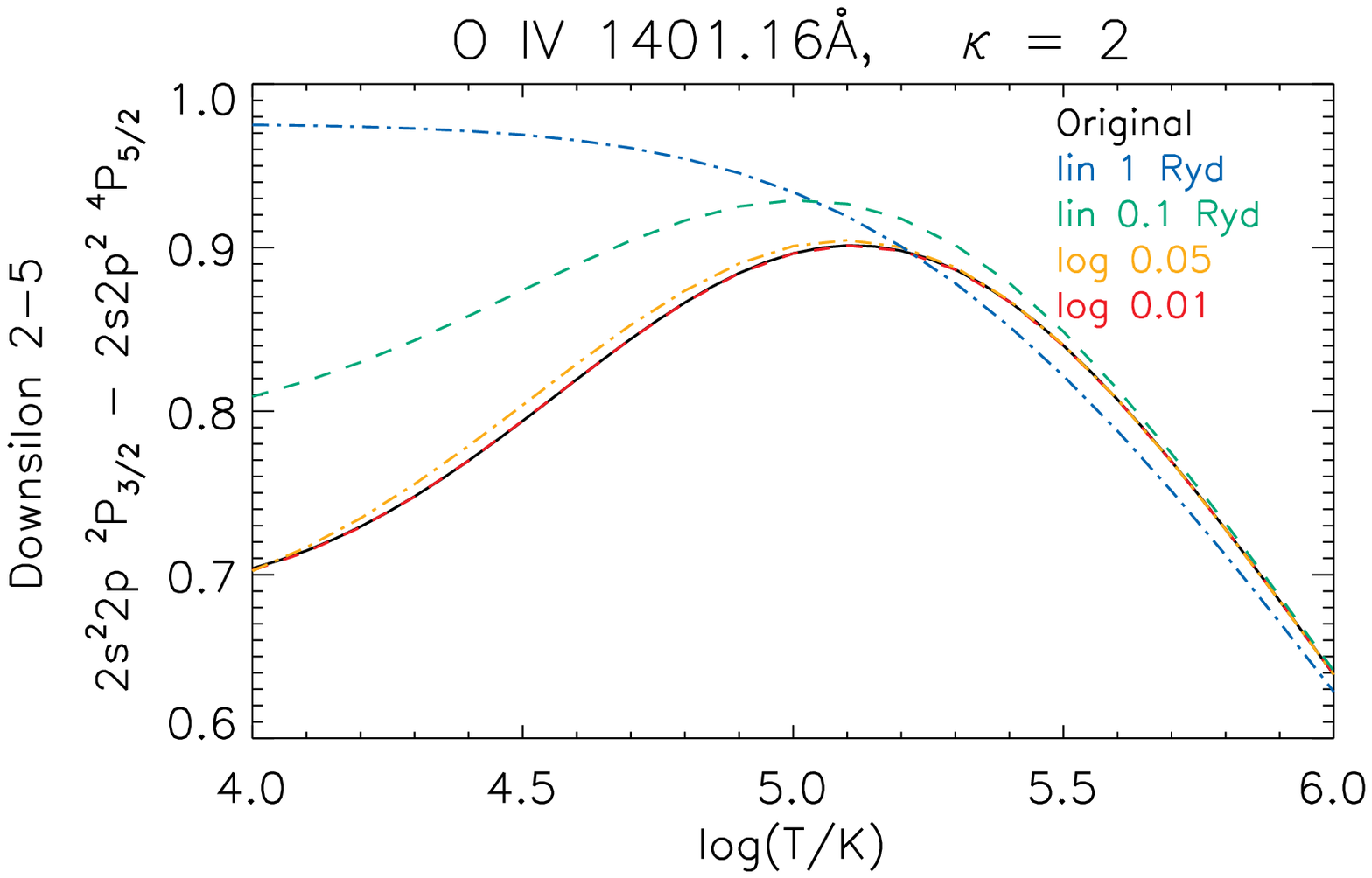}
   \includegraphics[width=8.8cm]{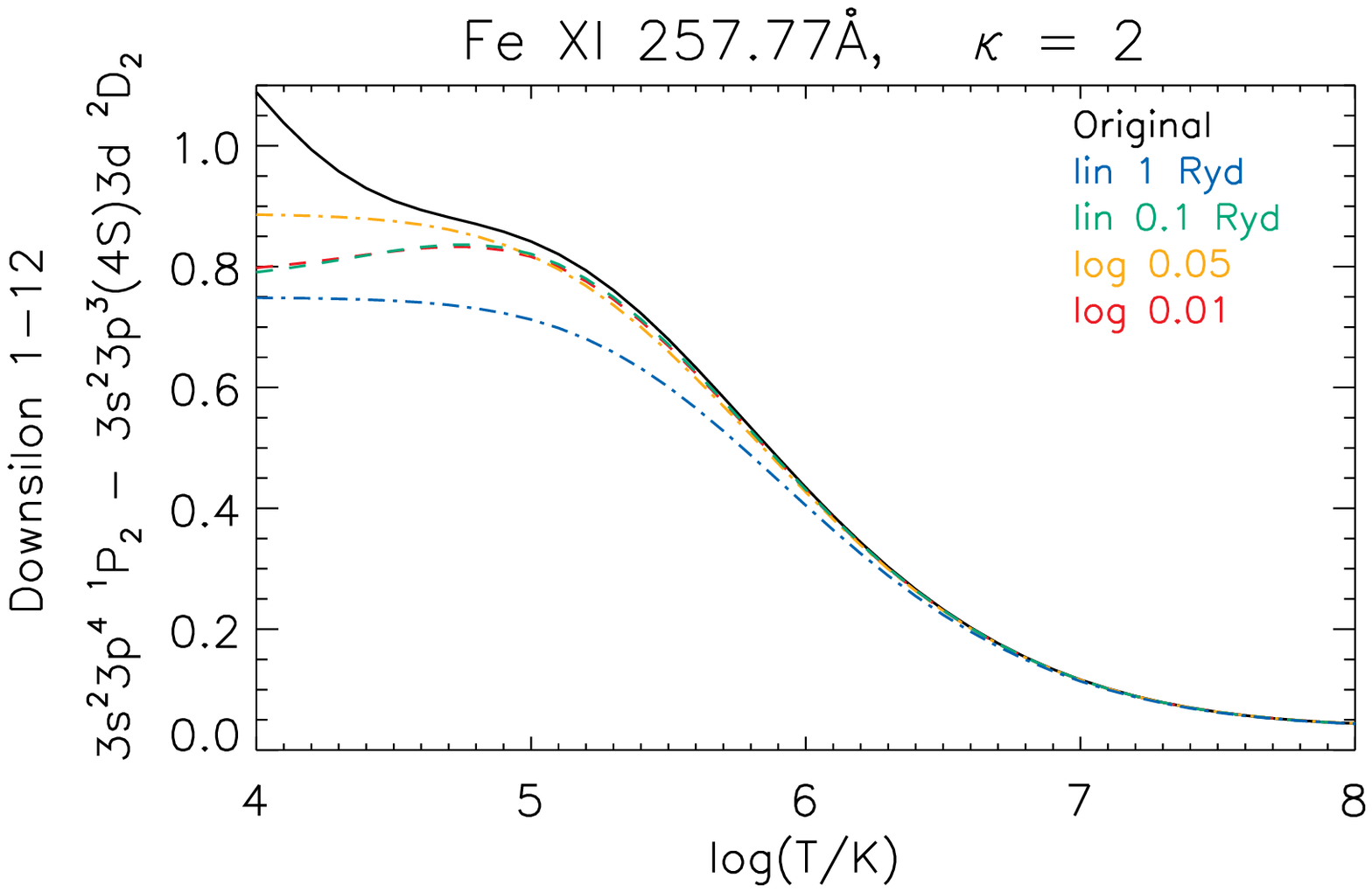}
   \caption{Examples of the collision strengths. \textit{Top}: Original collision strengths $\Omega_{ij}(E_i)$ (black) for the 1404.16\AA~transition in \ion{O}{IV} (\textit{left}) and the 257.77\AA~transition in \ion{Fe}{XI} (\textit{right}). The colored lines correspond to the four averaging methods (Appendix \ref{Appendix:Omega}). \textit{Middle}: Maxwellian $\Upsilon_{ji}$ for these transitions together with the $\Upsilon$s obtained using the averaged $\left<\Omega_{ij}(E_i)\right>$. \textit{Bottom}: The same as the middle panel for $\kappa$\,=\,2.}
   \label{Fig:Averaged}%
\end{figure*}

\begin{figure*}
   \centering
   \includegraphics[width=8.8cm]{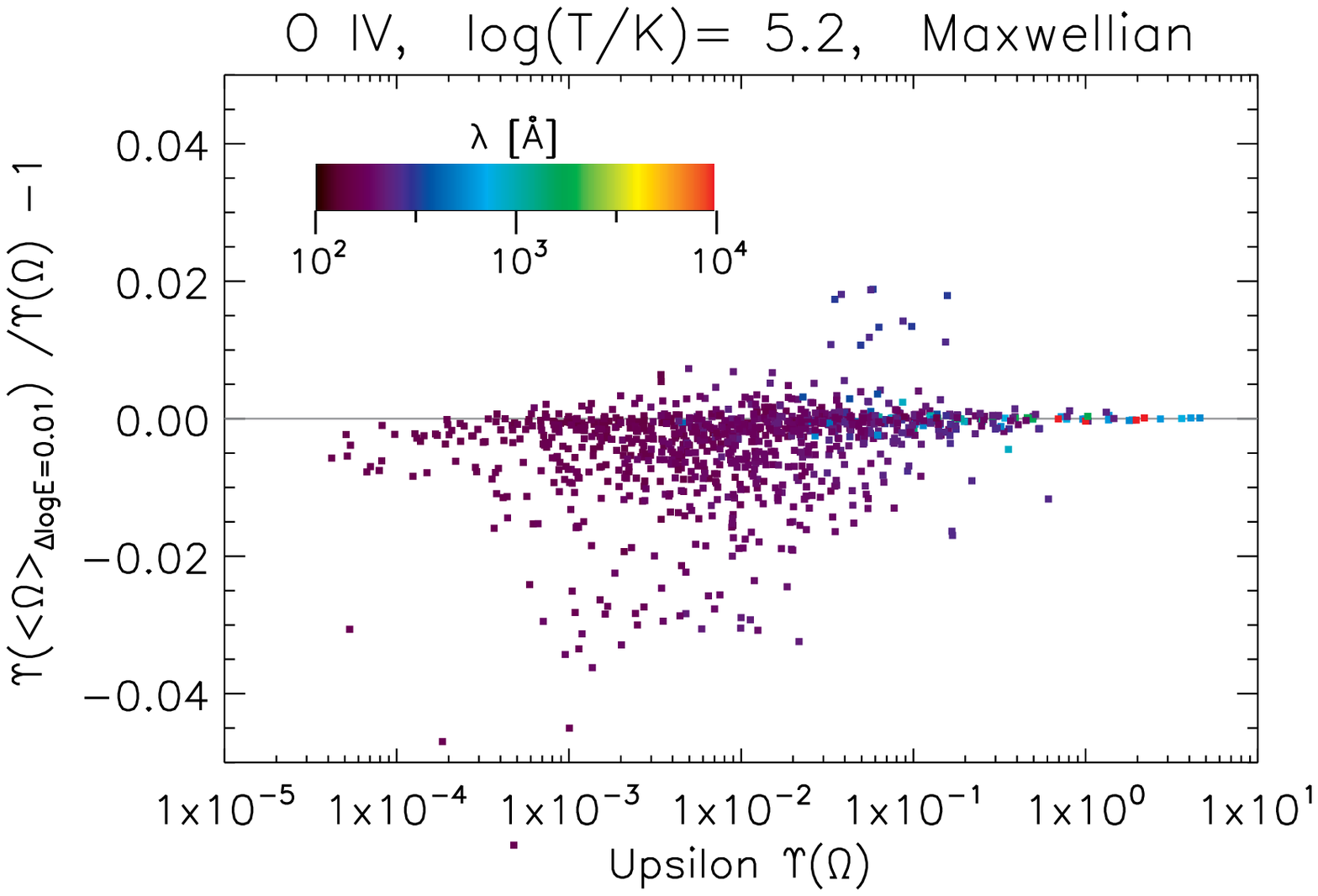}
   \includegraphics[width=8.8cm]{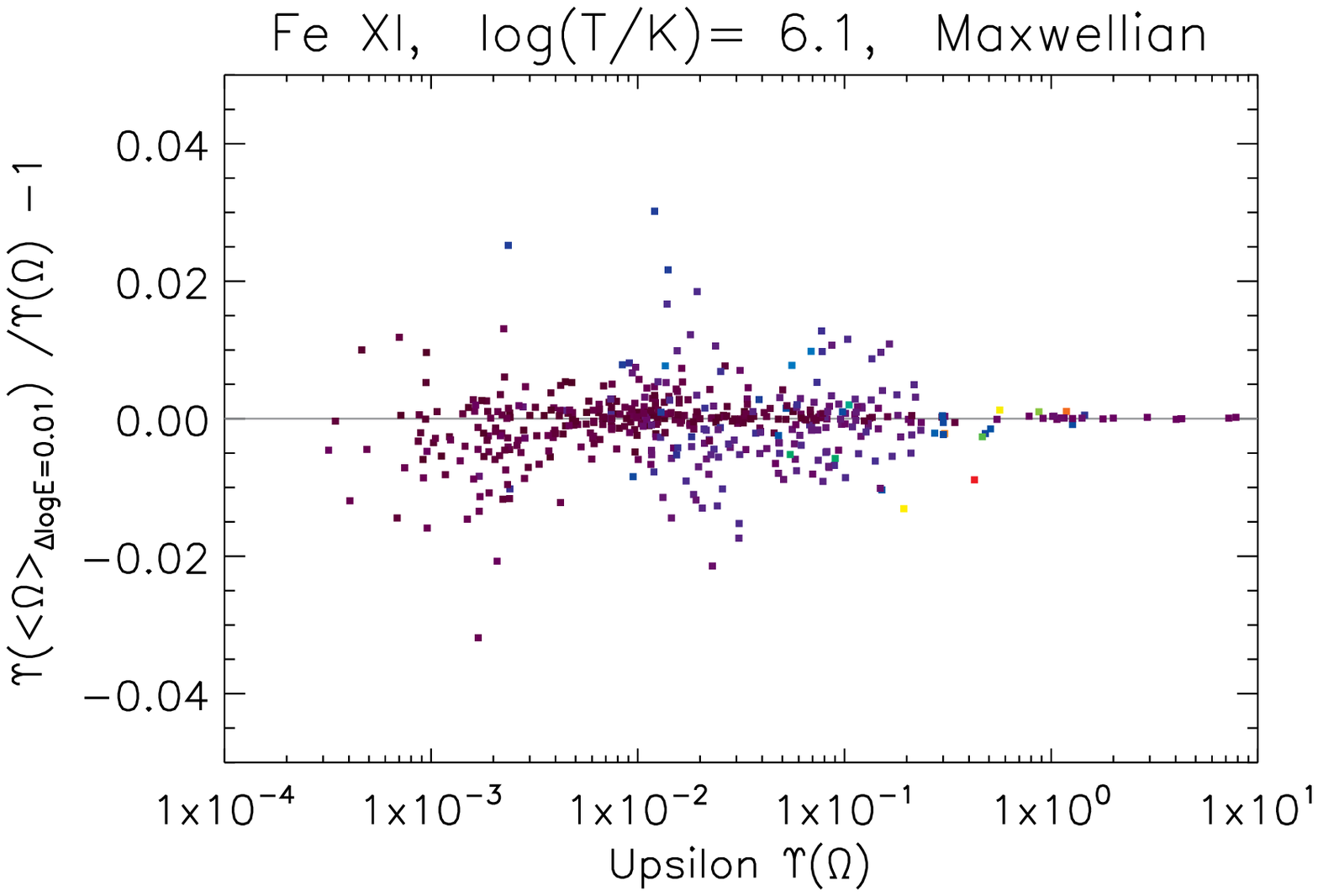}
   \includegraphics[width=8.8cm]{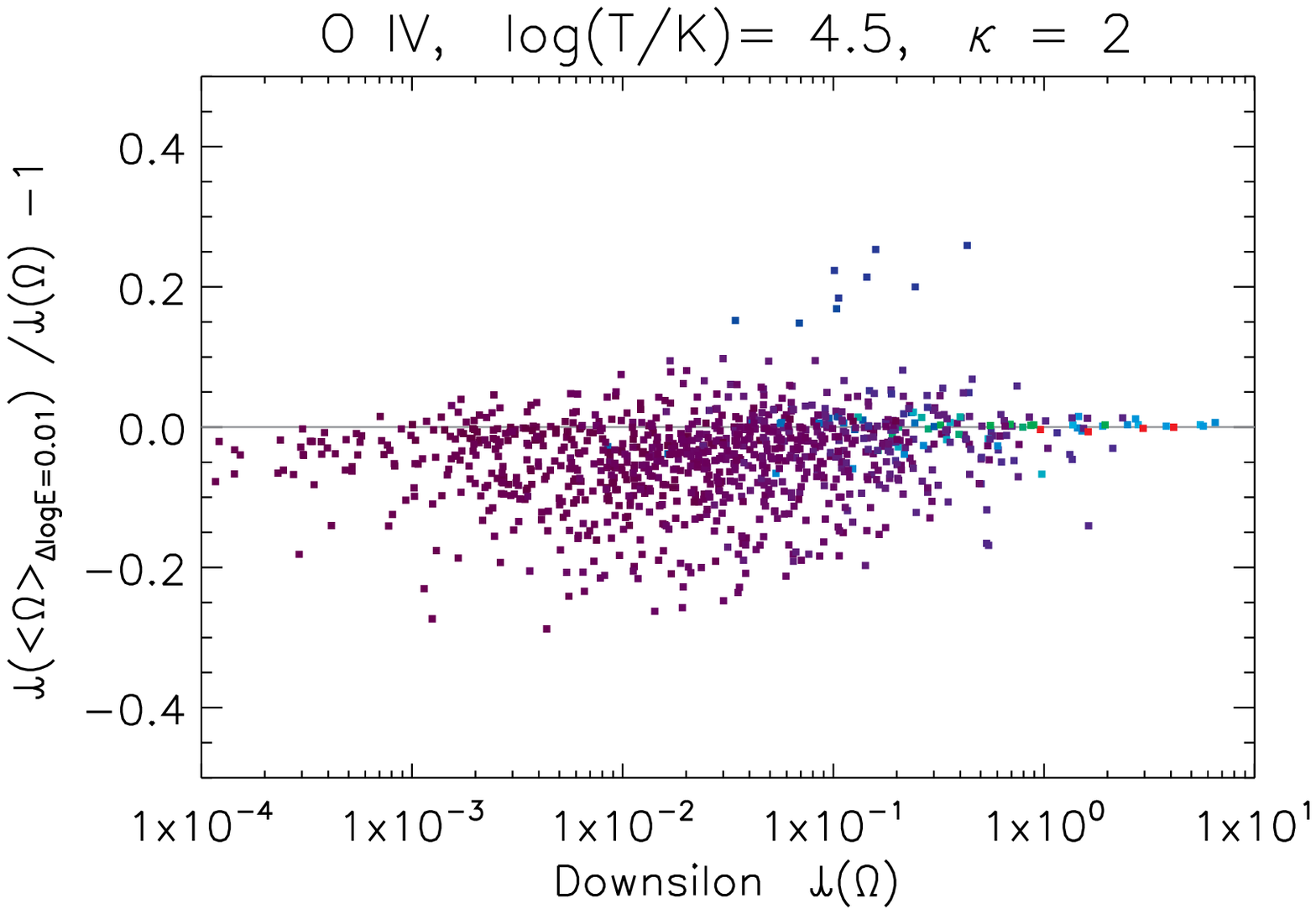}
   \includegraphics[width=8.8cm]{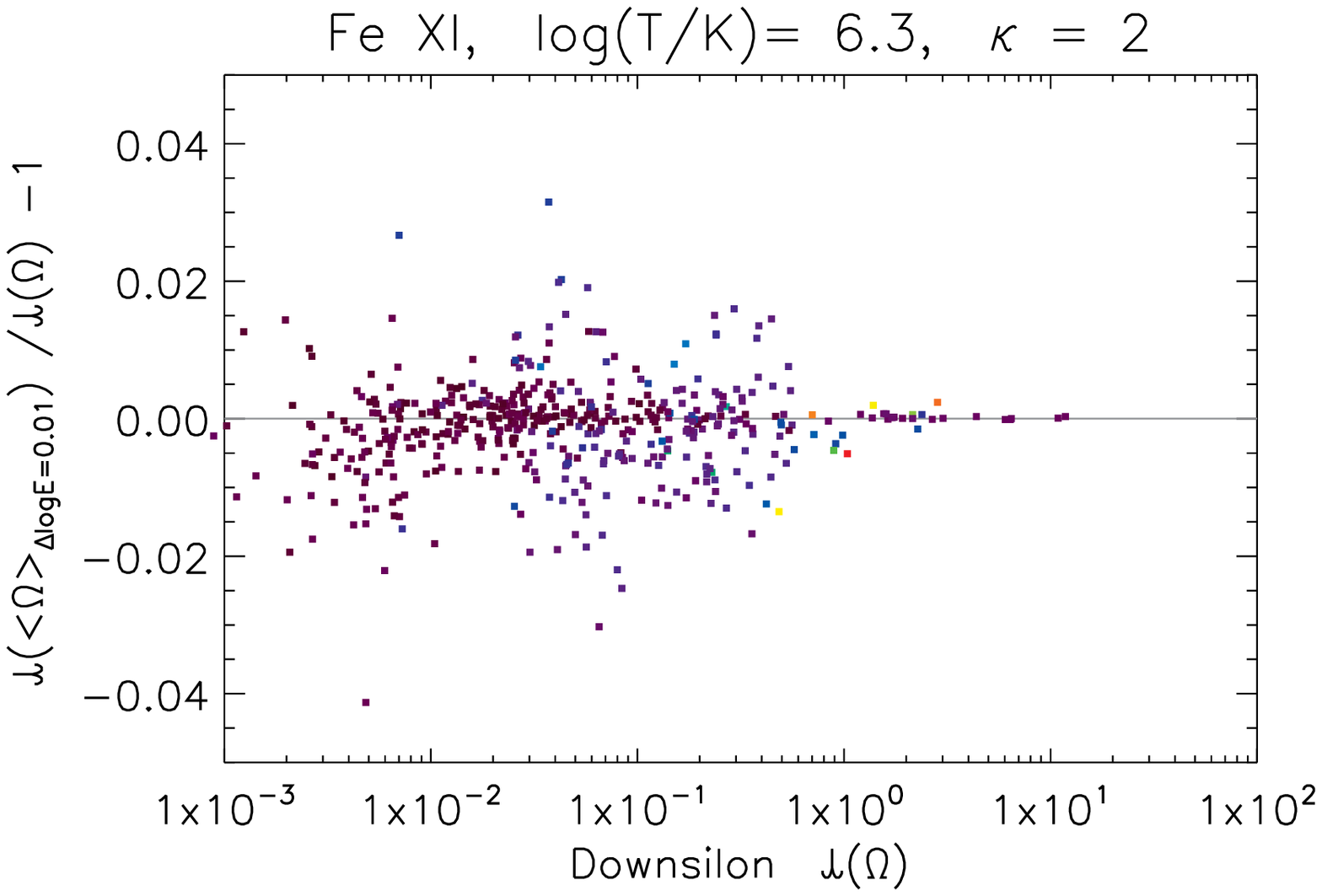}
   \caption{Relative error of the distribution-averaged collision strengths for \ion{O}{IV} and \ion{Fe}{XI}, produced using Method 4 (Appendix \ref{Appendix:Omega}), is shown for two extreme cases: Maxwellian distribution (\textit{top}) and $\kappa$\,=\,2 (\textit{bottom}). The temperature corresponds to the maximum of the relative ion abundance.}
   \label{Fig:Averaged_rel_error}%
\end{figure*}
%

Because of the large number of transitions in the Fe ions, the $\Omega_{ji}$ datafiles are several tens of GiB large. This hampers effective storage, handling, and publishing of such large datasets. Furthermore, calculations for some ions do not contain $\approx$10$^4$ $E_i^{(q)}$ points, but up to one order of magnitude more. An example is the \ion{O}{IV} calculations done by \citet{Liang12}, which contain $\approx$1.87\,$\times$10$^5$ points. We note that this ion was partially investigated already by \citet{Dudik14}.

In an effort to overcome the problems posed by the large data volume of the collison strength calculations, we tested whether coarser $E_i^{(q)}$ grids could be sufficient. To do so, we devised and tested several simple methods of averaging over the $E_i$ points:
\begin{enumerate}
 \item Averaging over uniformly spaced grid of $E_i^{(q)}$ points, with $\Delta E_i^{(q)}$\,=\,1\,Ryd
 \item Averaging over uniformly spaced grid of $E_i^{(q)}$ points, with $\Delta E_i^{(q)}$\,=\,0.1\,Ryd
 \item Averaging over uniformly spaced grid in log$(E_i^{(q)})$, with $\Delta$log($E_i^{(q)}/$Ryd)\,=\,0.05
 \item Averaging over uniformly spaced grid in log$(E_i^{(q)})$, with $\Delta$log($E_i^{(q)}/$Ryd)\,=\,0.01.
\end{enumerate}
An example of the $\left<\Omega_{ji}\right>$ obtained by these methods is shown in Fig. \ref{Fig:Averaged} (\textit{top row}) for \ion{O}{IV} and \ion{Fe}{XI}. The \ion{O}{IV} is used here since its $\Omega_{ij}(E_i)$ has more than one order of magnitude more energy points than the \ion{Fe}{IX}--\ion{Fe}{XIII} ions studied in this paper. We note that in Fig. \ref{Fig:Averaged}, the \rotatebox[origin=c]{180}{$\Upsilon$}$_{ji}(T,\kappa)$ are plotted for $\kappa$\,=\,2 instead of $\Upsilon_{ji}(T,\kappa)$ because of the exp($\Delta E_{ij}/k_\mathrm{B}T)$ factor present in Eq. (\ref{Eq:Upsilon_kappa}) reaches very large values for low log$(T/$K). The level of reduction of $E_i^{(q)}$ points depends strongly on the number of the original points, as well as the original interval covered by the $E_i^{(q)}$ points. Typically, the reduction is about an order of magnitude in case of the less coarse Methods 2 and 4, but can be much larger, e.g. a factor of $\approx$5$\times$10$^3$ for \ion{O}{IV} and Method 1.

Upon calculation of $\Upsilon_{ij}(T,\kappa)$ and \rotatebox[origin=c]{180}{$\Upsilon$}$_{ji}(T,\kappa)$, we found that Method 1 fails dramatically, since the $\Upsilon_{ij}\left(\left<\Omega_{ji}\right>_{\Delta E_i = 1 \mathrm{Ryd}}\right)$ do not match the $\Upsilon_{ij}(\Omega_{ji})$ calculated using the original data. Method 4 gives the best results in terms of most closely matching the $\Upsilon_{ij}(T,\kappa)$ and \rotatebox[origin=c]{180}{$\Upsilon$}$_{ji}(T,\kappa)$ (see Fig. \ref{Fig:Averaged}) calculated using original, non-averaged $\Omega_{ji}$ data. Differences can arise at low log$(T/$K) because at low log$(T/$K), the $\Upsilon_{ij}(T,\kappa)$ and \rotatebox[origin=c]{180}{$\Upsilon$}$_{ji}(T,\kappa)$ are dominated by the contribution from the energies near the excitation threshold at $E_i$\,=\,$\Delta E_{ij}$. The averaging of $\Omega_{ji}$ cannot be performed in the energy interval of the coarse $E_i$ grid containing the excitation threshold. As a result, inaccurate extrapolation of the first point of the $\left<\Omega_{ji}\right>$ to the energy $\Delta E_{ij}$ causes departure of the $\Upsilon_{ij}\left(\left<\Omega_{ji}\right>_{\Delta \mathrm{log}(E_i/\mathrm{Ryd}) = 0.01}\right)$ from the $\Upsilon_{ij}(\Omega_{ji})$ calculated using the original $\Omega_{ji}$ data.

For Method 4, the relative error $R$, defined as
	 \begin{equation}
		  R = \frac{\Upsilon_{ij}\left(\left<\Omega_{ji}\right>_{\Delta \mathrm{log}(E_i/\mathrm{Ryd}) = 0.01}\right)}{ \Upsilon_{ij}(\Omega_{ji})} -1\,
		  \label{Eq:rel_error}
	 \end{equation}
is typically only a few per cent (see Fig. \ref{Fig:Averaged_rel_error}). In this figure, we plot the relative error at temperatures corresponding to the temperature of the maximum relative ion abundance \citep{Dzifcakova13}. For the transitions having high rates (i.e. $\Upsilon_{ij}$ values), the relative error is very small, less than 0.5\% (Fig. \ref{Fig:Averaged_rel_error}). The relative error is larger for weaker transitions with lower $\Upsilon_{ij}$ values. For such transitions, the error is typically a few per cent. However, it can reach $\approx$20--30\% for $\kappa$\,=\,2 and weak (unobservable) transitions in \ion{O}{IV}.

The \ion{O}{IV} and $\kappa$\,=\,2 is an extreme case for two reasons. First, the collision strength $\Omega_{ji}$ for the weak transitions are steeply decreasing with $E_i$, and thus the values of $\Upsilon_{ij}\left(\left<\Omega_{ji}\right>_{\Delta \mathrm{log}(E_i/\mathrm{Ryd}) = 0.01}\right)$ are dominated by the first energy point. Second, the maximum of the relative ion abundance of the \ion{O}{IV} ion is, for $\kappa$-distributions, shifted to lower log$(T/$K) than for the Maxwellian distribution \citep{Dzifcakova13,Dudik14}, increasing the error of the calculation. We consider even this value of $R$\,=\,$\approx$20--30\% acceptable given the uncertainties in the atomic data calculations, particularly at the excitation threshold. We also note that the strongest transitions have small relative errors of only a few percentage points (Fig. \ref{Fig:Averaged_rel_error}, \textit{bottom left}).

\section{Selected transitions}
\label{Appendix:Transitions}

A list of the spectral lines selected by the procedure described in Sect. \ref{Sect:5.1} are listed in Tables \ref{Table:fe9}--\ref{Table:fe13} together with their selfblends. Wavelengths of the primary (strongest) transition within a selfblend are given together with the energy levels involved. Selfblending transitions are indicated together with their level numbers. Most of the selected spectral lines do not contain any selflblending transitions.

%
\begin{table*}[!ht]
  \begin{minipage}[t]{\linewidth}
\centering
\renewcommand{\footnoterule}{}  
    \caption[]{\ion{Fe}{IX} lines selected with the procedure described in Sect. \ref{Sect:5.1}. Strongest transitions are listed together with the wavelengths of their selfblends. Intensities relative to the strongest line are listed for the Maxwellian distribution, $\kappa$\,=\,5 and 2, for temperature and density corresponding to Fig. \ref{Fig:Spectra}. Blended transitions are indicated by ``bl''; ``th'' denotes that the wavelength is theoretical only; ``w'' denotes a line that is observed to be weak.}
         \label{Table:fe9}
    \vspace{-0.8cm}
    \centering
    \renewcommand{\footnoterule}{}  
    \scriptsize
    $$
\begin{tabular}{llclllll}
\hline
\hline
\noalign{\smallskip}
$\lambda~[\AA]$ & levels & level description					& $I$(Maxw) & $I$($\kappa$=5) & $I$($\kappa$=2)	&  selfblends	 & notes  		\\
\noalign{\smallskip}
\hline
\noalign{\smallskip}
157.764    & $6 - 121   $ & $3s^2 3p^5 3d ~{}^3F_3 - 3s^2 3p^4 3d^2  ~{}^3F_3 $ & 0.0158    & 0.0098    & 0.0021	&			&			\\
168.080    & $5 - 118   $ & $3s^2 3p^5 3d ~{}^3F_4 - 3s^2 3p^4 3d^2  ~{}^3F_4 $ & 0.0362    & 0.0223    & 0.0047	& $169.010~ (3 - 117) $	& th			\\
	   & 		  &							&	    &		&	& $169.048~ (8 - 131) $	& th			\\
159.949    & $12 - 133  $ & $3s^2 3p^5 3d ~{}^1F_3 - 3s^2 3p^4 3d^2  ~{}^1F_3 $ & 0.0047    & 0.0030    & 0.0008	& $159.986~ (3 - 112) $	&			\\
160.379    & $4 - 115   $ & $3s^2 3p^5 3d ~{}^3P_2 - 3s^2 3p^4 3d^2  ~{}^3P_2 $ & 0.0043    & 0.0029    & 0.0008	&			&			\\
162.945    & $8 - 124   $ & $3s^2 3p^5 3d ~{}^3D_3 - 3s^2 3p^4 3d^2  ~{}^3D_3 $ & 0.0089    & 0.0056    & 0.0013	& $162.901~ (9 - 125) $	&			\\
171.073    & $1 - 13    $ & $3s^2 3p^6 ~{}^1S_0 - 3s^2 3p^5 3d  ~{}^1P_1      $ & 1.0000    & 0.7430    & 0.3092	&            		&			\\
177.592    & $6 - 111   $ & $3s^2 3p^5 3d ~{}^3F_3 - 3s^2 3p^4 3d^2  ~{}^3D_2 $ & 0.0047    & 0.0031    & 0.0008	&             		&			\\
176.930    & $12 - 113  $ & $3s^2 3p^5 3d ~{}^1F_3 - 3s^2 3p^4 3d^2  ~{}^1G_4 $ & 0.0188    & 0.0117    & 0.0026	&            		&			\\
176.945    & $5 - 110   $ & $3s^2 3p^5 3d ~{}^3F_4 - 3s^2 3p^4 3d^2  ~{}^3D_3 $ & 0.0427    & 0.0262    & 0.0054	& $165.876~ (6 - 113) $	& bl \ion{Fe}{VII} 	\\
	   & 		  &							&	    &		&	& $165.880~ (8 - 118) $	& th			\\
	   & 		  &							&	    &		&	& $165.908~ (12 - 123)$ & th			\\	
183.851    & $9 - 99    $ & $3s^2 3p^5 3d ~{}^1D_2 - 3s^2 3p^4 3d^2  ~{}^1F_3 $ & 0.0055    & 0.0035    & 0.0008	& $183.877~ (12 - 105)$	& bl \ion{Fe}{VII}	\\
	   &		  &							&	    &		&	& $183.898~ (5 - 89)  $	& 			\\
188.497    & $5 - 96    $ & $3s^2 3p^5 3d ~{}^3F_4 - 3s^2 3p^4 3d^2  ~{}^3G_5 $	& 0.0500    & 0.0307    & 0.0064&      		 	&   			\\
189.941    & $6 - 95    $ & $3s^2 3p^5 3d ~{}^3F_3 - 3s^2 3p^4 3d^2  ~{}^3G_4 $ & 0.0299    & 0.0187    & 0.0041	&     			&       		\\
191.216    & $7 - 97    $ & $3s^2 3p^5 3d ~{}^3F_2 - 3s^2 3p^4 3d^2  ~{}^3G_3 $ & 0.0124    & 0.0079    & 0.0018	&			&			\\
194.784    & $4 - 85    $ & $3s^2 3p^5 3d ~{}^3_P2 - 3s^2 3p^4 3d^2  ~{}^3D_3 $ & 0.0082    & 0.0052    & 0.0013	&   			&			\\
197.862    & $13 - 148  $ & $3s^2 3p^5 3d ~{}^1P_1 - 3s^2 3p^5 4p  ~{}^1S_0   $ & 0.0233    & 0.0174    & 0.0072	&			&			\\
217.101    & $1 - 10    $ & $3s^2 3p^6 ~{}^1S_0 - 3s^2 3p^5 3d  ~{}^3D_1      $ & 0.0566    & 0.0439    & 0.0202	&			&			\\
218.937    & $1 - 9     $ & $3s^2 3p^6 ~{}^1S_0 - 3s^2 3p^5 3d  ~{}^1D_2      $ & 0.0308    & 0.0233    & 0.0100	&			&			\\
230.511    & $1 - 7     $ & $3s^2 3p^6 ~{}^1S_0 - 3s^2 3p^5 3d  ~{}^3F_2      $ & 0.0096    & 0.0071    & 0.0028	& $230.475~ (11 - 55)$ 	&			\\
	   &		  &						        &	    &		&	& $230.479~ (10 - 52)$	& 			\\
241.739    & $1 - 4     $ & $3s^2 3p^6 ~{}^1S_0 - 3s^2 3p^5 3d  ~{}^3P_2      $ & 0.1849    & 0.1386    & 0.0590	&			&			\\
244.909    & $1 - 3     $ & $3s^2 3p^6 ~{}^1S_0 - 3s^2 3p^5 3d  ~{}^3P_1      $ & 0.0998    & 0.0773    & 0.0369	&			&			\\
317.193    & $4 - 16    $ & $3s^2 3p^5 3d ~{}^3P_2 - 3s 3p^6 3d  ~{}^3D_3     $ & 0.0121    & 0.0083    & 0.0024	&			&			\\
329.897    & $5 - 16    $ & $3s^2 3p^5 3d ~{}^3F_4 - 3s 3p^6 3d  ~{}^3D_3     $ & 0.0153    & 0.0105    & 0.0031	&			&			\\
352.060    & $12 - 17   $ & $3s^2 3p^5 3d ~{}^1F_3 - 3s 3p^6 3d  ~{}^1D_2     $ & 0.0168    & 0.0122    & 0.0048	&			&			\\
\noalign{\smallskip}
\hline
\noalign{\smallskip}
2043.01    & $4 - 11    $ & $3s^2 3p^5 3d ~{}^3P_2 - 3s^2 3p^5 3d  ~{}^3D_2   $ & 0.0497    & 0.0375    & 0.0158	&			&			\\
2384.19    & $4 - 8     $ & $3s^2 3p^5 3d ~{}^3_P2 - 3s^2 3p^5 3d  ~{}^3D_3   $ & 0.0432    & 0.0318    & 0.0124	&			&			\\
2498.84    & $5 - 12    $ & $3s^2 3p^5 3d ~{}^3F_4 - 3s^2 3p^5 3d  ~{}^1F_3   $ & 0.0849    & 0.0649    & 0.0282	&			&			\\
3644.08    & $6 - 9     $ & $3s^2 3p^5 3d ~{}^3F_3 - 3s^2 3p^5 3d  ~{}^1D_2   $ & 0.0240    & 0.0182    & 0.0078	&			&			\\
3802.10    & $6 - 8     $ & $3s^2 3p^5 3d ~{}^3F_3 - 3s^2 3p^5 3d  ~{}^3D_3   $ & 0.0264    & 0.0194    & 0.0075	&			&			\\
22183.0    & $6 - 7     $ & $3s^2 3p^5 3d ~{}^3F_3 - 3s^2 3p^5 3d  ~{}^3F_2   $ & 0.0149    & 0.0111    & 0.0044	&			&			\\
28563.0    & $5 - 6     $ & $3s^2 3p^5 3d ~{}^3F_4 - 3s^2 3p^5 3d  ~{}^3F_3   $ & 0.0101    & 0.0073    & 0.0027	& 			&			\\
\noalign{\smallskip}
\hline
\hline
\end{tabular}
     $$
  \end{minipage}
\end{table*}
%
%
\begin{table*}
  \begin{minipage}[t]{\linewidth}
\centering
\renewcommand{\footnoterule}{}  
    \caption[]{Same as Table \ref{Table:fe9}, but for \ion{Fe}{X}.}
         \label{Table:fe10}
    \vspace{-0.8cm}
    \centering
    \renewcommand{\footnoterule}{}  
    \scriptsize
    $$
\begin{tabular}{llclllll}
\hline
\hline
\noalign{\smallskip}
$\lambda~[\AA]$ & levels & level description					   & $I$(Maxw) & $I$($\kappa$=5) & $I$($\kappa$=2)	&  selfblends	 & notes  \\
\noalign{\smallskip}
\hline
\noalign{\smallskip}
174.531    & $1 - 30$	& $3s^2 3p^5 ~{}^2P_{3/2} - 3s^2 3p^4 3d ~{}^2D_{5/2}   $  & 1.0000    & 0.8535    & 0.3956	& $174.482 ~(10-145)$	&	\\
175.263    & $2 - 31$	& $3s^2 3p^5 ~{}^2P_{1/2} - 3s^2 3p^4 3d ~{}^2D_{3/2}   $  & 0.1115    & 0.1042    & 0.0670	& $175.223 ~(18-155)$	&	\\
	   &		&						     	   &	       &	   &		& $175.280 ~(18-154)$	&	\\
175.475    & $1 - 29$ 	& $3s^2 3p^5 ~{}^2P_{3/2} - 3s^2 3p^4 3d ~{}^2P_{1/2}   $  & 0.0308    & 0.0273    & 0.0147	& $175.487 ~(17-153)$	&	\\ 
177.240    & $1 - 28$	& $3s^2 3p^5 ~{}^2P_{3/2} - 3s^2 3p^4 3d ~{}^2P_{3/2}   $  & 0.5731    & 0.4905    & 0.2293	&          		&	\\
180.441    & $2 - 29$	& $3s^2 3p^5 ~{}^2P_{1/2} - 3s^2 3p^4 3d ~{}^2P_{1/2}   $  & 0.0798    & 0.0707    & 0.0380	&          		& bl \ion{Fe}{XI}	\\
184.537    & $1 - 27$	& $3s^2 3p^5 ~{}^2P_{3/2} - 3s^2 3p^4 3d ~{}^2S_{1/2}   $  & 0.2291    & 0.1980    & 0.0958	& $184.509 ~(20-144)$	&	\\
190.037    & $2 - 27$	& $3s^2 3p^5 ~{}^2P_{1/2} - 3s^2 3p^4 3d ~{}^2S_{1/2}   $  & 0.0777    & 0.0671    & 0.0325	& $189.992 ~(23-153)$	& bl \ion{Fe}{XII}	\\
	   &		&							   &	       &	   &		& $189.996 ~(19-131)$	&	\\
193.715    & $1 - 26$	& $3s^2 3p^5 ~{}^2P_{3/2} - 3s^2 3p^4 3d ~{}^2D_{5/2}   $  & 0.0308    & 0.0288    & 0.0168	&         		&	\\
207.449    & $1 - 23$	& $3s^2 3p^5 ~{}^2P_{3/2} - 3s^2 3p^4 3d ~{}^2F_{5/2}   $  & 0.0359    & 0.0337    & 0.0202	&          		&	\\
220.247    & $1 - 21$	& $3s^2 3p^5 ~{}^2P_{3/2} - 3s^2 3p^4 3d ~{}^2F_{5/2}   $  & 0.0512    & 0.0476    & 0.0287	&          		&	\\
224.800    & $1 - 11$	& $3s^2 3p^5 ~{}^2P_{3/2} - 3s^2 3p^4 3d ~{}^2P_{3/2}   $  & 0.0327    & 0.0306    & 0.0185	&          		&	\\
225.856    & $1 - 19$	& $3s^2 3p^5 ~{}^2P_{3/2} - 3s^2 3p^4 3d ~{}^2D_{5/2}   $  & 0.0649    & 0.0602    & 0.0355	&          		&	\\
226.998    & $1 - 17$	& $3s^2 3p^5 ~{}^2P_{3/2} - 3s^2 3p^4 3d ~{}^4P_{5/2}   $  & 0.0451    & 0.0421    & 0.0250	&          		&	\\
234.315    & $1 - 12$	& $3s^2 3p^5 ~{}^2P_{3/2} - 3s^2 3p^4 3d ~{}^4F_{5/2}   $  & 0.0475    & 0.0456    & 0.0297	&          		& bl \ion{He}{II}	\\
256.398    & $1 - 6$	& $3s^2 3p^5 ~{}^2P_{3/2} - 3s^2 3p^4 3d ~{}^4D_{3/2}   $  & 0.0344    & 0.0336    & 0.0231	&          	 	& bl \ion{He}{II}	\\
257.263    & $1 - 5$	& $3s^2 3p^5 ~{}^2P_{3/2} - 3s^2 3p^4 3d ~{}^4D_{7/2}   $  & 0.2397    & 0.2148    & 0.1110	& $257.257 ~(24-81)$	&	\\
	   &		&							   &	       &	   &		& $257.259 ~(1-4)  $	& 	\\
345.738    & $1 - 3$	& $3s^2 3p^5 ~{}^2P_{3/2} - 3s 3p^6 ~{}^2S_{1/2}        $  & 0.1476    & 0.1390    & 0.0886	&			&	\\
365.560    & $2 - 3$	& $3s^2 3p^5 ~{}^2P_{1/2} - 3s 3p^6 ~{}^2S_{1/2}        $  & 0.0645    & 0.0608    & 0.0387	&          	 	&	\\
\noalign{\smallskip}
\hline
\noalign{\smallskip}
3454.95    & $5 - 8$	& $3s^2 3p^4 3d ~{}^4D_{7/2} - 3s^2 3p^4 3d ~{}^4F_{9/2}$  & 0.0582    & 0.0488    & 0.0191	&			&	\\
6376.29    & $1 - 2$	& $3s^2 3p^5 ~{}^2P_{3/2} - 3s^2 3p^5 ~{}^2P_{1/2}      $  & 0.6373    & 0.6969    & 0.8182	&			&	\\
\noalign{\smallskip}
\hline
\hline
\end{tabular}
     $$
  \end{minipage}
\end{table*}
%
%
\begin{table*}
  \begin{minipage}[t]{\linewidth}
\centering
\renewcommand{\footnoterule}{}  
    \caption[]{Same as Table \ref{Table:fe9}, but for \ion{Fe}{XI}.}
         \label{Table:fe11}
    \vspace{-0.8cm}
    \centering
    \renewcommand{\footnoterule}{}  
    \scriptsize
    $$
\begin{tabular}{llclllll}
\hline
\hline
\noalign{\smallskip}
$\lambda~[\AA]$ & levels & level description				& $I$(Maxw) & $I$($\kappa$=5) & $I$($\kappa$=2)	&  selfblends	 	& notes  \\
\noalign{\smallskip}
\hline
\noalign{\smallskip}
178.058    & $1 - 43$	& $3s^2 3p^4 ~{}^3P_2 - 3s^2 3p^3 3d ~{}^3D_2$	& 0.0496    & 0.0416    & 0.0202	&          		&	\\
179.758    & $4 - 46$	& $3s^2 3p^4 ~{}^1D_2 - 3s^2 3p^3 3d ~{}^1F_3$	& 0.0497    & 0.0444    & 0.0257	& $179.734 ~(12-140)$	&	\\
	   &		&						&	    &	   	&		& $179.772 ~(15-158)$	&	\\
	   &		&						&	    &	   	&		& $179.792 ~(13-141)$	&	\\
	   &		&						&	    &		&		& $179.820 ~(18-164)$	&	\\
180.401    & $1 - 42$	& $3s^2 3p^4 ~{}^3P_2 - 3s^2 3p^3 3d ~{}^3D_3$	& 1.0000    & 0.8171    & 0.3574	& $180.392 ~(8-100) $	& bl \ion{Fe}{X}	\\
180.594    & $2 - 44$	& $3s^2 3p^4 ~{}^3P_1 - 3s^2 3p^3 3d ~{}^3D_1$	& 0.0417    & 0.0361    & 0.0193	& $180.554 ~(17-160)$	&	\\
	   &		&						&	    &	   	&		& $180.643 ~(21-167)$	&	\\
181.130    & $3 - 44$	& $3s^2 3p^4 ~{}^3P_0 - 3s^2 3p^3 3d ~{}^3D_1$	& 0.0581    & 0.0503    & 0.0270	&          		&	\\
182.167    & $2 - 43$	& $3s^2 3p^4 ~{}^3P_1 - 3s^2 3p^3 3d ~{}^3D_2$	& 0.1810    & 0.1518    & 0.0739	&          		&	\\
184.410    & $5 - 48$	& $3s^2 3p^4 ~{}^1S_0 - 3s^2 3p^3 3d ~{}^1P_1$	& 0.0046    & 0.0041    & 0.0022	& $184.446 ~(36-216)$	&	\\
184.793    & $4 - 45$	& $3s^2 3p^4 ~{}^1D_2 - 3s^2 3p^3 3d ~{}^1D_2$	& 0.0311    & 0.0278    & 0.0161	& $184.805 ~(32-193)$	&	\\
188.216    & $1 - 38$	& $3s^2 3p^4 ~{}^3P_2 - 3s^2 3p^3 3d ~{}^3P_2$	& 0.5096    & 0.4196    & 0.1884	&          		&	\\
188.299    & $1 - 37$	& $3s^2 3p^4 ~{}^3P_2 - 3s^2 3p^3 3d ~{}^1P_1$	& 0.2901    & 0.2383    & 0.1061	&          		&	\\
188.997    & $2 - 40$	& $3s^2 3p^4 ~{}^3P_1 - 3s^2 3p^3 3d ~{}^3P_0$	& 0.0271    & 0.0238    & 0.0134	&          		&	\\
189.123    & $2 - 41$	& $3s^2 3p^4 ~{}^3P_1 - 3s^2 3p^3 3d ~{}^3P_1$	& 0.0239    & 0.0207    & 0.0110	&          		&	\\
189.711    & $3 - 41$	& $3s^2 3p^4 ~{}^3P_0 - 3s^2 3p^3 3d ~{}^3P_1$	& 0.0201    & 0.0174    & 0.0093	& $189.736 ~(30-184)$	& bl	\\
190.382    & $1 - 36$	& $3s^2 3p^4 ~{}^3P_2 - 3s^2 3p^3 3d ~{}^1F_3$	& 0.0221    & 0.0191    & 0.0099	& $190.390 ~(7-85)  $	&	\\
192.021    & $2 - 39$	& $3s^2 3p^4 ~{}^3P_1 - 3s^2 3p^3 3d ~{}^3S_1$	& 0.0219    & 0.0187    & 0.0097	&          		& bl	\\
192.813    & $2 - 38$	& $3s^2 3p^4 ~{}^3P_1 - 3s^2 3p^3 3d ~{}^3P_2$	& 0.1091    & 0.0898    & 0.0403	&          		& bl \ion{Ca}{XVII}	\\
	    &		&						&	    &	   	&		& 			& bl \ion{O}{V}		\\
192.900    & $2 - 37$	& $3s^2 3p^4 ~{}^3P_1 - 3s^2 3p^3 3d ~{}^1P_1$	& 0.0605    & 0.0497    & 0.0221	&          		& bl \ion{Ca}{XVII}	\\
	   &		&						&	    &	   	&		& 			& bl \ion{O}{V}		\\
198.538    & $4 - 41$	& $3s^2 3p^4 ~{}^1D_2 - 3s^2 3p^3 3d ~{}^3P_1$	& 0.0284    & 0.0246    & 0.0131	&          		& bl \ion{S}{VIII}	\\
201.112    & $1 - 35$	& $3s^2 3p^4 ~{}^3P_2 - 3s^2 3p^3 3d ~{}^3D_3$	& 0.0304    & 0.0263    & 0.0138	&          		& bl \ion{Fe}{XIII}	\\
201.734    & $4 - 39$	& $3s^2 3p^4 ~{}^1D_2 - 3s^2 3p^3 3d ~{}^3S_1$	& 0.0384    & 0.0328    & 0.0170	&          		& bl \ion{Fe}{XII}	\\
202.424    & $1 - 34$	& $3s^2 3p^4 ~{}^3P_2 - 3s^2 3p^3 3d ~{}^3P_2$	& 0.0575    & 0.0482    & 0.0226	& $202.388 ~(13-117)$	&	\\
	   &		&						&	    &	 	&		& $202.393 ~(1-31)  $	&	\\
	   &		&						&	    &	  	&		& $202.461 ~(14-118)$	&	\\
230.165    & $4 - 26$	& $3s^2 3p^4 ~{}^1D_2 - 3s^2 3p^3 3d ~{}^1D_2$	& 0.0187    & 0.0171    & 0.0103	&          		&	\\
234.730    & $1 - 20$	& $3s^2 3p^4 ~{}^3P_2 - 3s^2 3p^3 3d ~{}^3F_3$	& 0.0240    & 0.0217    & 0.0130	&          		&	\\
236.494    & $1 - 18$	& $3s^2 3p^4 ~{}^3P_2 - 3s^2 3p^3 3d ~{}^3F_2$	& 0.0281    & 0.0258    & 0.0161	&          		&	\\
239.787    & $4 - 22$	& $3s^2 3p^4 ~{}^1D_2 - 3s^2 3p^3 3d ~{}^3G_3$	& 0.0316    & 0.0289    & 0.0177	& $239.780 ~(4-22)  $	&	\\
240.717    & $1 - 16$	& $3s^2 3p^4 ~{}^3P_2 - 3s^2 3p^3 3d ~{}^3D_3$	& 0.0810    & 0.0701    & 0.0373	& $240.757 ~(9-75)  $	& bl \ion{Fe}{XIII}	\\
242.215    & $1 - 15$	& $3s^2 3p^4 ~{}^3P_2 - 3s^2 3p^3 3d ~{}^3D_2$	& 0.0354    & 0.0317    & 0.0185	&          		& bl 	\\
256.919    & $1 - 14$	& $3s^2 3p^4 ~{}^3P_2 - 3s^2 3p^3 3d ~{}^5D_4$	& 0.0796    & 0.0695    & 0.0351	&          		& bl \ion{Fe}{XII} 	\\
257.554    & $1 - 13$	& $3s^2 3p^4 ~{}^3P_2 - 3s^2 3p^3 3d ~{}^5D_3$	& 0.0940    & 0.0862    & 0.0528	& $257.538 ~(22-96) $	&	\\
	   &		&						&	    &	   	&		& $257.547 ~(4-20)  $	&	\\
	   &		&						&	    &	   	&		& $257.558 ~(42-147)$	&	\\
257.772    & $1 - 12$	& $3s^2 3p^4 ~{}^3P_2 - 3s^2 3p^3 3d ~{}^5D_2$  & 0.0365    & 0.0341    & 0.0220		& $257.725 ~(23-97) $	&	\\
264.772    & $4 - 16$	& $3s^2 3p^4 ~{}^1D_2 - 3s^2 3p^3 3d ~{}^3D_3$	& 0.0309    & 0.0267    & 0.0142	&          		&	\\
308.544    & $4 - 9 $	& $3s^2 3p^4 ~{}^1D_2 - 3s 3p^5 ~{}^1P_1  $	& 0.0384    & 0.0365    & 0.0251	&          		&	\\
341.113    & $1 - 7 $	& $3s^2 3p^4 ~{}^3P_2 - 3s 3p^5 ~{}^3P_1  $ 	& 0.0600    & 0.0542    & 0.0339	&          		&	\\
349.046    & $2 - 8 $	& $3s^2 3p^4 ~{}^3P_1 - 3s 3p^5 ~{}^3P_0  $ 	& 0.0189    & 0.0183    & 0.0137	&          		& bl u	\\
352.670    & $1 - 6 $	& $3s^2 3p^4 ~{}^3P_2 - 3s 3p^5 ~{}^3P_2  $  	& 0.2461    & 0.2188    & 0.1295	& $352.622 ~(35-78) $	&	\\
356.519    & $2 - 7 $	& $3s^2 3p^4 ~{}^3P_1 - 3s 3p^5 ~{}^3P_1  $ 	& 0.0312    & 0.0282    & 0.0176	&          		&	\\
358.613    & $3 - 7 $	& $3s^2 3p^4 ~{}^3P_0 - 3s 3p^5 ~{}^3P_1  $ 	& 0.0399    & 0.0361    & 0.0225	&          		&	\\
369.163    & $2 - 6 $	& $3s^2 3p^4 ~{}^3P_1 - 3s 3p^5 ~{}^3P_2  $ 	& 0.0766    & 0.0681    & 0.0403	&          		&	\\
\noalign{\smallskip}
\hline
\noalign{\smallskip}
2649.50    & $1 - 4 $	& $3s^2 3p^4 ~{}^3P_2 - 3s^2 3p64 ~{}^1D_2$    	& 0.4325    & 0.4315    & 0.4003	&          		&	\\
7894.03    & $1 - 2 $	& $3s^2 3p^4 ~{}^3P_2 - 3s^2 3p^4 ~{}^3P_1$	& 0.6523    & 0.6240    & 0.5295	&          		&	\\
\noalign{\smallskip}
\hline
\hline
\end{tabular}
     $$
  \end{minipage}
\end{table*}
%
%
\begin{table*}
  \begin{minipage}[t]{\linewidth}
\centering
\renewcommand{\footnoterule}{}  
    \caption[]{Same as Table \ref{Table:fe9}, but for \ion{Fe}{XII}.}
         \label{Table:fe12}
    \vspace{-0.8cm}
    \centering
    \renewcommand{\footnoterule}{}  
    \scriptsize
    $$
\begin{tabular}{llclllll}
\hline
\hline
\noalign{\smallskip}
$\lambda~[\AA]$ & levels & level description					& $I$(Maxw) & $I$($\kappa$=5) & $I$($\kappa$=2)	&  selfblends	 	& notes  \\
\noalign{\smallskip}
\hline
\noalign{\smallskip}
186.887    & $3 - 39$	&   $3s^2 3p^3 ~{}^2D_{5/2} - 3s^2 3p^2 3d ~{}^2F_{7/2}$& 0.3499    & 0.2888    & 0.0953	& $186.854 ~(2-36)  $	&	\\
	   &		&							&	    &		&	& $186.931 ~(22-135)$	&	\\
188.170    & $4 - 41$	&   $3s^2 3p^3 ~{}^2P_{1/2} - 3s^2 3p^2 3d ~{}^2D_{3/2}$& 0.0065    & 0.0057    & 0.0023	& $188.114 ~(3-38)  $	&	\\
	   &		&							&	    &		&	& $188.130 ~(10-99) $	&	\\
190.467    & $5 - 41$	&   $3s^2 3p^3 ~{}^2P_{3/2} - 3s^2 3p^2 3d ~{}^2D_{3/2}$& 0.0038    & 0.0032    & 0.0011	& $190.487 ~(17-130)$	&	\\
191.049    & $5 - 40$	&   $3s^2 3p^3 ~{}^2P_{3/2} - 3s^2 3p^2 3d ~{}^2D_{5/2}$& 0.0208    & 0.0174    & 0.0059	& $190.977 ~(40-203)$	&	\\
	   &		&							&	    &		&	& $191.017 ~(6-78)  $	&	\\
	   &		&							&	    &		&	& $191.059 ~(20-131)$	&	\\
	   &		&							&	    &		&	& $191.074 ~(33-178)$	&	\\
	   &		&							&	    &		&	& $191.096 ~(17-129)$	&	\\
192.394    & $1 - 30$	&   $3s^2 3p^3 ~{}^4S_{3/2} - 3s^2 3p^2 3d ~{}^4P_{1/2}$& 0.3143    & 0.2517    & 0.0783	&          		&	\\
193.509    & $1 - 29$	&   $3s^2 3p^3 ~{}^4S_{3/2} - 3s^2 3p^2 3d ~{}^4P_{3/2}$& 0.6713    & 0.5380    & 0.1675	& $193.500 ~(7-80)  $	&	\\
195.119    & $1 - 27$	&   $3s^2 3p^3 ~{}^4S_{3/2} - 3s^2 3p^2 3d ~{}^4P_{5/2}$& 1.0000    & 0.8026    & 0.2510	& $195.078 ~(31-151)$	& sbl	\\
195.179    & $2 - 33$	&   $3s^2 3p^3 ~{}^2D_{3/2} - 3s^2 3p^2 3d ~{}^2D_{3/2}$& 0.0373    & 0.0322    & 0.0122	& $195.221 ~(13-115)$	& sbl	\\
196.640    & $3 - 34$	&   $3s^2 3p^3 ~{}^2D_{5/2} - 3s^2 3p^2 3d ~{}^2D_{5/2}$& 0.0944    & 0.0776    & 0.0252	& $196.646 ~(23-131)$	&	\\
201.140    & $5 - 38$	&   $3s^2 3p^3 ~{}^2P_{3/2} - 3s^2 3p^2 3d ~{}^2P_{3/2}$& 0.0095    & 0.0080    & 0.0029	& $201.133 ~(11-104)$	& bl \ion{Fe}{XIII}	\\
	   &		& 							&	    &		&	& $201.186 ~(6-67)  $	&	\\
201.740    & $4 - 35$	&   $3s^2 3p^3 ~{}^2P_{1/2} - 3s^2 3p^2 3d ~{}^2P_{1/2}$& 0.0097    & 0.0085    & 0.0034	& $201.760 ~(5-37)  $	& 	\\
203.728    & $3 - 32$	&   $3s^2 3p^3 ~{}^2D_{5/2} - 3s^2 3p^2 3d ~{}^2D_{5/2}$& 0.0867    & 0.0711    & 0.0231	&          		& bl \ion{Fe}{XIII}	\\
211.732    & $2 - 28$	&   $3s^2 3p^3 ~{}^2D_{3/2} - 3s 3p^4 ~{}^2P_{1/2}     $& 0.0242    & 0.0203    & 0.0074	& $211.700 ~(40-179)$	&	\\
217.276    & $2 - 26$	&   $3s^2 3p^3 ~{}^2D_{3/2} - 3s 3p^4 ~{}^2P_{3/2}     $& 0.0345    & 0.0289    & 0.0100	&          		&	\\
219.437    & $3 - 26$	&   $3s^2 3p^3 ~{}^2D_{5/2} - 3s 3p^4 ~{}^2P_{3/2}     $& 0.0864    & 0.0724    & 0.0251	&          		&	\\
220.870    & $1 - 22$	&   $3s^2 3p^3 ~{}^4S_{3/2} - 3s^2 3p^2 3d ~{}^4D_{5/2}$& 0.0321    & 0.0273    & 0.0102	&          		& w	\\
222.306    & $1 - 18$	&   $3s^2 3p^3 ~{}^4S_{3/2} - 3s^2 3p^2 3d ~{}^2F_{5/2}$& 0.0384    & 0.0334    & 0.0133	&          		& w	\\
223.000    & $3 - 24$	&   $3s^2 3p^3 ~{}^2D_{5/2} - 3s^2 3p^2 3d ~{}^2G_{7/2}$& 0.0176    & 0.0156    & 0.0063	&          		& w	\\
249.388    & $3 - 20$	&   $3s^2 3p^3 ~{}^2D_{5/2} - 3s^2 3p^2 3d ~{}^4D_{7/2}$& 0.0739    & 0.0635    & 0.0242	& $249.384 ~(3-21)  $	&	\\
256.410    & $3 - 16$	&   $3s^2 3p^3 ~{}^2D_{5/2} - 3s^2 3p^2 3d ~{}^4F_{7/2}$& 0.0662    & 0.0572    & 0.0219	&          		& bl	\\
283.443    & $2 - 12$	&   $3s^2 3p^3 ~{}^2D_{3/2} - 3s^2 3p^2 3d ~{}^2P_{1/2}$& 0.0166    & 0.0151    & 0.0070	&          		&	\\
291.010    & $3 - 11$	&   $3s^2 3p^3 ~{}^2D_{5/2} - 3s^2 3p^2 3d ~{}^2P_{3/2}$& 0.0781    & 0.0681    & 0.0271	& $290.962 ~(32-104)$	&	\\
303.135    & $5 - 13$	&   $3s^2 3p^3 ~{}^2P_{3/2} - 3s 3p^4 ~{}^2S_{1/2}     $& 0.0096    & 0.0088    & 0.0041	&          		& bl	\\
335.380    & $2 - 9 $	&   $3s^2 3p^3 ~{}^2D_{3/2} - 3s 3p^4 ~{}^2D_{3/2}     $& 0.0458    & 0.0429    & 0.0214	&          		& bl \ion{Fe}{XVI}	\\
	   &		&							&	    &		&	&			& bl \ion{Mg}{VIII}	\\
338.263    & $3 - 10$	&   $3s^2 3p^3 ~{}^2D_{5/2} - 3s 3p^4 ~{}^2D_{5/2}     $& 0.0897    & 0.0802    & 0.0350	&          		&	\\
346.852    & $1 - 8 $	&   $3s^2 3p^3 ~{}^4S_{3/2} - 3s 3p^4 ~{}^4P_{1/2}     $& 0.1085    & 0.0930    & 0.0375	&          		&	\\
352.106    & $1 - 7 $	&   $3s^2 3p^3 ~{}^4S_{3/2} - 3s 3p^4 ~{}^4P_{3/2}     $& 0.2193    & 0.1893    & 0.0778	&          		&	\\
364.467    & $1 - 6 $	&   $3s^2 3p^3 ~{}^4S_{3/2} - 3s 3p^4 ~{}^4P_{5/2}     $& 0.3865    & 0.3327    & 0.1345	&          		&	\\
\noalign{\smallskip}
\hline
\noalign{\smallskip}
1242.01    & $1 - 5 $	&   $3s^2 3p^3 ~{}^4S_{3/2} - 3s^2 3p^3 ~{}^2P_{3/2}   $& 0.1287    & 0.1283    & 0.0855	&          		&	\\
1349.40    & $1 - 4 $	&   $3s^2 3p^3 ~{}^4S_{3/2} - 3s^2 3p^3 ~{}^2P_{1/2}   $& 0.0709    & 0.0721    & 0.0508	&          		&	\\
2169.76    & $1 - 3 $	&   $3s^2 3p^3 ~{}^4S_{3/2} - 3s^2 3p^3 ~{}^2D_{5/2}   $& 0.0782    & 0.0687    & 0.0310	&          		&	\\
2406.41    & $1 - 2 $	&   $3s^2 3p^3 ~{}^4S_{3/2} - 3s^2 3p^3 ~{}^2D_{3/2}   $& 0.4086    & 0.3880    & 0.2286	&          		&	\\
2566.77    & $2 - 5 $	&   $3s^2 3p^3 ~{}^2D_{3/2} - 3s^2 3p^3 ~{}^2P_{3/2}   $& 0.0772    & 0.0770    & 0.0513	&          		&	\\
22063.0    & $2 - 3 $	&   $3s^2 3p^3 ~{}^2D_{3/2} - 3s^2 3p^3 ~{}^2D_{5/2}   $& 0.0398    & 0.0350    & 0.0158	&          		&	\\
\noalign{\smallskip}
\hline
\hline
\end{tabular}
     $$
  \end{minipage}
\end{table*}
%
%
%
%
\begin{table*}
  \begin{minipage}[t]{\linewidth}
\centering
\renewcommand{\footnoterule}{}  
    \caption[]{Same as Table \ref{Table:fe9}, but for \ion{Fe}{XIII}.}
         \label{Table:fe13}
    \vspace{-0.8cm}
    \centering
    \renewcommand{\footnoterule}{}  
    \scriptsize
    $$
\begin{tabular}{llclllll}
\hline
\hline
\noalign{\smallskip}
$\lambda~[\AA]$ & levels & level description					& $I$(Maxw) & $I$($\kappa$=5) & $I$($\kappa$=2)	&  selfblends	 	& notes  \\
\noalign{\smallskip}
\hline
\noalign{\smallskip}
196.525    & $4 - 26$	&   $3s^2 3p^2 ~{}^1D_2 - 3s^2 3p 3d ~{}^1F_3$	& 0.0564    & 0.0479    & 0.0133	&          		&	\\
200.021    & $2 - 24$	&   $3s^2 3p^2 ~{}^3P_1 - 3s^2 3p 3d ~{}^3D_2$	& 0.1391    & 0.1142    & 0.0283	&          		&	\\
201.126    & $2 - 23$	&   $3s^2 3p^2 ~{}^3P_1 - 3s^2 3p 3d ~{}^3D_1$	& 0.2363    & 0.1885    & 0.0425	&          		& bl \ion{Fe}{XII}	\\
202.044    & $1 - 20$	&   $3s^2 3p^2 ~{}^3P_0 - 3s^2 3p 3d ~{}^3P_1$	& 0.6461    & 0.4987    & 0.0996	&          		&	\\
203.165    & $2 - 22$	&   $3s^2 3p^2 ~{}^3P_1 - 3s^2 3p 3d ~{}^3P_0$	& 0.0707    & 0.0579    & 0.0143	&          		&	\\
203.826    & $3 - 25$	&   $3s^2 3p^2 ~{}^3P_2 - 3s^2 3p 3d ~{}^3D_3$	& 0.6762    & 0.5533    & 0.1361	& $203.772 ~(7-60) $	& bl \ion{Fe}{XII}	\\
	   &		&						&	    &		&		& $203.795 ~(3-24) $	&	\\
	   &		&						&	    &		&		& $203.835 ~(8-60) $	&	\\
204.262    & $2 - 21$	&   $3s^2 3p^2 ~{}^3P_1 - 3s^2 3p 3d ~{}^1D_2$	& 0.0586    & 0.0485    & 0.0124	&          		&	\\
204.942    & $3 - 23$	&   $3s^2 3p^2 ~{}^3P_2 - 3s^2 3p 3d ~{}^3D_1$	& 0.0700    & 0.0558    & 0.0126	&          		&	\\
208.667    & $5 - 27$	&   $3s^2 3p^2 ~{}^1S_0 - 3s^2 3p 3d ~{}^1P_1$	& 0.0228    & 0.0185    & 0.0044	& $208.716 ~(11-67)$	&	\\
	   &		&						&	    &		&		& $208.754 ~(9-57) $	&	\\
209.619    & $2 - 19$	&   $3s^2 3p^2 ~{}^3P_1 - 3s^2 3p 3d ~{}^3P_2$	& 0.0970    & 0.0801    & 0.0203	& $209.654 ~(6-43) $	&	\\
209.916    & $3 - 20$	&   $3s^2 3p^2 ~{}^3P_2 - 3s^2 3p 3d ~{}^3P_1$	& 0.1159    & 0.0895    & 0.0179	&          		&	\\
213.768    & $3 - 19$	&   $3s^2 3p^2 ~{}^3P_2 - 3s^2 3p 3d ~{}^3P_2$	& 0.0958    & 0.0792    & 0.0201	&          		&	\\
216.835    & $4 - 24$	&   $3s^2 3p^2 ~{}^1D_2 - 3s^2 3p 3d ~{}^3D_2$	& 0.0457    & 0.0375    & 0.0093	& $216.870 ~(4-25) $	&	\\
221.828    & $4 - 21$	&   $3s^2 3p^2 ~{}^1D_2 - 3s^2 3p 3d ~{}^1D_2$	& 0.1217    & 0.1006    & 0.0256	& $221.824 ~(6-20) $	&	\\
228.160    & $4 - 19$	&   $3s^2 3p^2 ~{}^1D_2 - 3s^2 3p 3d ~{}^3P_2$	& 0.0836    & 0.0691    & 0.0176	&          		&	\\
239.030    & $3 - 16$	&   $3s^2 3p^2 ~{}^3P_2 - 3s^2 3p 3d ~{}^3F_3$	& 0.0785    & 0.0658    & 0.0173	&          		&	\\
240.696    & $1 - 14$	&   $3s^2 3p^2 ~{}^3P_0 - 3s 3p^3 ~{}^3S_1   $	& 0.0744    & 0.0609    & 0.0151	&          		&	\\
246.209    & $2 - 14$	&   $3s^2 3p^2 ~{}^3P_1 - 3s 3p^3 ~{}^3S_1   $	& 0.1856    & 0.1519    & 0.0376	& $246.241 ~(16-76) $	& bl \ion{Si}{VI}	\\
251.952    & $3 - 14$	&   $3s^2 3p^2 ~{}^3P_2 - 3s 3p^3 ~{}^3S_1   $	& 0.3557    & 0.2912    & 0.0721	&          		&	\\
256.400    & $4 - 17$	&   $3s^2 3p^2 ~{}^1D_2 - 3s 3p^3 ~{}^1P_1   $	& 0.0694    & 0.0581    & 0.0155	&          		& bl	\\
261.743    & $4 - 15$	&   $3s^2 3p^2 ~{}^1D_2 - 3s^2 3p 3d ~{}^3F_2$ 	& 0.0393    & 0.0325    & 0.0083	&          		&	\\
303.364    & $1 - 11$	&   $s^2 3p^2 ~{}^3P_0 - 3s 3p^3 ~{}^3P_1    $	& 0.0624    & 0.0521    & 0.0140	&          		& bl	\\
312.174    & $2 - 11$	&   $3s^2 3p^2 ~{}^3P_1 - 3s 3p^3 ~{}^3P_1   $	& 0.0912    & 0.0761    & 0.0205	&          		&	\\
312.868    & $2 - 10$	&   $3s^2 3p^2 ~{}^3P_1 - 3s 3p^3 ~{}^3P_0   $	& 0.0446    & 0.0386    & 0.0117	&          		&	\\
318.130    & $4 - 13$	&   $3s^2 3p^2 ~{}^1D_2 - 3s 3p^3 ~{}^1D_2   $	& 0.0621    & 0.0561    & 0.0188	&          		& bl?	\\
320.800    & $3 - 12$	&   $3s^2 3p^2 ~{}^3P_2 - 3s 3p^3 ~{}^3P_2   $	& 0.1566    & 0.1361    & 0.0415	&          		&	\\
348.183    & $1 - 7 $	&   $s^2 3p^2 ~{}^3P_0 - 3s 3p^3 ~{}^3D_1    $	& 0.2185    & 0.1793    & 0.0465	&          		&	\\
359.644    & $2 - 8 $	&   $3s^2 3p^2 ~{}^3P_1 - 3s 3p^3 ~{}^3D_2   $	& 0.2180    & 0.1886    & 0.0580	&          		&	\\
368.164    & $3 - 9 $	&   $3s^2 3p^2 ~{}^3P_2 - 3s 3p^3 ~{}^3D_3   $	& 0.2145    & 0.1877    & 0.0595	&          		& bl \ion{Mg}{IX}	\\
487.042    & $2 - 6 $	&   $3s^2 3p^2 ~{}^3P_1 - 3s 3p^3 ~{}^5S_2   $	& 0.0286    & 0.0274    & 0.0115	&          		&	\\
510.042    & $3 - 6 $	&   $3s^2 3p^2 ~{}^3P_2 - 3s 3p^3 ~{}^5S_2   $	& 0.0486    & 0.0466    & 0.0195	&          		&	\\
\noalign{\smallskip}
\hline
\noalign{\smallskip}
2579.54    & $2 - 4 $	&   $3s^2 3p^2 ~{}^3P_1 - 3s^2 3p^2 ~{}^1D_2 $	& 0.2933    & 0.2719    & 0.1157	&          		&	\\
3388.91    & $3 - 4 $	&   $3s^2 3p^2 ~{}^3P_2 - 3s^2 3p^2 ~{}^1D_2 $	& 0.3585    & 0.3323    & 0.1414	&          		&	\\
10749.0    & $1 - 2 $	&   $3s^2 3p^2 ~{}^3P_0 - 3s^2 3p^2 ~{}^3P_1 $	& 1.0000    & 0.8658    & 0.2871	&          		&	\\
10801.0    & $2 - 3 $	&   $3s^2 3p^2 ~{}^3P_1 - 3s^2 3p^2 ~{}^3P_2 $	& 0.7816    & 0.6789    & 0.2282	&          		&	\\
\noalign{\smallskip}
\hline
\hline
\end{tabular}
     $$
  \end{minipage}
\end{table*}
%
%
%
\end{appendix}

\end{document}